\documentclass[11pt,english,a4paper]{article}
\pdfoutput=1
\usepackage{amsfonts} 
\usepackage{amsmath}
\usepackage{amsthm}
\usepackage{graphicx}
\usepackage{latexsym}
\usepackage{bbm}
\usepackage[british]{babel}
\usepackage[margin=2.75cm]{geometry}
\usepackage{fancyhdr}
\usepackage{lastpage}
\usepackage{youngtab}
\usepackage{url}
\usepackage{subfigure}

\usepackage{color}
\definecolor{lightblue}{rgb}{0,0.25,0.75}
\definecolor{dullblue}{rgb}{0.15,0.25,0.45}
\definecolor{darkblue}{rgb}{0,0,0.55}
\definecolor{bluegreen}{rgb}{0,0.3,0.35}
\definecolor{darkgreen}{rgb}{0,0.35,0}
\definecolor{darkred}{rgb}{0.5,0,0}

\usepackage[linktocpage=true]{hyperref}
\hypersetup{
colorlinks=true,
citecolor=darkblue,
linkcolor=black,
urlcolor=darkblue,
pdfauthor={},
pdftitle={},
breaklinks=true
}

\newcommand{\p}{\partial}
\renewenvironment{abstract}{\begin{quote}
{\bf \textsf{Abstract:}} \\ \vspace{-0.1cm} \\ 
}{ \end{quote} \vspace{5mm} \thispagestyle{empty} }

\begin{document}

\begin{flushright} 
\small{DMUS-MP-14/06}
\end{flushright}

\begin{center}

{\bf \Large \textsf{D1 and D5-brane giant gravitons on $AdS_{3} \times S^{3} \times S^{3} \times S^{1}$}}

\vspace{0.5cm}

{\large Andrea Prinsloo} 

\vspace{0.2cm}

\small{\it Department of Mathematics, \\
University of Surrey, \\
Guildford, GU2 7XH, United Kingdom} 

\vspace{-0.1cm}
\small{ \texttt{a.prinsloo@surrey.ac.uk} }

\vspace{1.5cm}

\end{center}

\begin{abstract}
We construct various examples of $\tfrac{1}{4}$-BPS giant gravitons embedded into the type IIB supergravity background 
$AdS_{3}\times S_{+}^{3} \times S_{-}^{3} \times S^{1}$ with pure R-R flux: two D1-brane giants wrapping 1-cycles in $AdS_{3}$ and $S^{3}_{+}\times S^{3}_{-}$, and one D5-brane giant wrapping a 4-cycle in $S^{3}_{+} \times S^{3}_{-}$ and the $S^{1}$.  These D-branes are 
supported by angular momenta $\alpha \, P$ on one 3-sphere and $(1-\alpha) \, P$ on the other.  We then construct a general class of $\tfrac{1}{8}$-BPS D5-brane giant gravitons wrapping 4-cycles $\Sigma$ in $S_{+}^{3} \times S_{-}^{3}$ and the $S^{1}$.  Here $\Sigma$ is the intersection of a holomorphic surface $\mathcal{C}$ in $\mathbb{C}^{2}_{+} \times \mathbb{C}^{2}_{-}$ with the $S^{3}_{+} \times S^{3}_{-}$ submanifold. The
holomorphic surface $\mathcal{C}$ is defined by $f(y_{1}z_{1},y_{1}z_{2},y_{2}z_{1},y_{2}z_{2})=0$, with $y_{a}$ and $z_{a}$ the $\mathbb{C}^{2}_{\pm}$ complex coordinates. There is supersymmetry enhancement to $\tfrac{1}{4}$-BPS in the special case $f(y_{1}z_{1})=0$ of which our original D5-brane giant graviton is an example.  
\end{abstract} 

\newpage
\vspace{0.75cm}

\pagenumbering{roman}
\tableofcontents

\pagestyle{fancyplain}

\renewcommand{\headrulewidth}{0pt}

\lhead{}
\chead{}
\rhead{}
\lfoot{}
\cfoot{\thepage}
\rfoot{}

\vspace{0.75cm}

\pagenumbering{arabic}

\newpage

%%%%%%%%%%%%%%%%%%%%%%%%%%%%%%%%%%%%%%%%%%%%%%%%%%%%%%%%%%%%%%%%%%%%%%%%%%%%%%%%%%%%%%%%%%%%%%
\section{Introduction} \label{section-introduction}
%%%%%%%%%%%%%%%%%%%%%%%%%%%%%%%%%%%%%%%%%%%%%%%%%%%%%%%%%%%%%%%%%%%%%%%%%%%%%%%%%%%%%%%%%%%%%%

AdS/CFT correspondences provide a non-perturbative reformulation of string theory on anti-de Sitter spacetimes in terms of a large $N$ gauge theory.  The original AdS$_{5}$/CFT$_{4}$ correspondence \cite{Maldacena:1997} between type IIB superstring theory on $AdS_{5}\times S^{5}$ and $\mathcal{N}=4$ Super Yang-Mills (SYM) theory has been the subject of extensive study. Among the many remarkable features uncovered is an integrable structure which gives rise to an infinite tower of conserved charges associated with closed IIB superstrings and their dual single trace operators \cite{Minahan:2003,Beisert:2010,Frolov:2009}.  Integrable open IIB superstrings require integrability preserving boundary conditions which are provided by certain supersymmetric D-branes, such as maximal giant gravitons, embedded into $AdS_{5}\times S^{5}$ \cite{Mann:2006,Dekel:2011,Zoubos:2012}.  Giant gravitons are a special class of supersymmetric probe D-branes, which are dynamically stable as a result of their angular momentum and their coupling to the supergravity background R-R potentials. AdS and sphere giant gravitons on $AdS_{5}\times S^{5}$ were originally constructed in \cite{McGreevy:2000,Grisaru:2000,Hashimoto:2000} and shown to be $\tfrac{1}{2}$-BPS.  These D3-brane giants have a microscopic description in terms of gravitational waves \cite{Janssen1:2003,Janssen2:2003}. They are dual to Schur polynomial operators $\chi_{R}(Z_{1})$ in $\mathcal{N}$=4 SYM theory \cite{Balasubramanian:2001,Corley:2001}, built from $n \sim O(N)$ of one of the three scalar fields $Z_{a}$ which transform in the adjoint representation of the $SU(N)$ gauge group.  A general class of $\tfrac{1}{8}$-BPS D3-brane giant gravitons embedded into and moving on $S^{5}$ was  constructed in \cite{Mikhailov:2000} from holomorphic surfaces $f(z_{1},z_{2},z_{3})=0$ in the complex manifold $\mathbb{C}^{3}$. These are rigidly rotating 3-manifolds with non-trivial topology \cite{AMPR}.  The dual operators are not known, but  progress has been made \cite{Kimura:2010,RdMK:2011,Ramgoolam:2011} in studying $\tfrac{1}{4}$-BPS and $\tfrac{1}{8}$-BPS operators in $\mathcal{N}=4$ SYM theory.

There has been considerable recent interest \cite{Babichenko:2010,Sax-Stefanski:2011,Borsato-et-al:2012,Sax-et-al:2013} in integrable structures in the AdS$_{3}$/CFT$_{2}$ correspondence\footnote{For a detailed recent review, see \cite{Sfondrini:2014}.} between type IIB superstring theory on $AdS_{3}\times S^{3}_{+} \times S^{3}_{-} \times S^{1}$ and a 2-dimensional $\mathcal{N}=(4,4)$ superconformal field theory with two $SU(2)_{\pm}$ $\mathcal{R}$-symmetry groups, which remains imperfectly understood.  
$AdS_{3} \times S^{3}_{+} \times S^{3}_{-} \times S^{1}$ is a $\tfrac{1}{2}$-BPS type IIB supergravity background containing a free parameter $\alpha$, which controls the relative size of the two 3-spheres. The size of one or other of these 3-spheres goes to infinity in the $\alpha \rightarrow 0$ and $\alpha \rightarrow 1$ limits, in which the string theory (after a compactification) becomes type IIB superstring theory on $AdS_{3}\times S^{3}\times T^{4}$.
This $\tfrac{1}{2}$-BPS type IIB supergravity background $AdS_{3}\times S^{3}\times T^{4}$ arises as the near-horizon geometry of a stack of $Q_{1}$ D1-branes coincident with a stack of $Q_{5}$ D5-branes \cite{Cowdall:1998}. The worldvolume gauge theory on this stack of D1-branes is a 2-dimensional $\mathcal{N}=(4,4)$ supersymmetric theory with one $SU(2)$ $\mathcal{R}$-symmetry group which is thought to flow to a $\mathrm{Sym}^{Q_{1}Q_{5}}(T^{4})$ orbifold conformal field theory in the infrared \cite{Strominger:2004,Seiberg:1999,Pakman1:2009,Pakman2:2009}.  This CFT$_{2}$ is conjectured to be the holographic dual of type IIB superstring theory on $AdS_{3}\times S^{3}\times T^{4}$. This D1-D5 brane system can be generalized to a D1-D5-D5$'$ system consisting of  
a stack of $Q_{1}$ D1-branes coincident with two orthogonal stacks of $Q^{+}_{5}$ and $Q^{-}_{5}$ D5-branes, which intersect only along the line of the D1s.
The near-horizon geometry is $AdS_{3} \times S^{3}_{+} \times S^{3}_{-} \times \mathbb{R}$ \cite{Cowdall:1998,deBoer:1999,Donos:2009}.  The worldvolume gauge theory on the D1-branes is a $\mathcal{N}=(0,4)$ supersymmetric field theory \cite{Tong:2014} containing two pairs of complex scalar fields, $Y_{a}$ and $Z_{a}$, charged under different $SU(2)_{\pm}$ $\mathcal{R}$-symmetry groups and transforming in the adjoint of the $U(Q_{1})$ gauge group.  These scalars are associated with the directions transverse to the two stacks of D5 and D5$'$-branes. The $\mathbb{R}$ instead of $S^{1}$ factor in the near-horizon geometry was suggested in \cite{Tong:2014} to be an artefact of the D1-branes being smeared over the transverse directions, and it 
was conjectured that this worldvolume gauge theory flows in the infrared to a CFT$_{2}$ with enhanced $\mathcal{N}=(4,4)$ supersymmetry which is the holographic dual of type IIB superstring theory on $AdS_{3} \times S^{3}_{+} \times S^{3}_{-} \times S^{1}$.

The goal of this work is to study D1 and D5-brane giant gravitons on $AdS_{3} \times S_{+}^{3} \times S_{-}^{3} \times S^{1}$.  Our motivation is two-fold:
While the integrability of closed superstrings on this background has been investigated extensively, open string integrability has not been considered, perhaps for lack of suitable D-brane boundary conditions.  Maximal giant gravitons are good candidates for integrable boundary conditions for open IIB superstrings on $AdS_{3} \times S_{+}^{3} \times S_{-}^{3} \times S^{1}$.  A better understanding of giant gravitons on 
$AdS_{3} \times S_{+}^{3} \times S_{-}^{3} \times S^{1}$ is also likely to provide insight into the dual long operators  in the unknown $\mathcal{N}=(4,4)$ supersymmetric CFT$_{2}$ - which, being protected, it should theoretically be possible to build from the scalars $Y_{a}$ and $Z_{a}$ in the $\mathcal{N}=(0,4)$ supersymmetric gauge theory of \cite{Tong:2014}.

Giant gravitons on $AdS_{3}\times S^{3}\times T^{4}$ were studied in 
\cite{McGreevy:2000,Grisaru:2000,Hashimoto:2000,Janssen:2005,Raju-et-al:2008} and several oddities observed.  Both D1 and D5-brane giant gravitons exist, wrapping a 1-cycle in $AdS_{3} \times S^{3}$, with the D1-brane point-like in the $T^{4}$ and the D5-branes wrapping the entire $T^{4}$ space.  The AdS and sphere giant gravitons wrap a circle in the $AdS_{3}$ and $S^{3}$, respectively, with angular momentum on the $S^{3}$ in both cases.  These $\tfrac{1}{2}$-BPS giants have a flat potential with their energy independent of the size of the circle.  These D1 and D5-branes may be thought of as having separated, at no cost in energy, from the D1-D5-branes setting up the geometry.  The microscopic description of giant gravitons on $AdS_{3}\times S^{3}\times T^{4}$ in terms of gravitational waves was found in \cite{Janssen:2005}.  A general class of $\tfrac{1}{4}$-BPS giant gravitons wrapping 1-cycles in $AdS_{3}\times S^{3}$ was shown to exist in \cite{Raju-et-al:2008}. 
We expect D1 and D5-brane giant gravitons on $AdS_{3} \times S_{+}^{3} \times S_{-}^{3} \times S^{1}$ to  interpolate between those on $AdS_{3} \times S_{-}^{3} \times T^{4}$ and $AdS_{3} \times S_{+}^{3} \times T^{4}$ 
in the $\alpha \rightarrow 0$ and $\alpha \rightarrow 1$ limits.  We thus expect these giants to carry angular momentum on both 3-spheres for intermediate values of  $\alpha$.  Indeed, we will show that the angular momenta on $S^{3}_{+}$ and $S^{3}_{-}$ are
$\alpha \, P$ and $(1-\alpha) \, P$, respectively, with $P$ the total angular momentum.  

We begin, in section \ref{section - background}, with a brief description of the type IIB supergravity background $AdS_{3} \times S_{+}^{3} \times S_{-}^{3} \times S^{1}$  with pure R-R flux.
In section \ref{section - examples}, we construct two examples of D1-brane giant gravitons wrapping 1-cycles in $AdS_{3}$ and $S^{3}_{+} \times S^{3}_{-}$, and one example of a D5-brane giant graviton wrapping a 4-cycle in $S^{3}_{+} \times S^{3}_{-}$ and the $S^{1}$.  These will be shown to be  $\tfrac{1}{4}$-BPS D-branes by a study of the kappa symmetry conditions for worldvolume supersymmetry.  In section \ref{section - holomorphic surfaces} we turn our attention to the construction of a more general class of $\tfrac{1}{8}$-BPS D5-brane giant gravitons wrapping a 4-cycle $\Sigma$ in $S^{3}_{+}\times S^{3}_{-}$ and the $S^{1}$.  Here $\Sigma$ is the intesection of a holomorphic surface $\mathcal{C}$ in the complex manifold 
$\mathbb{C}^{2}_{+} \times \mathbb{C}^{2}_{-}$ with the $S^{3}_{+}\times S^{3}_{-}$ submanifold. Crucial to this holomorphic surface construction is our choice of holomorphic surface $\mathcal{C}$  to be
\begin{equation} \hspace{-1.0cm} \nonumber
\mathcal{C}: \hspace{0.2cm} f(y_{1}z_{1},y_{1}z_{2},y_{2}z_{1},y_{2}z_{2}) = 0
\end{equation}
with $y_{a}$ and $z_{a}$ the $\mathbb{C}^{2}_{+}$ and $\mathbb{C}^{2}_{-}$ complex coordinates\footnote{We note that this holomorphic surface $\mathcal{C}$ resembles the holomorphic surface in $\mathbb{C}^{4}$ used in \cite{LMP:2013} to construct D4-brane giant gravitons embedded into $AdS_{4}\times \mathbb{CP}^{3}$, which wrap a 4-cycle in the complex projective space.}.  Key also is our $\alpha$-dependent choice of the preferred direction $\mathbf{e}^{\parallel}$ in $\mathrm{T}(S^{3}_{+}\times S^{3}_{-})$ along which $\Sigma$ is boosted into motion. A discussion of our results is presented in section \ref{section - discussion}. Technical details of the type IIB supergravity background and D-brane supersymmetry analyses are included in appendices \ref{appendix - KSEs} and \ref{appendix - kappa symmetry}.

%%%%%%%%%%%%%%%%%%%%%%%%%%%%%%%%%%%%%%%%%%%%%%%%%%%%%%%%%%%%%%%%%%%%%%%%%%%%%%%%%%%%%%%%%%%%%%
\section{$AdS_{3} \times S_{+}^{3} \times S_{-}^{3} \times S^{1}$ background with pure R-R flux}  \label{section - background}
%%%%%%%%%%%%%%%%%%%%%%%%%%%%%%%%%%%%%%%%%%%%%%%%%%%%%%%%%%%%%%%%%%%%%%%%%%%%%%%%%%%%%%%%%%%%%%

The $\tfrac{1}{2}$-BPS type IIB supergravity background $AdS_{3} \times S_{+}^{3} \times S_{-}^{3} \times S^{1}$ with 3-form fluxes $F_{(3)}$ and $H_{(3)}$, which mix under S-duality, preserves 16 of 32 supersymmetries (see appendix \ref{subappendix - KSEs - 10D}).
We shall focus here on the case of pure R-R flux $F_{(3)}$.

%---------------------------------------------------------------------------------------------
\subsection{Type IIB supergravity background $AdS_{3} \times S_{+}^{3} \times S_{-}^{3} \times S^{1}$} \label{subsection - background - 10D}
%---------------------------------------------------------------------------------------------

Here we summarize the relevant details of the $AdS_{3} \times S_{+}^{3} \times S_{-}^{3} \times S^{1}$ background with pure R-R flux. The metric is given by
\begin{eqnarray} \label{metric}
\nonumber && \hspace{-0.75cm} ds^{2} = 
L^{2} \left( - \cosh^{2}{\rho} \hspace{0.1cm} dt^{2} + d\rho^{2} + \sinh^{2}{\rho} \hspace{0.1cm} d\varphi^{2} \right)
+  L^{2} \sec^{2}{\beta}  \left( d\theta_{+}^{2} + \cos^{2}{\theta_{+}} \, d\chi_{+}^{2} + \sin^{2}{\theta_{+}} \, d\phi_{+}^{2} \right) \\
&& \hspace{-0.75cm} \hspace{0.7cm}  + \hspace{0.05cm} L^{2} \csc^{2}{\beta} \left( d\theta_{-}^{2} + \cos^{2}{\theta_{-}} \, d\chi_{-}^{2} + \sin^{2}{\theta_{-}} \, d\phi_{-}^{2} \right) + \ell^{2} \hspace{0.1cm} d\xi^{2},
\end{eqnarray}
The axion and dilaton vanish, $C_{(0)} = \Phi = 0$, and we set to zero the self-dual 5-form flux:
\begin{equation} \hspace{-0.5cm}
\nonumber \tilde{F}_{(5)} \equiv F_{(5)} - \tfrac{1}{2} C_{(2)} \wedge H_{(3)} + \tfrac{1}{2} B_{(2)} \wedge F_{(3)} 
= d \left[C_{(4)} - \tfrac{1}{2} C_{(2)} \wedge B_{(2)} \right] = \ast \tilde{F}_{(5)} = 0.
\end{equation}
We take the NS-NS 3-form flux to vanish, $H_{(3)} = dB_{(2)} = 0$, to obtain the background with pure R-R 3-form flux $F_{(3)} = dC_{(2)}$ given by
\begin{eqnarray}  \label{R3}
&& \hspace{-0.65cm} F_{(3)} = 2 L^{2} \hspace{0.15cm} dt \wedge (\sinh{\rho} \cosh{\rho} \hspace{0.1cm} d\rho) \wedge d\varphi \\
\nonumber && \hspace{-0.65cm} \hspace{0.8cm} + \hspace{0.075cm} 
2L^{2} \sec^{2}{\beta} \, (\sin{\theta_{+}} \cos{\theta_{+}} \, d\theta_{+}) \wedge d\chi_{+} \wedge d\phi_{+}
+ 2L^{2}\csc^{2}{\beta} \, (\sin{\theta_{-}} \cos{\theta_{-}} \, d\theta_{-}) \wedge d\chi_{-} \wedge d\phi_{-},
\end{eqnarray}
with Hodge dual 7-form field strength $F_{(7)} = dC_{(6)} = \ast \, F_{(3)}$ computed to be
\begin{eqnarray} \label{R7}
\nonumber && \hspace{-0.65cm} F_{(7)}  = - \,2L^{5} \ell \, \sec^{3}{\beta} \, \csc^{3}{\beta} \hspace{0.1cm} (\sin{\theta_{+}} \cos{\theta_{+}} \hspace{0.1cm} d\theta_{+}) \wedge d\chi_{+} \wedge d\phi_{+} \wedge   (\cos{\theta_{-}} \sin{\theta_{-}} \hspace{0.1cm}  d\theta_{-}) \wedge d\chi_{-} \wedge d\phi_{-} \wedge d\xi \\
\nonumber && \hspace{-0.65cm} \hspace{1.16cm} - \, 2L^{5} \ell \, \cos{\beta} \, \csc^{3}{\beta} \hspace{0.15cm} 
dt \wedge  (\sinh{\rho} \, \cosh{\rho} \hspace{0.1cm} d\rho) \wedge d\varphi \wedge  (\sin{\theta_{-}} \cos{\theta_{-}} \hspace{0.1cm} d\theta_{-}) \wedge d\chi_{-} \wedge d\phi_{-}\wedge d\xi   \\
&& \hspace{-0.65cm} \hspace{1.16cm} + \, 2L^{5} \ell \, \sin{\beta} \, \sec^{3}{\beta} \hspace{0.15cm} 
dt \wedge (\sinh{\rho} \, \cosh{\rho} \hspace{0.1cm}  d\rho) \wedge d\varphi \wedge  (\sin{\theta_{+}} \cos{\theta_{+}}  \hspace{0.1cm} d\theta_{+}) \wedge d\chi_{+} \wedge d\phi_{+}\wedge d\xi.
\end{eqnarray}
We must insist that the charges\footnote{Here we work in units of $\ell_{s}=1$.}
\begin{eqnarray} 
&& \hspace{-0.65cm} Q_{5}^{+} = \frac{1}{(2\pi)^{2}} \int_{S^{3}_{+}}{F_{(3)}} =  L^{2} \, \sec^{2}{\beta} = \frac{\, L^{2}}{\alpha} \hspace{1.0cm}
Q_{5}^{-} = \frac{1}{(2\pi)^{2}} \int_{S^{3}_{-}}{F_{(3)}} = L^{2} \, \csc^{2}{\beta} = \frac{L^{2}}{(1-\alpha)} \hspace{0.75cm} \label{charges+-} \\
&& \hspace{-0.65cm}
Q_{1} = \frac{1}{(2\pi)^{6}} \, \int_{S^{3}_{+}\times S^{3}_{-} \times S^{1}}{F_{(7)}} = \frac{ L^{5} \ell}{4 \pi} \hspace{0.1cm} \sec^{3}{\beta} \, \csc^{3}{\beta} \label{chargeQ1}
\end{eqnarray}
are all integers \cite{Strominger:2004,Donos:2009}.  We may then express the radii in the metric through these charges:
\begin{eqnarray} \nonumber \hspace{-0.2cm}
L^{2} = \frac{Q_{5}^{+} Q_{5}^{-}}{(Q_{5}^{+} + Q_{5}^{-})} = \hspace{0.15cm} Q_{5}^{+}  \hspace{0.05cm} \cos^{2}{\beta}
= Q_{5}^{-} \hspace{0.05cm} \sin^{2}{\beta} \hspace{1.2cm}
\ell^{2} = \frac{16 \pi^{2}  \, \, Q_{1}^{2}}{(Q_{5}^{+})^{2} (Q_{5}^{-})^{2} (Q_{5}^{+} + Q_{5}^{-})}. 
\end{eqnarray}
The parameter $\alpha \equiv \cos^{2}{\beta}$ controls the relative size of the two 3-spheres $S^{3}_{\pm}$.  Taking $\alpha \rightarrow 1-\alpha$ and interchanging the  $S_{+}^{3}$ and $S_{-}^{3}$ leaves this type IIB supergravity background invariant.  
\vspace{-0.1cm} 
 
%---------------------------------------------------------------------------------------------
\subsection{Embedding $AdS_{3} \times S_{+}^{3} \times S_{-}^{3} \times S^{1}$ into $\mathbb{R}^{2+2} \times \mathbb{C}_{+}^{2} \times \mathbb{C}_{-}^{2} \times S^{1}$} \label{subsection - background - 13D}
%---------------------------------------------------------------------------------------------

It will prove useful in subsequent sections to embed this 1+9 dimensional $AdS_{3}\times S_{+}^{3} \times S_{-}^{3} \times S^{1}$ supergravity geometry  into a 2+11 dimensional $\mathbb{R}^{2+2} \times \mathbb{C}_{+}^{2} \times \mathbb{C}_{-}^{2} \times S^{1}$ spacetime.  The complex $\mathbb{C}_{\pm}^{2}$ coordinates are denoted $y_{a}$ and $z_{a}$ in terms of which the
$\mathbb{R}^{2+2} \times \mathbb{C}_{+}^{2} \times \mathbb{C}_{-}^{2} \times S^{1}$ metric is
\begin{eqnarray}  \label{metric-flat-complex-coords}
&& \hspace{-1.0cm} ds^{2} = \eta_{ij} \hspace{0.15cm} dx^{i} dx^{j} + \sec^{2}{\beta} \hspace{0.15cm} dy^{a} d\bar{y}_{a} + \csc^{2}{\beta} \hspace{0.15cm} dz^{a} d\bar{z}_{a} + \ell^{2} \hspace{0.05cm} d\xi^{2},
\end{eqnarray}
with all dependence on $\alpha \equiv \cos^{2}{\beta}$ shown explicitly for convenience.

Let us exchange the $x^{i}$ coordinates of $\mathbb{R}^{2+2}$ for the usual global coordinates $(t,\rho,\varphi)$ of $AdS_{3}$ and an extra radial coordinate $\hat{R}$.  We use the radii and phases of the $\mathbb{C}^{2}_{+}$ and $\mathbb{C}^{2}_{-}$ complex coordinates $y_{a} = r_{y,a} \, e^{i\psi_{y,a}}$ and $z_{a} = r_{z,a} \, e^{i\psi_{z,a}}$.  The metric of $\mathbb{R}^{2+2} \times \mathbb{C}_{+}^{2} \times \mathbb{C}_{-}^{2} \times S^{1}$ is then
\begin{eqnarray} \label{metric-flat-radii-and-phases}
\nonumber && \hspace{-1.0cm} ds^{2} =  \hat{R}^{2} \left(  - \, \cosh^{2}{\rho} \hspace{0.1cm} dt^{2} + d\rho^{2} + \sinh^{2}{\rho} \hspace{0.1cm} d\varphi^{2} \right)  - d\hat{R}^{2} \\
&& \hspace{-1.0cm} \hspace{0.65cm} + \, \sec^{2}{\beta} \hspace{0.1cm} \sum_{a} \hspace{0.05cm} \left\{ dr_{y,a}^{2} + r_{y,a}^{2} \hspace{0.1cm} d\psi_{y,a}^{2} \right\} 
+ \, \csc^{2}{\beta} \hspace{0.1cm}  \sum_{a} \hspace{0.05cm}  \left\{ dr_{z,a}^{2} + r_{z,a}^{2} \hspace{0.1cm} d\psi_{z,a}^{2} \right\} 
+ \ell^{2} \hspace{0.05cm} d\xi^{2}. \hspace{0.5cm}
\end{eqnarray}
The complex coordinates of $\mathbb{C}_{+}^{2}\times \mathbb{C}_{-}^{2}$ can now also be parameterized by
\begin{eqnarray} \label{complex-coords}
\nonumber && \hspace{-0.5cm} y_{1} = R_{+} \cos{\theta_{+}} \hspace{0.15cm} e^{i \, \chi_{+}} \hspace{1.2cm} 
z_{1} = R_{-} \cos{\theta_{-}} \hspace{0.15cm} e^{i \, \chi_{-}} \hspace{1.0cm} \\
\nonumber && \hspace{-0.5cm} y_{2} = R_{+} \sin{\theta_{+}} \hspace{0.15cm} e^{i \, \phi_{+}} \hspace{1.26cm} 
z_{2} = R_{-} \sin{\theta_{-}} \hspace{0.15cm} e^{i \, \phi_{-}} \hspace{1.0cm}
\end{eqnarray}
in terms of which the $\mathbb{R}^{2+2} \times \mathbb{C}_{+}^{2} \times \mathbb{C}_{-}^{2} \times S^{1}$ metric becomes
\begin{eqnarray} \label{metric-flat}
\nonumber && \hspace{-0.5cm} ds^{2} = \hat{R}^{2} \left(  - \, \cosh^{2}{\rho} \, dt^{2} + d\rho^{2} + \sinh^{2}{\rho} \, d\varphi^{2} \right)  - d\hat{R}^{2} \\
\nonumber && \hspace{-0.5cm} \hspace{0.65cm} + \, \sec^{2}{\beta} \left\{ dR_{+}^{2} + R_{+}^{2} \left(d\theta_{+}^{2} + \cos^{2} {\theta_{+}} \, d\chi_{+}^{2} + \sin^{2}{\theta_{+}} \, d\phi_{+}^{2} \right) \right\} \hspace{1.0cm} \\
&& \hspace{-0.5cm} \hspace{0.65cm} + \, \csc^{2}{\beta} \left\{ dR_{-}^{2} + R_{-}^{2} \left(d\theta_{-}^{2} + \cos^{2}{\theta_{-}} \, d\chi_{-}^{2} + \sin^{2}{\theta_{-}} \, d\phi_{-}^{2} \right) \right\} +  \ell^{2} \, d\xi^{2}.
\hspace{0.75cm}
\end{eqnarray} 
This reduces to the metric (\ref{metric}) of $AdS_{3} \times S_{+}^{3} \times S_{-}^{3} \times S^{1}$ when $\hat{R} = R_{+} = R_{-} \equiv L$ are constant. 

We define new radial coordinates $(R,\tilde{R})$ mixing the overall radial coordinates $R_{\pm}$ of the $\mathbb{C}^{2}_{\pm}$ complex manifolds:
\begin{eqnarray}
\nonumber &&  \hspace{-1.0cm} R_{+} = (\cos^{2}{\beta}) \, R - \tilde{R} \hspace{0.75cm} \Longrightarrow \hspace{0.5cm}   R = R_{+} + R_{-} \hspace{3.0cm} \\
&&  \hspace{-1.0cm} R_{-} = (\sin^{2}{\beta}) \, R + \tilde{R}  \hspace{2.15cm} \tilde{R} = -(\sin^{2}{\beta}) \, R_{+} + (\cos^{2}{\beta}) \, R_{-} 
\end{eqnarray}
which we shall need in section \ref{section - holomorphic surfaces}. Note that these new mixed radial coordinates are orthogonal:
\begin{equation} \hspace{-0.5cm}
\nonumber
\sec^{2}{\beta} \hspace{0.15cm} dR_{+}^{2} + \csc^{2}{\beta} \hspace{0.15cm} dR_{-}^{2} = dR^{2} + \sec^{2}{\beta} \csc^{2}{\beta} \hspace{0.15cm} d\tilde{R}^{2}. 
\end{equation}
We also define the mixed $S_{\pm}^{3}$ phases $(\chi, \tilde{\chi})$ and $(\phi,\tilde{\phi})$ in a similar way:
\begin{eqnarray} \label{mixed-phases-chi}
\nonumber &&  \hspace{-1.0cm} \chi_{+} = (\cos^{2}{\beta}) \, \chi - \tilde{\chi} \hspace{0.75cm} \Longrightarrow \hspace{0.5cm}  \chi = \chi_{+} + \chi_{-} \hspace{3.0cm} \\
&&  \hspace{-1.0cm} \chi_{-} = (\sin^{2}{\beta}) \, \chi + \tilde{\chi}  \hspace{2.15cm} \tilde{\chi} = -(\sin^{2}{\beta}) \, \chi_{+} + (\cos^{2}{\beta}) \, \chi_{-} 
\end{eqnarray}
\vspace{-0.75cm}
\begin{eqnarray} \label{mixed-phases-phi}
\nonumber &&  \hspace{-1.0cm} \phi_{+} = (\cos^{2}{\beta}) \, \phi - \tilde{\phi} \hspace{0.75cm} \Longrightarrow \hspace{0.5cm}   \phi = \phi_{+} + \phi_{-} \hspace{3.0cm} \\
&&  \hspace{-1.0cm} \phi_{-} = (\sin^{2}{\beta}) \, \phi + \tilde{\phi}  \hspace{2.15cm} \tilde{\phi} = -(\sin^{2}{\beta}) \, \phi_{+} + (\cos^{2}{\beta}) \, \phi_{-}
\end{eqnarray}
which we use in section \ref{section - examples}. Observe that $\chi$ and $\phi$ are the phases of the composite complex coordinates $y_{1}z_{1}$ and
$y_{2}z_{2}$, respectively:
\begin{equation} \hspace{-0.5cm}
\nonumber
y_{1}z_{1} = R_{+}R_{-} \, \cos{\theta_{+}} \cos{\theta_{-}} \hspace{0.1cm} e^{i\chi}  \hspace{1.2cm}
y_{2}z_{2} = R_{+}R_{-} \, \sin{\theta_{+}} \sin{\theta_{-}} \hspace{0.1cm} e^{i\phi}.
\end{equation}

%%%%%%%%%%%%%%%%%%%%%%%%%%%%%%%%%%%%%%%%%%%%%%%%%%%%%%%%%%%%%%%%%%%%%%%%%%%%%%%%%%%%%%%%%%%%%%
\section{Examples of D1 and D5-brane giant gravitons} \label{section - examples}
%%%%%%%%%%%%%%%%%%%%%%%%%%%%%%%%%%%%%%%%%%%%%%%%%%%%%%%%%%%%%%%%%%%%%%%%%%%%%%%%%%%%%%%%%%%%%%

Here we construct $\tfrac{1}{4}$-BPS giant gravitons embedded into the type IIB supergravity background 
$AdS_{3}\times S_{+}^{3} \times S_{-}^{3} \times S^{1}$ with pure R-R flux. These are probe D1 and D5-branes supported by their motion on both the 3-spheres, and by their coupling to the R-R potentials, $C_{(2)}$ and $C_{(6)}$. 

Curiously, we shall observe for all our examples that the ratio of the angular momenta on the 3-spheres,  $\alpha \, P$  on $S^{3}_{+}$ and $(1-\alpha) \, P$ on $S^{3}_{-}$, must be fixed to be  $\alpha/(1-\alpha) = Q_{5}^{-}/Q_{5}^{+}$, which is a rational number as a result of the flux quantization condition
(\ref{charges+-}) .
Indeed, all the D1 and D5-brane giant gravitons in this section will be moving in the direction
\begin{equation} 
\nonumber \hspace{-0.5cm}
\frac{\p}{\p\chi} \, \equiv \, \cos^{2}{\beta} \hspace{0.15cm} \frac{\p}{\p\chi_{+}} + \sin^{2}{\beta} \hspace{0.1cm} \frac{\p}{\p\chi_{-}}
\, = \,\alpha \hspace{0.1cm} \frac{\p}{\p\chi_{+}} + (1-\alpha) \hspace{0.1cm} \frac{\p}{\p\chi_{-}}
\end{equation}
in terms of the mixed phases (\ref{mixed-phases-chi}). The total angular momentum is
$P_{\chi} = \alpha \, P_{\chi_{+}} \hspace{-0.05cm} + (1-\alpha) \, P_{\chi_{-}}$, with $P_{\chi_{+}} = P_{\chi_{-}} = P_{\chi}$, when decomposed into components on each of the (rescaled) 3-spheres.

%---------------------------------------------------------------------------------------------
\subsection{Examples of D1-brane giant gravitons in $AdS_{3}$ and $\mathbb{R} \times S_{+}^{3}\times S_{-}^{3}$} \label{subsection - examples - D1}
%---------------------------------------------------------------------------------------------

We construct two examples of D1-brane giant gravitons, both in motion along the $\chi(t)$ direction in $S^{3}_{+}\times S^{3}_{-}$.  The  giant graviton in $AdS_{3}$ will wrap the $\varphi$ circle in $AdS_{3}$, whereas the giant graviton in  $\mathbb{R} \times S_{+}^{3} \times S_{-}^{3}$ will wrap the $\phi$ path wound around a $S^{1} \times S^{1}$ torus in $S_{+}^{3} \times S_{-}^{3}$.  

These D1-branes couple to the 2-form potential:
\begin{equation}  \label{C2}
\hspace{-0.1cm} C_{(2)} = - \, L^{2} \sinh^{2}{\rho} \hspace{0.1cm} dt \wedge d\varphi + L^{2}\sec^{2}{\beta} \, \sin^{2}{\theta_{+}} \hspace{0.1cm} d\chi_{+} \wedge d\phi_{+} + L^{2} \csc^{2}{\beta} \, \sin^{2}{\theta_{-}} \hspace{0.1cm} d\chi_{-} \wedge d\phi_{-}
\end{equation}
through the D1-brane action
\begin{equation} 
\hspace{-1.0cm} S_{\text{D}1} = -\frac{1}{2\pi} \int_{\mathbb{R} \times \gamma} d^{2}\sigma \hspace{0.1cm} \sqrt{-\det {g}}  \hspace{0.2cm} 
\pm \hspace{0.1cm}\frac{1}{2\pi} \int_{\mathbb{R} \times \gamma} \, C_{(2)},
\end{equation}
with $\gamma$ the circle in $AdS_{3}$ or the curve in $S_{+}^{3} \times S_{-}^{3}$ wrapped by the D1-brane. The kappa symmetry condition is
\begin{equation} \label{kappa-D1}
\Gamma \, \varepsilon = \mp \, i (C \varepsilon)^{\ast}
\hspace{0.75cm} \text{with} \hspace{0.5cm}
\Gamma = \frac{1}{2!} \, \frac{\epsilon^{a_{0} a_{1}}}{\sqrt{-\det{g}}} \hspace{0.1cm} (\p_{a_{0}}X^{\mu_{0}}) (\p_{a_{1}}X^{\mu_{1}}) \, \, \Gamma_{\mu_{0} \mu_{1}}.
\end{equation}

%%%%%%%%%%
\subsubsection*{D1-brane giant graviton embedded into $AdS_{3}$}
%%%%%%%%%%

Let us consider a D1-brane, with worldvolume coordinates $\sigma^{a} = (t,\varphi)$, wrapping the curve $\gamma$ in $AdS_{3}$, which is the $\varphi$ circle of radius $L \sinh{\rho}$ with $\rho$ constant, and point-like in the compact space.  Here we set $\theta_{+} = \theta_{-} = 0$ and $\xi=0$.
Motion is along the $\chi(t)$ direction in $S^{3}_{+} \times S ^{3}_{-}$ in terms of the mixed phases (\ref{mixed-phases-chi}).  Hence
$\dot{\chi}_{+} = \cos^{2}{\beta} \hspace{0.15cm} \dot{\chi} = \alpha \hspace{0.1cm} \dot{\chi}$ and $\dot{\chi}_{-}=\sin^{2}{\beta} \hspace{0.15cm} \dot{\chi} = (1-\alpha) \hspace{0.1cm} \dot{\chi}$.
\begin{figure}[htb!]
\begin{center}
\includegraphics[scale=0.35]{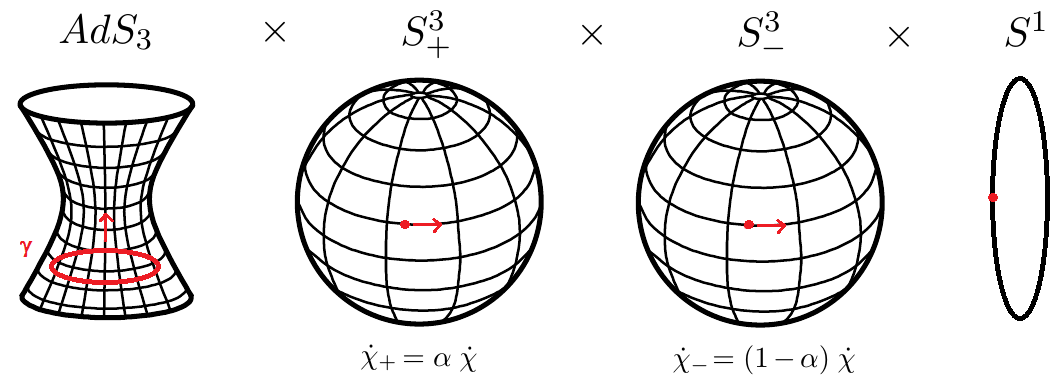}
\vspace{-0.3cm}
\caption{Diagram of the D1-brane giant graviton wrapping a 1-cycle $\gamma$ in $AdS_{3}$.  }
\end{center}
\end{figure}
\vspace{-0.5cm}

Now, the induced metric on the $\mathbb{R} \times \gamma$ worldvolume of this D1-brane is
\begin{eqnarray}
\nonumber && \hspace{-0.5cm}  ds^{2}  = - \, L^{2} \left( \cosh^{2}{\rho} - \dot{\chi}^{2} \right) dt^{2} \, + \, L^{2} \, \sinh^{2}{\rho} \hspace{0.15cm} d\varphi^{2} 
\end{eqnarray}
and the pull-back of the 2-form potential (\ref{C2}) is given by
\begin{equation}
\nonumber \hspace{-0.5cm} C_{(2)} = -  L^{2} \hspace{0.05cm} \sinh^{2}{\rho} \hspace{0.15cm} dt \wedge d\varphi.
\end{equation}
The D1-brane action is hence
\begin{equation}
\hspace{-0.5cm}  S_{\text{D}1}
= - \, L^{2} \int dt \hspace{0.2cm} \left\{ \sinh{\rho} \, \sqrt{ \cosh^{2}{\rho} - \dot{\chi}^{2} } \hspace{0.2cm} 
- \hspace{0.1cm} \sinh^{2}{\rho} \right\}  
\end{equation}
choosing the lower sign in the WZ action associated with an anti-brane solution. The momentum conjugate to $\chi$ is
\begin{equation}
\hspace{-0.5cm}
P_{\chi} = L^{2} \left\{ \frac{ \sinh{\rho} \hspace{0.2cm} \dot{\chi} }
{\sqrt{ \cosh^{2}{\rho} - \dot{\chi}^{2} }}  \right\} \equiv   L^{2} \hspace{0.15cm} p.
\end{equation}
The energy $H = P_{\chi} \, \dot{\chi} - L$ of the D1-brane is given by
\begin{equation}
\hspace{-0.5cm}
H = L^{2} \left\{  \frac{ \sinh{\rho}  \cosh^{2}{\rho}}
{\sqrt{ \cosh^{2}{\rho} - \dot{\chi}^{2} }} \hspace{0.1cm} - \hspace{0.1cm} \sinh^{2}{\rho} \right\} 
= L^{2}\left\{ \cosh{\rho}\sqrt{p^{2}+\sinh^{2}{\rho}} \hspace{0.1cm} - \hspace{0.1cm} \sinh^{2}{\rho} \right\}.
\end{equation}
We observe BPS solutions when $\dot{\chi} = \pm1$ and $p = \pm 1$.  These D1-(anti-)brane giant gravitons in $AdS_{3}$ (with opposite directions of motion) have energy and angular momentum 
\begin{equation} \hspace{-0.5cm}
\boxed{H = \pm \hspace{0.05cm} P_{\chi} = \pm \left[ \alpha \hspace{0.08cm} P_{\chi_{+}} + (1-\alpha) \hspace{0.08cm} P_{\chi -} \right] = L^{2}
= \frac{Q_{5}^{+} Q_{5}^{-}}{(Q_{5}^{+} + Q_{5}^{-})}}
\end{equation}
Notice that the energy and angular momentum are independent of the radius $L\sinh{\rho}$ of the $\varphi$ circle in $AdS_{3}$ (the size of the 1-cycle $\gamma$).  The potential $H(\rho)$ is flat when $p=\pm 1$.

The kappa symmetry condition (\ref{kappa-D1}) can be written as 
\begin{eqnarray} \label{kappa-AdS}
&& \hspace{-1.0cm}
\Gamma \, \varepsilon =  i \, (C\varepsilon)^{\ast} \hspace{0.75cm} \text{with} \hspace{0.5cm} \Gamma =  \frac{\left( \cosh{\rho} \hspace{0.1cm} \gamma_{0} + \dot{\chi} \hspace{0.05cm} \cos{\beta} \hspace{0.1cm} \gamma_{5} + \dot{\chi} \hspace{0.05cm} \sin{\beta} \hspace{0.1cm} \gamma_{9}\right)  \gamma_{2} }{\sinh{\rho}}   \hspace{0.5cm}
\end{eqnarray}
imposed on the pullback of the Killing spinor (\ref{Killing-Spinor}) to the giant's worldvolume: $\varepsilon = \varepsilon^{+} + \varepsilon^{-}$ with
$\varepsilon^{\pm} = \mathcal{M}^{\pm}(t,\varphi) \hspace{0.1cm} \varepsilon^{\pm}_{0}$ (from the gravitino KSEs) further satisfying $\mathcal{O} \, \varepsilon^{\pm} = \varepsilon^{\pm}$ where we define
$\mathcal{O} = \cos{\beta} \hspace{0.1cm} \hat{\gamma} \gamma_{+} + \sin{\beta} \hspace{0.1cm} \hat{\gamma} \gamma_{-}$ (from the dilatino KSE).
Here  $i \, (C\varepsilon^{\pm})^{\ast} = \pm \, \varepsilon^{\pm}$.  The kappa symmetry condition becomes $\Gamma \, \varepsilon^{\pm} = \pm \, \varepsilon^{\pm}$.
There is an additional consistency condition $[\Gamma,\mathcal{O}] \, \varepsilon^{\pm} = 0$ needed for the dilatino condition to be consistent with worldvolume supersymmetry. We show in appendix \ref{subappendix - kappa symmetry - D1} that these kappa symmetry, dilatino and consistency conditions imply the following three conditions on the constant spinor $\varepsilon^{\pm}_{0}$:
\begin{eqnarray}
&& \hspace{-0.2cm} 
\dot{\chi} \hspace{0.1cm} \gamma_{1}\gamma_{2} \hspace{0.1cm} \varepsilon^{\pm}_{0} = \gamma_{4}\gamma_{6} \hspace{0.1cm} \varepsilon_{0}^{\pm} 
= \gamma_{8}\gamma_{10} \hspace{0.1cm} \varepsilon^{\pm}_{0}   
\hspace{1.2cm}  \gamma_{0} \left(\cos{\beta} \hspace{0.1cm} \gamma_{5} + \sin{\beta} \hspace{0.1cm} \gamma_{9}\right) \, \varepsilon^{\pm}_{0} 
= \dot{\chi} \hspace{0.1cm} \varepsilon^{\pm}_{0}.   \hspace{1.0cm}
\end{eqnarray}
We conclude that the Weyl spinors $\varepsilon^{+}_{0}$ and $\varepsilon^{-}_{0}$ each contain 2 degrees of freedom which we may label by the $i r_{1}$ eigenvalues of $\gamma_{1}\gamma_{2}$, where $r_{1} = \pm$.  The D1-brane giant graviton in $AdS_{3}$ therefore preserves 4 of the original 16 supersymmetries and is hence 
$\tfrac{1}{4}$-BPS.

In the $\alpha \rightarrow 0$ or $\alpha \rightarrow 1$ limits, this D1-brane giant graviton has angular momentum $P_{\chi} = P_{\chi_{-}}$ or $P_{\chi} = P_{\chi_{+}}$ on only one 3-sphere. There is now enhanced $\tfrac{1}{2}$-BPS supersymmetry, since the consistency condition becomes trivial. These limits therefore give rise to AdS D1-brane giant gravitons on $AdS_{3}\times S^{3}_{\mp} \times T^{4}$ after a compactification.

%%%%%%%%%%
\subsubsection*{D1-brane giant graviton embedded into $\mathbb{R}\times S_{+}^{3}\times S_{-}^{3}$}
%%%%%%%%%%

Let us consider a D1-brane, with worldvolume coordinates $\sigma^{a} = (t,\phi)$, wrapping the curve $\gamma$ in $S^{3}_{+} \times S^{3}_{-}$ parameterized by the mixed phase $\phi$ defined in (\ref{mixed-phases-phi}). Here $\theta_{+} = \theta_{-} = \theta$ is constant, and $\p_{\phi}\phi_{+} = \cos^{2}{\beta} = \alpha$ and $\p_{\phi}\phi_{-} = \sin^{2}{\beta} = (1-\alpha)$.  The curve $\gamma$ is wound around a torus $S^{1} \times S^{1}$ in 
$S^{3}_{+} \times S^{3}_{-}$. Note that $\gamma$ is closed due to rational $\alpha/(1-\alpha) = Q_{5}^{-}/Q_{5}^{+}$ as a result of the flux quantization condition (\ref{charges+-}). This D1-brane is point-like in $AdS_{3}$ and $S^{1}$, with $\rho=0$ and $\xi=0$.  Motion is along the $\chi(t)$ direction in the $S_{+}^{3} \times S_{-}^{3}$ compact space.
\begin{figure}[htb!]
\begin{center}
\subfigure[]{\includegraphics[scale=0.35]{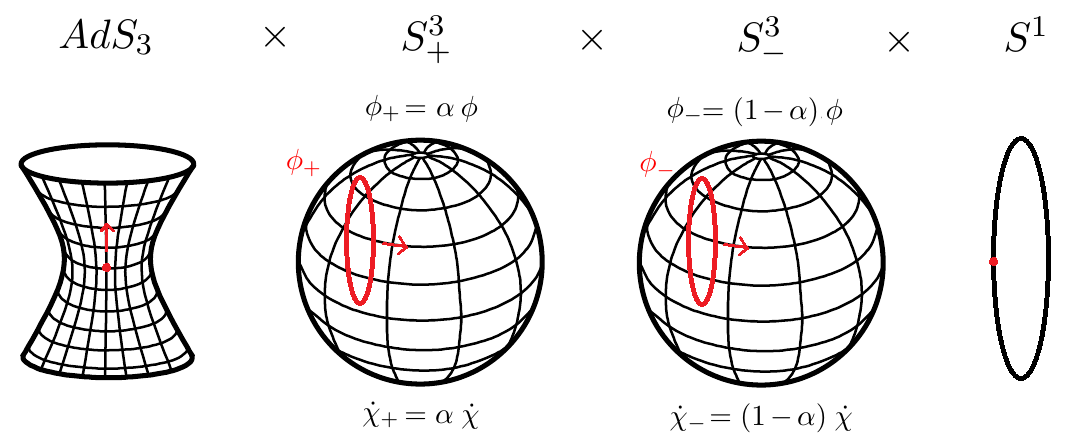} }  \hspace{1.2cm}
\subfigure[]{\includegraphics[scale=0.35]{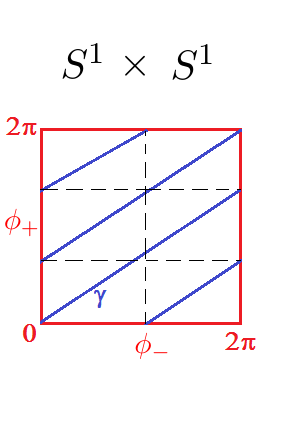} }  
\vspace{-0.3cm}
\caption{Diagram of the D1-brane giant graviton wrapping a 1-cycle $\gamma$ wound around a torus $S^{1} \times S^{1}$ in $S^{3}_{+}\times S^{3}_{-}$: (a) shows the torus $S^{1} \times S^{1}$ in red and (b) shows the 1-cycle $\gamma$ in blue.}
\end{center}
\end{figure}
\vspace{-0.5cm}

The induced metric on the $\mathbb{R} \times \gamma$ worldvolume of this D1-brane is
\begin{eqnarray}
\nonumber && \hspace{-0.5cm} 
ds^{2} =- \, L^{2} \left(1 -  \cos^{2}{\theta} \hspace{0.15cm} \dot{\chi}^{2} \right) dt^{2} \, + \, L^{2} \sin^{2}{\theta} \hspace{0.15cm} d\phi^{2} 
\end{eqnarray}
and the pullback of the 2-form potential (\ref{C2}) to the worldvolume is
\begin{eqnarray}
\nonumber && \hspace{-1.0cm} C_{(2)} 
=  L^{2}  \sin^{2}{\theta}\hspace{0.2cm} \dot{\chi} \hspace{0.15cm} dt \wedge d\phi.
\end{eqnarray}
The D1-brane action is therefore given by 
\begin{equation}
\hspace{-0.5cm} S_{\text{D}1}
= - \, L^{2} \int dt \hspace{0.2cm}  \left\{ \sin{\theta} \, \sqrt{1 -  \cos^{2}{\theta} \hspace{0.15cm} \dot{\chi}^{2} } \hspace{0.2cm}  \mp  \hspace{0.1cm} \sin^{2}{\theta} \hspace{0.15cm} \dot{\chi} \right\}.
\end{equation}
The angular momentum conjugate to $\chi$ is
\begin{equation}
\hspace{-0.5cm}
P_{\chi} = L^{2} \left\{ \frac{\sin{\theta} \, \cos{\theta} \hspace{0.15cm} \dot{\chi}}{ \sqrt{1 -  \cos^{2}{\theta} \hspace{0.15cm} \dot{\chi}^{2} }} 
\, \pm \, \sin^{2}{\theta}\right\} \equiv   L^{2} \hspace{0.10cm} p.
\end{equation}
The energy  $H = P_{\chi} \, \dot{\chi} - L$ of this D1-brane then takes the form
\begin{equation}
\hspace{-0.5cm}
H = L^{2} \left\{ \frac{\sin{\theta}  }{ \sqrt{1 -  \cos^{2}{\theta} \hspace{0.15cm} \dot{\chi}^{2} }} \right\}
= L^{2}\hspace{0.10cm} p \hspace{0.25cm} \sqrt{1 + \tan^{2}{\theta}\left( 1 \mp \frac{1}{p} \right)^{2} } \,.
\end{equation}
There is a BPS solution with $\dot{\chi} = \pm 1$ and $p = \pm 1$ associated with D1-brane (and anti-brane) giant gravitons in $\mathbb{R} \times S_{+}^{3}\times S_{-}^{3}$ with opposite directions of motion.  These giant gravitons have energy and angular momentum 
\begin{equation} \hspace{-0.5cm}
\boxed{H = \pm \, P_{\chi} = \pm \left[ \alpha \hspace{0.08cm} P_{\chi_{+}} + (1-\alpha) \hspace{0.08cm} P_{\chi -}\right] = L^{2} 
= \frac{Q_{5}^{+} Q_{5}^{-}}{(Q_{5}^{+} + Q_{5}^{-})}}
\end{equation}  
The energy and angular momentum are independent of the radii $L \sec{\beta} \sin{\theta}$ and $L \csc{\beta} \sin{\theta}$ of the $S^{1}\times S^{1}$ torus on which  $\gamma$ is wound (the size of the 1-cycle $\gamma$ wrapping this torus).  The potential $H(\theta)$ is again flat when $p = \pm 1$.

The kappa symmetry condition (\ref{kappa-D1}) can be written as 
\begin{eqnarray} \label{kappa-S3xS3}
&& \hspace{-0.65cm}
\Gamma \, \varepsilon = - \dot{\chi}  \hspace{0.1cm} i \hspace{0.025cm} (C\varepsilon)^{\ast} \hspace{0.3cm} \text{with} \hspace{0.25cm} 
\Gamma =  \frac{\left( \gamma_{0} + \dot{\chi} \hspace{0.05cm} \cos{\theta} \cos{\beta} \hspace{0.1cm} \gamma_{5} + \dot{\chi} \hspace{0.05cm} \cos{\theta}\sin{\beta} \hspace{0.1cm} \gamma_{9}\right)  \left( \cos{\beta} \hspace{0.1cm} \gamma_{6} + \sin{\beta} \hspace{0.1cm} \gamma_{10} \right)}{\sin{\theta}}, \hspace{0.8cm}
\end{eqnarray}
where $\dot{\chi} = \pm 1$ for the brane/anti-brane. The pullback of the Killing spinor (\ref{Killing-Spinor}) to the giant's worldvolume:  
$\varepsilon = \varepsilon^{+} + \varepsilon^{-}$, with $\varepsilon^{\pm} = \mathcal{M}^{\pm}(t,\phi) \hspace{0.1cm} \varepsilon^{\pm}_{0}$,
 must again satisfy $\mathcal{O} \, \varepsilon^{\pm} = \varepsilon^{\pm}$.  The consistency of this dilatino condition with the kappa symmetry condition $\Gamma\, \varepsilon^{\pm} = \mp \, \dot{\chi} \hspace{0.1cm} \varepsilon^{\pm}$  requires also $[\Gamma,\mathcal{O}] \, \varepsilon^{\pm} = 0$.
In appendix \ref{subappendix - kappa symmetry - D1} we show that these conditions are all satisfied if we impose the same three conditions on the constant spinor $\varepsilon^{\pm}_{0}$:
\begin{eqnarray} 
\nonumber && \hspace{-0.2cm} \dot{\chi} \hspace{0.1cm} \gamma_{1}\gamma_{2} \hspace{0.1cm} \varepsilon^{\pm}_{0}  = \gamma_{4}\gamma_{6} \hspace{0.1cm} \varepsilon^{\pm}_{0} 
= \gamma_{8}\gamma_{10} \hspace{0.1cm} \varepsilon^{\pm}_{0} \hspace{1.2cm} \gamma_{0} \left(\cos{\beta} \hspace{0.1cm} \gamma_{5} + \sin{\beta \hspace{0.1cm} \gamma_{0} \gamma_{9}}\right) \hspace{0.1cm} \varepsilon^{\pm}_{0} 
= \dot{\chi} \hspace{0.1cm} \varepsilon^{\pm}_{0}.   \hspace{1.0cm}
\end{eqnarray}
Once more, $\varepsilon^{\pm}_{0}$ are labeled by the $ir_{1}$ eigenvalue of $\gamma_{1}\gamma_{2}$.  The D1-brane giant graviton in $\mathbb{R} \times S^{3}_{+} \times S^{3}_{-}$ is therefore $\tfrac{1}{4}$-BPS, preserving 4 of the 16 background supersymmetries. 

In the $\alpha \rightarrow 0$ or $\alpha \rightarrow 1$ limits, the 1-cycle $\gamma$ unwraps from the circle on one of the 3-spheres and becomes simply a D1-brane wrapping an $S^{1}$ in $S^{3}_{\mp}$, with angular momentum $P_{\chi} = P_{\chi_{-}}$ or $P_{\chi} = P_{\chi_{+}}$ on the same 3-sphere. There is again enhanced $\tfrac{1}{2}$-BPS supersymmetry. This gives rise to the sphere D1-brane giant graviton on $AdS_{3}\times S^{3}_{\mp} \times T^{4}$ after a compactification.

%---------------------------------------------------------------------------------------------
\subsection{Example of a D5-brane giant graviton in  $\mathbb{R} \times S_{+}^{3} \times S_{-}^{3} \times S^{1}$} \label{subsection - examples - D5}
%---------------------------------------------------------------------------------------------

Here we construct an example of a D5-brane giant graviton in motion along the $\chi(t)$ direction in the $S^{3}_{+} \times S^{3}_{-}$ compact space.  This D5-brane  giant graviton in $\mathbb{R} \times S^{3}_{+} \times S^{3}_{-} \times S^{1}$  will wrap a 4-cycle $\Sigma$ in $S^{3}_{+} \times S^{3}_{-}$ and the $S^{1}$.

This D5-brane couples to the 6-form potential: 
\begin{eqnarray} \label{C6}
\nonumber && \hspace{-0.5cm} C_{(6)}  = L^{5} \, \ell \hspace{0.15cm} \sec^{3}{\beta} \,\csc^{3}{\beta} \hspace{0.1cm} 
\left[\cos^{2}{\beta} \sin^{2}{\theta_{+}} \left( \cos{\theta_{-}} \sin{\theta_{-}} \, d\theta_{-} \right)  
- \sin^{2}{\beta} \sin^{2}{\theta_{-}} \left( \cos{\theta_{+}} \sin{\theta_{+}} \, d\theta_{+}\right) \right] \\
\nonumber && \hspace{-0.5cm} \hspace{4.5cm}  \wedge \hspace{0.1cm}
 d\chi_{+} \wedge d\chi_{-} \wedge d\phi_{+} \wedge d\phi_{-} \wedge d\xi \\
\nonumber && \hspace{-0.5cm} \hspace{0.85cm} + \, L^{5} \, \ell \hspace{0.15cm} \cos{\beta} \, \csc^{3}{\beta} \hspace{0.1cm} \sinh^{2}{\rho} \hspace{0.2cm} 
dt \wedge d\varphi \wedge \left( \sin{\theta_{-}} \cos{\theta_{-}}  \, d\theta_{-} \right) \wedge d\chi_{-} \wedge d\phi_{-}\wedge d\xi   \\
 && \hspace{-0.5cm} \hspace{0.85cm} - \, L^{5} \, \ell \hspace{0.15cm} \sin{\beta} \, \sec^{3}{\beta} \hspace{0.1cm}  \sinh^{2}{\rho} \hspace{0.2cm} 
dt \wedge d\varphi \wedge \left( \sin{\theta_{+}} \cos{\theta_{+}} \, d\theta_{+} \right) \wedge d\chi_{+} \wedge d\phi_{+}\wedge d\xi  \hspace{1.5cm}
\end{eqnarray}
through the D5-brane action
\begin{equation}
\hspace{-1.0cm} S_{\text{D}5} = -\frac{1}{(2\pi)^{5}} \int_{\mathbb{R}\times \Sigma \times S^{1}} d^{6}\sigma \hspace{0.2cm} \sqrt{- \det g}  \hspace{0.2cm} 
\pm \hspace{0.1cm} \frac{1}{(2\pi)^{5}} \int_{\mathbb{R}\times \Sigma \times S^{1}} \, C_{(6)}.
\end{equation}
The kappa symmetry condition is now given by
\begin{equation} \label{kappa-D5} \hspace{-1.0cm}
\Gamma \, \varepsilon = \mp \, i (C \varepsilon)^{\ast}
\hspace{0.75cm} \text{with} \hspace{0.5cm}
\Gamma = \frac{1}{6!} \, \frac{\epsilon^{a_{0} \hspace{0.025cm}\cdots \hspace{0.05cm} a_{5}}}{\sqrt{-\det{g}}} \hspace{0.1cm} (\p_{a_{0}}X^{\mu_{0}}) 
\hspace{0.075cm} \cdots \hspace{0.1cm} (\p_{a_{5}}X^{\mu_{5}}) \, \, \Gamma_{\mu_{0} \hspace{0.025cm} \cdots \hspace{0.05cm} \mu_{5}}.
\end{equation}

%%%%%%%%%%
\subsubsection*{D5-brane giant graviton embedded into $\mathbb{R} \times S_{+}^{3}\times S_{-}^{3} \times S^{1}$}
%%%%%%%%%%

Let us consider a D5-brane, with worldvolume coordinates $\sigma^{a} = (t,\tilde{\theta},\phi,\tilde{\phi},\tilde{\chi},\xi)$, wrapping a 4-cycle $\Sigma$ in $S^{3}_{+} \times S^{3}_{-}$ parameterized by $(\tilde{\theta},\phi,\tilde{\phi},\tilde{\chi})$ and the $S^{1}$ coordinate $\xi$, and point-like in $AdS_{3}$ with $\rho = 0$.   We leave $\theta_{+}(\tilde{\theta})$ and $\theta_{-}(\tilde{\theta})$ which define $\Sigma$ to be specified later. We shall choose $\Sigma$ to be in motion in $S^{3}_{+}\times S^{3}_{-}$ along the $\chi(t)$ direction.
\begin{figure}[htb!]
\begin{center}
\includegraphics[scale=0.35]{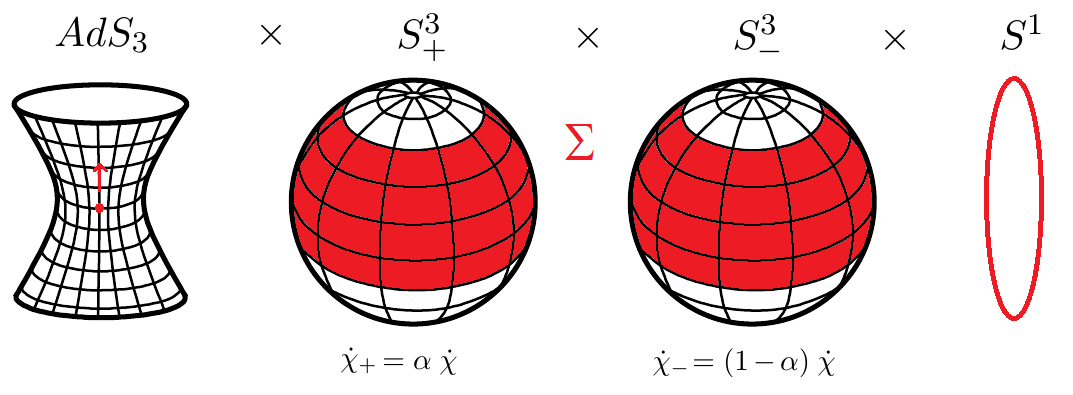}
\vspace{-0.3cm}
\caption{Diagram of the D5-brane giant wrapping a 4-cycle $\Sigma$ in $S^{3}_{+}\times S^{3}_{-}$ and the $S^{1}$.  }
\end{center}
\end{figure}
\vspace{-0.5cm}

The induced metric on the $\mathbb{R}\times \Sigma \times S^{1}$ worldvolume of this D5-brane is
\begin{eqnarray}
\nonumber && \hspace{-0.65cm} 
ds^{2} = - \, L^{2} \left[1 - \left( \cos^{2}{\beta} \cos^{2}{\theta_{+}} + \sin^{2}{\beta} \cos^{2}{\theta_{-}} \right) \dot{\chi}^{2} \right] dt^{2} \, 
- 2L^{2} \left( \cos^{2}{\theta_{+}} - \cos^{2}{\theta_{-}} \right) \dot{\chi} \hspace{0.15cm} dt \, d\tilde{\chi}  \\ 
\nonumber && \hspace{0.42cm} + \, L^{2} \hspace{0.05cm} \sec^{2}{\beta} \csc^{2}{\beta} \left( \sin^{2}{\beta} \cos^{2}{\theta_{+}} + \cos^{2}{\beta} \cos^{2}{\theta_{-}} \right)  d\tilde{\chi}^{2} \\
\nonumber && \hspace{0.42cm} + \, L^{2} \hspace{0.05cm} \sec^{2}{\beta} \csc^{2}{\beta} \left[ \sin^{2}{\beta}\hspace{0.1cm} (\p_{\tilde{\theta}} \theta_{+})^{2} + \cos^{2}{\beta} \hspace{0.1cm} (\p_{\tilde{\theta}} \theta_{-})^{2}  \right] d\tilde{\theta}^{2} \\
\nonumber && \hspace{0.42cm} + \, L^{2} \left( \cos^{2}{\beta} \sin^{2}{\theta_{+}} + \sin^{2}{\beta} \sin^{2}{\theta_{-}} \right)  d\phi^{2} - 2L^{2} \left(\sin^{2}{\theta_{+}} - \sin^{2}{\theta_{-}}\right) d\phi \, d\tilde{\phi}\\
\nonumber && \hspace{0.42cm} + \, L^{2} \hspace{0.05cm} \sec^{2}{\beta}\csc^{2}{\beta} \left( \sin^{2}{\beta} \sin^{2}{\theta_{+}} + \cos^{2}{\beta} \sin^{2}{\theta_{-}} \right) d\tilde{\phi}^{2} + \ell^{2} \hspace{0.05cm} d\xi^{2}
\end{eqnarray}
with
\small
\begin{eqnarray}
\nonumber && \hspace{-0.75cm} \sqrt{- \text{det} \, g} = L^{5} \, \ell\, \sec^{3}{\beta} \, \csc^{3}{\beta} \, \sin{\theta_{+}} \, \sin{\theta_{-}} \\
\nonumber && \hspace{-0.75cm} \hspace{1.8cm} \times \, \sqrt{ \left[ \sin^{2}{\beta}\hspace{0.1cm} (\p_{\tilde{\theta}} \theta_{+})^{2} + \cos^{2}{\beta} \hspace{0.1cm} (\p_{\tilde{\theta}} \theta_{-})^{2}  \right]
\left[ (\sin^{2}{\beta}\cos^{2}{\theta}_{+} + \cos^{2}{\beta}\cos^{2}{\theta_{-}}) - \cos^{2}{\theta_{+}} \, \cos^{2}{\theta_{-}} \right]} \\
\nonumber && \hspace{-0.75cm} \hspace{1.8cm} \times \, \sqrt{1 + \frac{\cos^{2}{\theta_{+}} \cos^{2}{\theta_{-}} \left(1-\dot{\chi}^{2}\right)}{\left[ (\sin^{2}{\beta}\cos^{2}{\theta}_{+} + \cos^{2}{\beta}\cos^{2}{\theta_{-}}) - \cos^{2}{\theta_{+}} \, \cos^{2}{\theta_{-}} \right]}}
\end{eqnarray}
\normalsize
and the pullback of the 6-form potential is
\small
\begin{eqnarray} \label{C6}
\nonumber && \hspace{-0.65cm} C_{(6)}  = - \, L^{5} \, \ell \hspace{0.15cm} \sec^{3}{\beta} \,\csc^{3}{\beta} \hspace{0.1cm} 
\left[\cos^{2}{\beta} \sin^{2}{\theta_{+}} \left( \cos{\theta_{-}} \sin{\theta_{-}} \, (\p_{\tilde{\theta}}\theta_{-}) \right)  
- \sin^{2}{\beta} \sin^{2}{\theta_{-}} \left( \cos{\theta_{+}} \sin{\theta_{+}} \, (\p_{\tilde{\theta}}\theta_{+})\right) \right]\\
\nonumber && \hspace{-0.65cm} \hspace{1.1cm} \times \hspace{0.1cm}  \dot{\chi} \hspace{0.1cm} dt \wedge  d\tilde{\theta} \wedge d\phi \wedge d\tilde{\phi} \wedge d\tilde{\chi} \wedge d\xi 
\end{eqnarray}
\normalsize

Let us now make the ansatz 
\begin{equation} \label{D5-ansatz} \hspace{-0.75cm}
\boxed{\cos{\theta_{+}} \cos{\theta_{-}} = \text{constant} \equiv \cos{\theta}} \hspace{0.5cm} \Longrightarrow \hspace{0.5cm}
\cos{\theta_{-}} = \frac{\cos{\theta}\hspace{0.1cm}}{\hspace{0.1cm}\cos{\theta_{+}}},
\end{equation}
which is associated with the holomorphic surface $f(y_{1}z_{1})=y_{1}z_{1} - \cos{\theta} = 0$ in $\mathbb{C}^{2}_{+} \times \mathbb{C}^{2}_{-}$, yielding $\Sigma$ when restricted to the submanifold $S^{3}_{+} \times S^{3}_{-}$. (A general holomorphic surface construction is presented in section \ref{section - holomorphic surfaces}).  This surface is boosted into motion along the overall phase of $y_{1}z_{1}$ which is the $\chi(t)$ direction.  We shall choose $\tilde{\theta} = \theta_{+}$ to be a worldvolume coordinate.
This yields the D5-brane action
\begin{eqnarray}
\nonumber && \hspace{-0.7cm} S_{\text{D5}} = -\frac{L^{5} \, \ell}{2 \pi} \, \sec^{3}{\beta} \csc^{3}{\beta} \int{dt} \int_{0}^{\theta}{d\theta_{+}} \, \frac{\sin{\theta_{+}}}{\cos{\theta_{+}}} \, 
\left[ (\sin^{2}{\beta}\cos^{2}{\theta}_{+} + \cos^{2}{\beta}\cos^{2}{\theta_{-}}) - \cos^{2}{\theta_{+}} \, \cos^{2}{\theta_{-}} \right] \\
&& \hspace{-0.7cm} \hspace{4.9cm} \times \left\{ \sqrt{1 + \frac{\cos^{2}{\theta_{+}} \cos^{2}{\theta_{-}} \left(1-\dot{\chi}^{2}\right)}{\left[ (\sin^{2}{\beta}\cos^{2}{\theta}_{+} + \cos^{2}{\beta}\cos^{2}{\theta_{-}}) - \cos^{2}{\theta_{+}} \, \cos^{2}{\theta_{-}} \right]}} \, \mp \dot{\chi} \right\} \hspace{0.85cm}
\end{eqnarray}
with $\theta_{-}(\theta_{+})$ determined through (\ref{D5-ansatz}).  The angular momentum conjugate to $\chi$ is
\begin{eqnarray}
\nonumber && \hspace{-0.75cm} P_{\chi} = \frac{L^{5} \, \ell}{2 \pi} \, \sec^{3}{\beta} \csc^{3}{\beta} \int_{0}^{\theta}{d\theta_{+}} \hspace{0.1cm} \frac{\sin{\theta_{+}}}{\cos{\theta_{+}}} \, 
\left[ (\sin^{2}{\beta}\cos^{2}{\theta}_{+} + \cos^{2}{\beta}\cos^{2}{\theta_{-}}) - \cos^{2}{\theta_{+}} \, \cos^{2}{\theta_{-}} \right] \\
&& \hspace{-0.75cm} \hspace{4.5cm} \times \, \left\{ \frac{\frac{\cos^{2}{\theta_{+}} \cos^{2}{\theta_{-}} \, \dot{\chi}}{\left[ (\sin^{2}{\beta}\cos^{2}{\theta}_{+} + \cos^{2}{\beta}\cos^{2}{\theta_{-}}) - \cos^{2}{\theta_{+}} \, \cos^{2}{\theta_{-}} \right]}}{\sqrt{1 + \frac{\cos^{2}{\theta_{+}} \cos^{2}{\theta_{-}} \left(1-\dot{\chi}^{2}\right)}{\left[ (\sin^{2}{\beta}\cos^{2}{\theta}_{+} + \cos^{2}{\beta}\cos^{2}{\theta_{-}}) - \cos^{2}{\theta_{+}} \, \cos^{2}{\theta_{-}} \right]}}} \, \pm 1 \right\} \hspace{0.5cm}
\end{eqnarray}
and the energy $H = P_{\chi}\, \dot{\chi} - L$ of the D5-brane is given by
\begin{eqnarray}
&& \hspace{-0.65cm} H = \frac{L^{5} \, \ell}{2 \pi} \, \sec^{3}{\beta} \csc^{3}{\beta} \int_{0}^{\theta}{d\theta_{+}} \hspace{0.1cm} \frac{\sin{\theta_{+}}}{\cos{\theta_{+}}} \, 
 \frac{ (\sin^{2}{\beta}\cos^{2}{\theta}_{+} + \cos^{2}{\beta}\cos^{2}{\theta_{-}}) }{\sqrt{1 + \frac{\cos^{2}{\theta_{+}} \cos^{2}{\theta_{-}} \left(1-\dot{\chi}^{2}\right)}
{\left[ (\sin^{2}{\beta}\cos^{2}{\theta}_{+} + \cos^{2}{\beta}\cos^{2}{\theta_{-}}) - \cos^{2}{\theta_{+}} \, \cos^{2}{\theta_{-}} \right]}}}. \hspace{1.0cm}
\end{eqnarray}
There are BPS solutions when $\dot{\chi} = \pm 1$  associated with D5-brane (and anti-brane) giant gravitons in $\mathbb{R} \times S_{+}^{3}\times S_{-}^{3} \times S^{1}$ with opposite directions of motion.  These giant gravitons have energy and angular momentum
\begin{eqnarray}
 && \hspace{-0.65cm} \boxed {H = \pm \, P_{\chi} = \pm \left[ \alpha \hspace{0.08cm} P_{\chi_{+}} + (1-\alpha) \hspace{0.08cm} P_{\chi -}\right] 
 = Q_{1} \, \sin^{2}{\theta}} \hspace{1.0cm}
\end{eqnarray}
with the quantized flux $Q_{1}$ given by (\ref{chargeQ1}).  
The energy and angular momentum of this D5-brane giant graviton do exhibit a dependence on $\theta$, which controls the size and shape of the 4-cycle $\Sigma$ in $S_{+}^{3} \times S_{-}^{3}$.  At maximum size, $H = \pm P_{\chi}= Q_{1}$, indicating a stringy exclusion principle.

The kappa symmetry condition (\ref{kappa-D5}) can be written as
\begin{eqnarray} 
\nonumber && \hspace{-0.75cm}
\Gamma \, \varepsilon = - \dot{\chi} \hspace{0.1cm} i (C\varepsilon)^{\ast}  \hspace{0.45cm} 
\text{with} \hspace{0.35cm}
\Gamma = \left[\left( \sin{\beta} \cos{\theta_{+}} \, \gamma_{0}\gamma_{5} - \cos{\beta} \cos{\theta_{-}} \, \gamma_{0}\gamma_{9} - \dot{\chi} \, \cos{\theta_{+}}\cos{\theta_{-}} \, \gamma_{5}\gamma_{9}\right) \right. \\
 && \hspace{-0.75cm} \hspace{4.7cm} \frac{ \times \left( \sin{\beta}\cos{\theta_{+}} \sin{\theta_{-}} \, \gamma_{4} + \cos{\beta} \cos{\theta_{-}} \sin{\theta_{+}}\, \gamma_{8} \right) \, \gamma_{6} \gamma_{10} \gamma_{12} ]}
{\left[ (\sin^{2}{\beta}\cos^{2}{\theta}_{+} + \cos^{2}{\beta}\cos^{2}{\theta_{-}}) - \cos^{2}{\theta_{+}} \, \cos^{2}{\theta_{-}} \right]}   \hspace{1.0cm}
\end{eqnarray}
\normalsize
where $\varepsilon = \varepsilon^{+} + \varepsilon^{-}$ is the pullback of the  Killing spinor (\ref{Killing-Spinor}) to the giant's worldvolume. 
Here
$\varepsilon^{\pm} = \mathcal{M}^{\pm}(t,\theta_{+},\phi,\tilde{\phi},\tilde{\chi},\xi) \hspace{0.1cm} \varepsilon^{\pm}_{0}$ again
 satisfies $\mathcal{O} \, \varepsilon^{\pm} = \varepsilon^{\pm}$ and $\Gamma \, \varepsilon^{\pm} = \mp \, \dot{\chi} \hspace{0.1cm} \varepsilon^{\pm}$ with consistency condition $[\Gamma,\mathcal{O}] \, \varepsilon^{\pm} = 0$.
We show in appendix \ref{subappendix - kappa symmetry - D1} that the consistency of the dilatino and kappa symmetry conditions once more implies
\begin{eqnarray}
\nonumber && \hspace{-0.2cm} 
\dot{\chi} \hspace{0.1cm} \gamma_{1}\gamma_{2} \hspace{0.1cm} \varepsilon^{\pm}_{0} = \gamma_{4}\gamma_{6} \hspace{0.1cm} \varepsilon_{0}^{\pm} 
= \gamma_{8}\gamma_{10} \hspace{0.1cm} \varepsilon^{\pm}_{0}   
\hspace{1.2cm}  \gamma_{0} \left(\cos{\beta} \hspace{0.1cm} \gamma_{5} + \sin{\beta} \hspace{0.1cm} \gamma_{9}\right) \, \varepsilon^{\pm}_{0} 
= \dot{\chi} \hspace{0.1cm} \varepsilon^{\pm}_{0}.   \hspace{1.0cm}
\end{eqnarray}
The Weyl spinors $\varepsilon^{+}_{0}$ and $\varepsilon^{-}_{0}$ each contain 2 degrees of freedom which are again labeled by the $i r_{1}$ eigenvalues of $\gamma_{1}\gamma_{2}$.  This D5-brane giant graviton in $\mathbb{R} \times S^{3}_{+} \times S^{3}_{-} \times S^{1}$ thus preserves 4 of the original 16 supersymmetries and is hence $\tfrac{1}{4}$-BPS.

\vspace{0.1cm}
\begin{figure}[htb!]
\begin{center}
\includegraphics[scale=0.255]{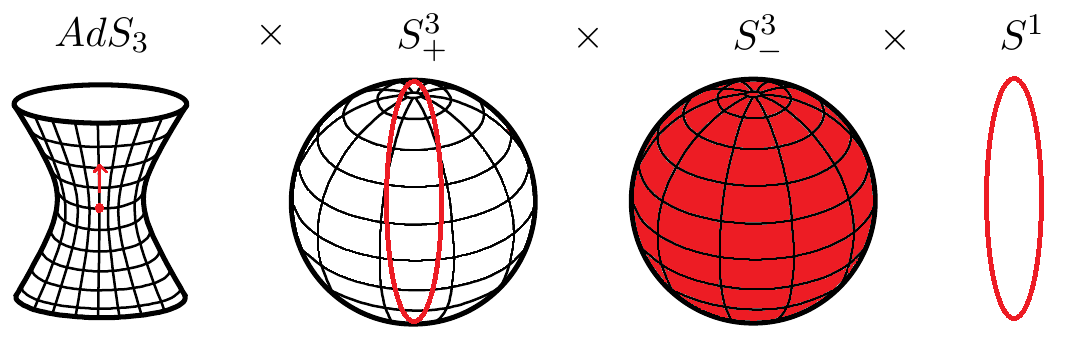} \, \raisebox{1.0cm}{+} \,
\includegraphics[scale=0.255]{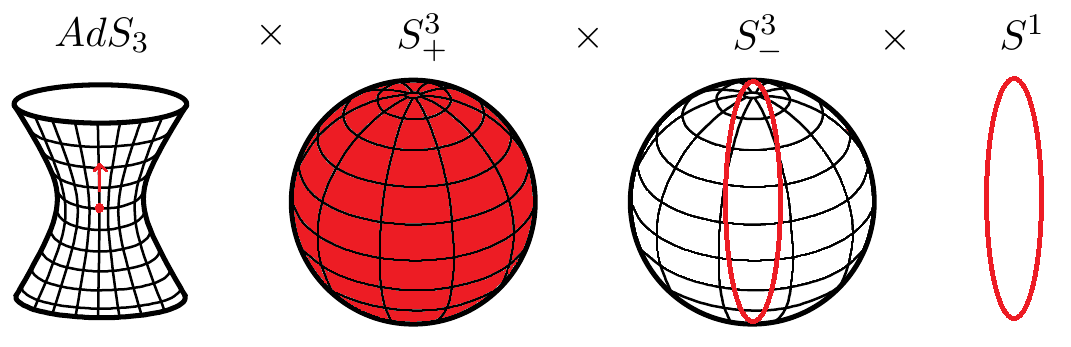}
%\vspace{-0.3cm}
\caption{Diagram of both halves of the maximal D5-brane giant which wrap $S^{1} \times S^{3}_{\mp} \times S^{1}$.}
\end{center}
\end{figure}
\vspace{-0.3cm}
The maximal D5-brane giant graviton with $\sin{\theta}=1$ has ansatz $\cos{\theta_{+}} \cos{\theta_{-}} = 0$. This splits into two D5-branes, $\cos{\theta_{+}}=0$ and $\cos{\theta_{-}}=0$, which we parameterize separately by $\tilde{\theta} = \theta_{-}$ and $\tilde{\theta} = \theta_{+}$. Each of these is a $\tfrac{1}{2}$-BPS D5-brane wrapping an $S^{1} \times S^{3}_{\mp} \times S^{1}$.

It is not immediately clear what happens to this D5-brane giant graviton in the $\alpha \rightarrow 0$ or $\alpha \rightarrow 1$ limits, in which one of the 3-spheres $S^{3}_{\mp}$ blows up to infinite size (and is then compactified to $T^{3}$). The constraint 
$y_{1}z_{1} = \cos{\theta}$ depends on the coordinates of both 3-spheres $S^{3}_{\mp}$ even when $\alpha \rightarrow 0$ or $\alpha \rightarrow 1$.  We therefore do not expect this D5-brane solution to survive the compactification (except at maximal size where half the maximal giant graviton will become the maximal sphere D5-brane giant graviton on $AdS_{3}\times S^{3}_{\mp} \times T^{4}$).

We anticipate that it may be possible to build the dual $\tfrac{1}{4}$-BPS operators in the $\mathcal{N}=(4,0)$ supersymmetric gauge theory defined in \cite{Tong:2014} on the worldvolume of the D1-D5-D5$'$-branes.  These operators will be protected under renormalization group flow to the $\mathcal{N}=(4,4)$ supersymmetric CFT$_{2}$ conjectured to be the holographic dual of type IIB superstring theory on $AdS_{3}\times S_{+}^{3}\times S_{-}^{3} \times S^{1}$. The dependence of the angular momentum $P_{\chi} = Q_{1} \, \sin^{2}{\theta}$ of our D5-brane giant graviton on the size parameter $\theta$, with maximum size occuring when $P_{\chi} = Q_{1}$, suggests a dual $\tfrac{1}{4}$-BPS operator built from an equal number $n \leq Q_{1}$ of scalar fields $Y_{1}$ and $Z_{1}$ in the adjoint representation of the $U(Q_{1})$ gauge group.  
At maximal length, $n = Q_{1}$, we expect these operators to split into two $\tfrac{1}{2}$-BPS operators, each built out of $Q_{1}$ of a single scalar field, $Y_{1}$ or $Z_{1}$, and dual to half the maximal D5-brane giant graviton.

%%%%%%%%%%%%%%%%%%%%%%%%%%%%%%%%%%%%%%%%%%%%%%%%%%%%%%%%%%%%%%%%%%%%%%%%%%%%%%%%%%%%%%%%%%%%%%
\section{D5-brane giant gravitons from holomorphic surfaces} \label{section - holomorphic surfaces} 
%%%%%%%%%%%%%%%%%%%%%%%%%%%%%%%%%%%%%%%%%%%%%%%%%%%%%%%%%%%%%%%%%%%%%%%%%%%%%%%%%%%%%%%%%%%%%%

We now construct a general class of $\tfrac{1}{8}$-BPS D5-brane giant gravitons on $AdS_{3}\times S^{3}_{+} \times S^{3}_{-} \times S^{1}$, by means of the embedding into $\mathbb{R}^{2+2} \times \mathbb{C}^{2}_{+} \times \mathbb{C}^{2}_{-} \times S^{1}$ (see section \ref{subsection - background - 13D}), making use of methods similar to those of \cite{Mikhailov:2000}.  This holomorphic surface construction is presented in section \ref{subsection - holomorphic surfaces - construction} while an analysis of the kappa symmetry conditions is given in section \ref{subsection - holomorphic surfaces - kappa symmetry}.

A D5-brane giant graviton at fixed time $t=0$ wraps a 4-cycle $\Sigma$ in  $S^{3}_{+}\times S^{3}_{-}$ and the  $S^{1}$, where $\Sigma$ is the intersection of a holomorphic surfaces $\mathcal{C}$ in the complex manifold $\mathbb{C}^{4} = \mathbb{C}^{2}_{+} \times \mathbb{C}^{2}_{-}$ with the $S^{3}_{+}\times S^{3}_{-}$ submanifold.  The holomorphic surface must take the form
\begin{equation} \nonumber \hspace{-1.0cm}
\mathcal{C}:  \hspace{0.2cm} f(y_{1}z_{1},y_{1}z_{2},y_{2}z_{1},y_{2}z_{2}) = 0.
\end{equation}
This 4-cycle $\Sigma$ is boosted into motion along a preferred direction $\mathbf{e}^{\parallel} = \mathrm{I} \, \mathbf{e}^{\perp}$ 
induced by the action of the complex structure $\mathrm{I}$ of $\mathbb{C}^{4}$.  This holomorphic surface construction generally gives rise to an $\tfrac{1}{8}$-BPS probe D5-brane. The D5-brane giant graviton in $\mathbb{R} \times S^{3}_{+} \times S^{3}_{-} \times S^{1}$ example of section \ref{subsection - examples - D5} is a special case in which $f(y_{1}z_{1}) = 0$ with enhanced $\tfrac{1}{4}$-BPS supersymmetry.

Essential to this construction is the choice of preferred direction.  At any point in $S^{3}_{+}\times S^{3}_{-}$, there are two natural preferred directions in $\mathrm{T}S_{\pm}^{3}$ respectively:
\begin{eqnarray}
\nonumber && \hspace{-1.0cm} \mathbf{e}^{\parallel +} \equiv \mathrm{I} \, \mathbf{e}^{\perp +} 
= \frac{\cos{\beta}}{L} \, \sum_{a} \, \p_{\psi_{a}^{y}}
= \frac{\cos{\beta}}{L} \, \sum_{a} i \left( y_{a} \, \p_{y_{a}} - \bar{y}_{a} \, \p_{\bar{y}_a} \right) \hspace{0.25cm} \\
&& \hspace{-1.0cm} \mathbf{e}^{\parallel -} \equiv \mathrm{I}\, \mathbf{e}^{\perp -} 
= \frac{\sin{\beta}}{L} \, \sum_{a} \, \p_{\psi_{a}^{z}}
= \frac{\sin{\beta}}{L} \, \sum_{a} i \left( z_{a} \, \p_{z_{a}} - \bar{z}_{a} \, \p_{\bar{z}_a} \right)
\end{eqnarray} 
induced by the action of the complex structure $\mathrm{I}$ on the directions in $\mathrm{T}\mathbb{C}^{2}_{\pm}$  orthogonal to $\mathrm{T}S^{3}_{\pm}$:
\begin{eqnarray}  \label{orthogonal+-}
\nonumber && \hspace{-1.0cm} \mathbf{e}^{\perp +} = \cos{\beta} \hspace{0.15cm} \p_{R_{+}} 
= \frac{\cos{\beta}}{L}  \sum_{a}\left( y_{a} \, \p_{y_{a}} + \bar{y}_{a} \, \p_{\bar{y}_a} \right) \hspace{0.25cm} \\
&& \hspace{-1.0cm} \mathbf{e}^{\perp -} = \sin{\beta} \hspace{0.15cm} \p_{R_{-}} 
= \frac{\sin{\beta}}{L}  \sum_{a}\left( z_{a} \, \p_{z_{a}} + \bar{z}_{a} \, \p_{\bar{z}_a} \right).
\end{eqnarray}
We introduce an alternative pair of mutually orthogonal unit vectors, also orthogonal to $\mathrm{T}(S^{3}_{+} \times S^{3}_{-})$, given by
\begin{eqnarray}  
&& \hspace{-0.5cm} \mathbf{e}^{\perp} \equiv \p_{R} = \cos{\beta}  \hspace{0.15cm}  \mathbf{e}^{\perp +} + \sin{\beta} \hspace{0.15cm} \mathbf{e}^{\perp -} 
= \alpha \hspace{0.15cm} \p_{R_{+}} + (1-\alpha) \hspace{0.1cm} \p_{R_{-}} \\
&& \nonumber \hspace{-0.5cm} \tilde{\mathbf{e}}^{\perp} \equiv \cos{\beta} \sin{\beta}  \hspace{0.15cm} \p_{\tilde{R}} = -\sin{\beta}  \hspace{0.15cm}  \mathbf{e}^{\perp +} + \cos{\beta}  \hspace{0.15cm}  \mathbf{e}^{\perp -},
\end{eqnarray}
which have an interpretation in terms of the mixed radii defined in section \ref{subsection - background - 13D}.
The associated preferred directions in $\mathrm{T}(S^{3}_{+} \times S^{3}_{-})$ are 
\begin{eqnarray}  
&& \hspace{-2.0cm}  \mathbf{e}^{\parallel} \equiv \mathrm{I} \, \mathbf{e}^{\perp} =\cos{\beta}  \hspace{0.15cm}  \mathbf{e}^{\parallel +} + \sin{\beta}  \hspace{0.15cm}  \mathbf{e}^{\parallel -}
= \frac{\alpha}{L}  \hspace{0.15cm} \text{$\sum_{a}$} \hspace{0.1cm} \p_{\psi_{y,a}} + \frac{(1-\alpha)}{L}  \hspace{0.15cm} \text{$\sum_{a}$} \hspace{0.1cm} \p_{\psi_{z,a}} \label{preferred+-} \\
&& \nonumber  \hspace{-2.0cm}  \tilde{\mathbf{e}}^{\parallel} \equiv \mathrm{I} \, \tilde{\mathbf{e}}^{\perp} = -\sin{\beta}  \hspace{0.15cm}  \mathbf{e}^{\parallel +} + \cos{\beta} \hspace{0.15cm}  \mathbf{e}^{\parallel -},
\end{eqnarray}
where $\mathbf{e}^{\parallel}$ has components $\alpha \, (L^{-1} \, \p_{\psi_{y,\text{total}}})$ and $(1-\alpha) \, (L^{-1}\, \p_{\psi_{z,\text{total}}})$ in $\mathrm{T}S^{3}_{\pm}$.  
It will be crucial to boost along this $\alpha$-dependent preferred direction $\mathbf{e}^{\parallel}$ to obtain supersymmetric D5-brane giant gravitons. 
This boost is implemented by taking 
$y_{a} \rightarrow y_{a} \, e^{- \hspace{0.025cm} i \alpha \hspace{0.025cm} t}$ and
$z_{a} \rightarrow z_{a} \, e^{- \hspace{0.025cm} i \hspace{0.025cm} (1-\alpha) \hspace{0.025cm} t}$ in the holomorphic function.  The rigidly rotating 
4-cycle $\Sigma(t)$ is then $S^{3}_{+} \times S^{3}_{-}$ intersected with
\begin{equation} \nonumber \hspace{-1.0cm}
\mathcal{C}(t):  \hspace{0.2cm} f(y_{1}z_{1} \hspace{0.05cm} e^{- \hspace{0.025cm} i \hspace{0.025cm} t}, \hspace{0.1cm} 
y_{1}z_{2} \hspace{0.05cm} e^{- \hspace{0.025cm} i \hspace{0.025cm} t}, \hspace{0.1cm} 
y_{2}z_{1} \hspace{0.05cm} e^{- \hspace{0.025cm} i \hspace{0.025cm} t}, \hspace{0.1cm} 
y_{2}z_{2} \hspace{0.05cm} e^{- \hspace{0.025cm} i \hspace{0.025cm} t} ) = 0.
\end{equation}

The type IIB Killing spinor $\varepsilon$ in 1+9 dimensions can be projected out of a covariantly constant spinor $\Psi$ in 2+11 dimensions\footnote{We  make use of the notation of \cite{Mikhailov:2000} in which $\Gamma({\bf v})$ is the gamma matrix corresponding to the vector ${\bf v}$.}:
\begin{equation} \hspace{-1.0cm}
\varepsilon(x^{\mu}) = \mathcal{P}_{\text{dilatino}} \hspace{0.1cm} \mathcal{P}_{\text{Weyl}} \hspace{0.15cm} \Psi(x^{\mu}) 
=
\frac{1}{2} \left( 1 + \Gamma(\mathbf{e}^{\hat{R}}) \, \Gamma(\mathbf{e}^{\perp}) \right)
\hspace{0.1cm} \frac{1}{2} \left( 1 + \Gamma_{10\text{D}} \right) \hspace{0.1cm} \Psi(x^{\mu}) 
\end{equation} 
with $\Gamma_{10\text{D}}$ the 1+9 dimensional chirality matrix (see appendix \ref{appendix - KSEs}). The gravitino KSEs on $\varepsilon$ arise from the covariantly constant condition on $\Psi$. The dilatino condition comes about as a result of our choice of the projection operator $\mathcal{P}_{\text{dilatino}}$ involving the $\alpha$-dependent orthogonal direction $\mathbf{e}^{\perp}$ which induces precisely the preferred direction $\mathbf{e}^{\parallel}$.
Note that $\Psi$ must satisfy three additional independent conditions (\ref{conditions-psi}).
The kappa symmetry conditions on the pullback $\varepsilon(\sigma^{a})$ to the worldvolume $\mathbb{R}\times \Sigma \times S^{1}$ of the D5-brane will be reinterpreted in section \ref{subsection - holomorphic surfaces - kappa symmetry} as conditions on the pullback $\Psi(\sigma^{a})$ to $\mathbb{R} \times \mathcal{C} \times S^{1}$.

%---------------------------------------------------------------------------------------------
\subsection{Holomorphic surface construction}  \label{subsection - holomorphic surfaces - construction}
%---------------------------------------------------------------------------------------------

Let us begin by considering a general holomorphic surface $\mathcal{C}$ defined by $f(y_{1},y_{2},z_{1},z_{2}) = 0$ in $\mathbb{C}^{4}$. The intersection with $S^{3}_{+} \times S^{3}_{-}$ gives the 4-cycle $\Sigma$.

The orthogonal space $(\mathrm{T}\mathcal{C})^{\perp}$ is spanned by the unit vectors
\begin{eqnarray}
\nonumber && \hspace{-0.65cm} \mathbf{v}_{1} 
= \frac{ \cos^{2}{\beta} \sum_{a} \left[ (\p_{ \bar{y}_{a}} \bar{f}) \, \p_{y_{a}} + (\p_{y_{a}} f) \, \p_{\bar{y}_{a}} \right] 
+ \sin^{2}{\beta} \sum_{a} \left[ (\p_{ \bar{z}_{a}} \bar{f}) \, \p_{z_{a}} + (\p_{z_{a}} f) \, \p_{\bar{z}_{a}} \right]}
{\sqrt{\cos^{2}{\beta} \sum_{b} | \p_{y_{b}}f |^{2} + \sin^{2}{\beta} \sum_{b} | \p_{z_{b}}f |^{2}} } \\
&& \hspace{-0.65cm} \mathbf{v}_{2} = \frac{ i \, \cos^{2}{\beta} \sum_{a} \left[ (\p_{ \bar{y}_{a}} \bar{f}) \, \p_{y_{a}} - (\p_{y_{a}} f) \, \p_{\bar{y}_{a}} \right] 
+ i \, \sin^{2}{\beta} \sum_{a} \left[ (\p_{ \bar{z}_{a}} \bar{f}) \, \p_{z_{a}} - (\p_{z_{a}} f) \, \p_{\bar{z}_{a}} \right]}
{\sqrt{\cos^{2}{\beta} \sum_{b} | \p_{y_{b}}f |^{2} + \sin^{2}{\beta} \sum_{b} | \p_{z_{b}}f |^{2}} } = \mathrm{I} \, \mathbf{v}_{1} \hspace{1.0cm}
\end{eqnarray}
Note that $\mathrm{T}\mathcal{C}$ and $(\mathrm{T}\mathcal{C})^{\perp}$ are closed under the action of the complex structure $\mathrm{I}$.

We define $\mathrm{T}_{0}\Sigma$ to be the maximal subspace of $\mathrm{T}\Sigma = \mathrm{T}\mathcal{C} \cap \mathrm{T}(S^{3} \times S^{3})$ closed under the action of the complex structure $\mathrm{I}$, i.e. satisfying $\mathrm{I} \, (\mathrm{T}_{0}\Sigma) = \mathrm{T}_{0}\Sigma$.  Hence $\mathrm{T}_{0}\Sigma$ consists of all those vectors in $\mathrm{T}\Sigma$ with no components along $\mathbf{e}^{\parallel +}$ or $\mathbf{e}^{\parallel -}$. We also define the unit vectors $\mathbf{e}^{\psi +}$ and $\mathbf{e}^{\psi -}$ to be the (normalized) components of $\mathbf{e}^{\parallel +}$ or $\mathbf{e}^{\parallel -}$ in $\mathrm{T}\Sigma$:
\begin{equation} \nonumber \hspace{-1.0cm}
\mathbf{e}^{\parallel \pm} = \lambda_{\pm} \, \mathbf{e}^{\psi \pm} \, + \, \ldots
\end{equation}
with the extra terms being vectors in $(\mathrm{T}\Sigma)^{\perp} \cap \mathrm{T}(S^{3}\times S^{3})$. Here $\mathbf{e}^{\psi \pm}$ span $(\mathrm{T}_{0}\Sigma)^{\perp} \cap \mathrm{T}\Sigma$.

The direction of motion $\mathbf{e}^{\phi}$ must be a unit vector in 
$(\mathrm{T}\mathcal{C})^{\perp} \cap \mathrm{T}(S^{3} \times S^{3}) \subset (\mathrm{T} \Sigma)^{\perp} \cap \mathrm{T}(S^{3} \times S^{3})$. 
We require $(\mathrm{T}\mathcal{C})^{\perp} \cap \mathrm{T}(S^{3} \times S^{3})$ to be a 1 dimensional space for this holomorphic construction to go through,
which we shall find places constraints on the holomorphic surface $\mathcal{C}$.

Let us now solve for $\mathbf{e}^{\phi}$ explicitly.  A vector in $(\mathrm{T}\mathcal{C})^{\perp}$ takes the form
\begin{equation}   \nonumber
A \hspace{0.1cm} \sum_{a} \left[ \cos^{2}{\beta} \hspace{0.1cm} (\p_{ \bar{y}_{a}} \bar{f}) \, \p_{y_{a}} + \sin^{2}{\beta} \hspace{0.1cm} (\p_{\bar{z}_{a}} \bar{f}) \, \p_{z_{a}} \right] 
+ B \hspace{0.1cm} \sum_{a} \left[ \cos^{2}{\beta} \hspace{0.1cm} (\p_{ y_{a}} f) \, \p_{\bar{y}_{a}} + \sin^{2}{\beta} \hspace{0.1cm} (\p_{z_{a}} f) \, \p_{\bar{z}_{a}} \right] 
\end{equation}
and must be orthogonal to both $\mathbf{e}^{\perp +}$ and $\mathbf{e}^{\perp -}$ given in (\ref{orthogonal+-}). This implies
\begin{equation} \hspace{-0.5cm} \nonumber
A \hspace{0.1cm} \sum_{a} \, (\p_{ \bar{y}_{a}} \bar{f}) \, \bar{y}_{a} = -B \hspace{0.1cm} \sum_{a} \, (\p_{ y_{a}} f) \, y_{a}  \hspace{1.0cm}
A \hspace{0.1cm} \sum_{a} \, (\p_{ \bar{z}_{a}} \bar{f}) \, \bar{z}_{a} = -B \hspace{0.1cm} \sum_{a} \, (\p_{ z_{a}} f) \, z_{a}
\end{equation}
which has solution
\begin{equation} \hspace{-0.5cm} \nonumber
A = N \hspace{0.1cm} \sum_{a} \, (\p_{ y_{a}} f) \, y_{a} = k N \hspace{0.1cm} \sum_{a} \, (\p_{ z_{a}} f) \, z_{a} \hspace{0.8cm}
B = N \hspace{0.1cm} \sum_{a} \, (\p_{ \bar{y}_{a}} \bar{f}) \, \bar{y}_{a} = k N \hspace{0.1cm} \sum_{a} \, (\p_{ \bar{z}_{a}} \bar{f}) \, \bar{z}_{a}
\end{equation}
with $N$ a normalization constant.  It is therefore clear that for $(\mathrm{T}\mathcal{C})^{\perp} \cap \mathrm{T}(S^{3} \times S^{3})$ to be a 1 dimensional space containing $\mathbf{e}^{\phi}$, we must restrict to holomorphic functions $f$ satisfying
\begin{equation}
\sum_{a} \, (\p_{ y_{a}} f) y_{a} = k \, \sum_{a} \, (\p_{ z_{a}} f) z_{a}
\end{equation}
which takes the form $f(y_{1}^{m}z_{1}^{n},y_{1}^{m}z_{2}^{n},y_{2}^{m}z_{1}^{n},y_{2}^{m}z_{2}^{n})$ with $ k \equiv m/n$ in terms of $m,n \in \mathbb{Z}^{+}$. 

We wish to obtain a supersymmetric probe D5-brane giant graviton wrapping $\Sigma \times S^{1}$ which, when boosted into motion along the preferred direction $\mathbf{e}^{\parallel}$, results in motion along the direction $\mathbf{e}^{\phi}$ orthogonal to $\Sigma$.  For the supersymmetry analysis to go through along the lines of \cite{Mikhailov:2000}, it will also prove necessary to set $k=1$.  Our holomorphic surface is then given by
\begin{equation} \hspace{-1.0cm}
\boxed{\mathcal{C}:  \hspace{0.2cm} f(w_{1},w_{2},w_{3},w_{4}) = f(y_{1}z_{1},y_{1}z_{2},y_{2}z_{1},y_{2}z_{2}) = 0}
\end{equation}
in terms of the composite complex fields $w_{1} \equiv y_{1}z_{1}$, $w_{2} = y_{1}z_{2}$, $w_{3} = y_{2} z_{1}$ and $w_{4} = y_{2} z_{2}$. In this case,
\begin{equation} \hspace{-1.0cm} \nonumber
\sum_{a} \, (\p_{ y_{a}} f) \, y_{a} = \sum_{a} \, (\p_{ z_{a}} f) \, z_{a} = \sum_{a} \, (\p_{ w_{a}} f) \, w_{a}.
\end{equation}
The direction of motion $\mathbf{e}^{\phi}$ and the orthogonal direction $\mathrm{I} \, \mathbf{e}^{\phi}$ are given by
\small
\begin{eqnarray} \label{directions-phi-Iphi}
\nonumber && \hspace{-0.65cm} \mathbf{e}^{\phi} \hspace{-0.05cm} = \frac{ _{i \hspace{0.1cm} \left\{ \left[ \sum_{b} (\p_{w_{b}}f) \, w_{b} \right] 
 \left[ \cos^{2}{\beta} \hspace{0.1cm} \sum_{a} (\p_{ \bar{y}_{a}} \bar{f}) \, \p_{y_{a}} +  
 \sin^{2}{\beta} \hspace{0.1cm} \sum_{a} (\p_{\bar{z}_{a}} \bar{f}) \, \p_{z_{a}} \right]
- \left[ \sum_{b} (\p_{\bar{w}_{b}}\bar{f}) \, \bar{w}_{b} \right] 
\left[ \cos^{2}{\beta} \hspace{0.1cm} \sum_{a} (\p_{ y_{a}} f) \, \p_{\bar{y}_{a}} 
+ \sin^{2}{\beta} \hspace{0.1cm} \sum_{a}(\p_{z_{a}} f) \, \p_{\bar{z}_{a}} \right] \right\} }}
{^{| \sum_{c}(\p_{w_{c}}f) \, w_{c} | \hspace{0.1cm} 
\sqrt{\cos^{2}{\beta} \sum_{d} | \p_{y_{d}}f |^{2} + \sin^{2}{\beta} \sum_{d} | \p_{z_{d}}f |^{2}}}} \\
\nonumber && \\
\nonumber && \hspace{-0.65cm} \mathrm{I} \, \mathbf{e}^{\phi} \hspace{-0.05cm} = \frac{ _{- \left\{ \left[ \sum_{b} (\p_{w_{b}}f) \, w_{b} \right] 
 \left[ \cos^{2}{\beta} \hspace{0.1cm} \sum_{a} (\p_{ \bar{y}_{a}} \bar{f}) \, \p_{y_{a}} +   \sin^{2}{\beta} \hspace{0.1cm} 
 \sum_{a} (\p_{\bar{z}_{a}} \bar{f}) \, \p_{z_{a}} \right] 
+ \left[ \sum_{b} (\p_{\bar{w}_{b}}\bar{f}) \, \bar{w}_{b} \right] 
\left[ \cos^{2}{\beta} \hspace{0.1cm} \sum_{a} (\p_{ y_{a}} f) \, \p_{\bar{y}_{a}} 
+ \sin^{2}{\beta} \hspace{0.1cm} \sum_{a}(\p_{z_{a}} f) \, \p_{\bar{z}_{a}} \right] \right\}}}
{^{| \sum_{c}(\p_{w_{c}}f) \, w_{c} | \hspace{0.1cm} 
\sqrt{\cos^{2}{\beta} \sum_{d} | \p_{y_{d}}f |^{2} + \sin^{2}{\beta} \sum_{d} | \p_{z_{d}}f |^{2}}}}. \\
\end{eqnarray}
\normalsize
Let us now define the unit vector $\mathbf{e}^{n}$ to be a direction orthogonal to $\mathbf{e}^{\phi}$ in $ (\mathrm{T} \Sigma)^{\perp} \cap \mathrm{T}(S^{3} \times S^{3})$. Since $\mathbf{e}^{\phi}$ is a vector in $(\mathrm{T}\mathcal{C})^{\perp}$ (and $\mathrm{T}(S^{3} \times S^{3})$), which is closed under the action of the complex structure $\mathrm{I}$, the unit vector $\mathrm{I} \, \mathbf{e}^{\phi}$ is also  in $(\mathrm{T}\mathcal{C})^{\perp}$ as well as being orthogonal to $ \mathbf{e}^{\phi}$. Hence
\begin{equation}  \label{direction-phi-mu}     
\mathrm{I} \, \mathbf{e}^{\phi} = \cos{\mu} \hspace{0.15cm} ( \cos{\nu} \hspace{0.15cm} \mathbf{e}^{\perp +} + \sin{\nu} \hspace{0.15cm} \mathbf{e}^{\perp -} ) \, + \, \sin{\mu} \hspace{0.15cm} \mathbf{e}^{n}.
\end{equation}
Since $(\mathrm{I} \, \mathbf{e}^{\phi}) \cdot \mathbf{e}^{\parallel \pm}  = - \, \mathbf{e}^{\phi} \cdot \mathbf{e}^{\perp \pm} = 0$, it follows that $\mathbf{e}^{n} \cdot \mathbf{e}^{\parallel \pm}$ = 0. Therefore $\mathbf{e}^{\parallel \pm}$ are linear combinations of $\mathbf{e}^{\phi}$ and the vectors $\mathbf{e}^{\psi \pm}$. It is possible to deduce that
\begin{eqnarray} \label{relation-parallel-phi}
\nonumber && \hspace{-0.75cm} \mathbf{e}^{\parallel +} = -\cos{\mu} \cos{\nu} \hspace{0.15cm}  \mathbf{e}^{\phi} \pm \sqrt{1-\cos^{2}{\mu} \cos^{2}{\nu}} \hspace{0.2cm} \mathbf{e}^{\psi +} \\
&& \hspace{-0.75cm} \mathbf{e}^{\parallel -} = -\cos{\mu} \sin{\nu} \hspace{0.15cm}  \mathbf{e}^{\phi} \mp \sqrt{1-\cos^{2}{\mu} \sin^{2}{\nu}} \hspace{0.2cm}  \mathbf{e}^{\psi -}.
\end{eqnarray}
Using the definitions of $\mathbf{e}^{\parallel \pm}$ given in (\ref{preferred+-}) and the explicit expressions for $\mathbf{e}^{\phi}$ and $\mathrm{I} \, \mathbf{e}^{\phi}$ shown in (\ref{directions-phi-Iphi}), as well as the expressions (\ref{relation-parallel-phi}) relating $\mathbf{e}^{\parallel \pm}$ and $\mathbf{e}^{\phi}$ in more general terms, we obtain
\begin{eqnarray}
\nonumber && \hspace{-0.75cm} \mathbf{e}^{\parallel +} \cdot \, \mathbf{e}^{\phi} = \frac{\cos{\beta}}{L}  \hspace{0.15cm} 
\frac{_{| \sum_{b} (\p_{w_{b}}f) \, w_{b} |^{2}}}
{^{\sqrt{\cos^{2}{\beta} \sum_{d} | \p_{y_{d}}f |^{2} + \sin^{2}{\beta} \sum_{d} | \p_{z_{d}}f |^{2}}}} = -\cos{\mu} \cos{\nu} \\
\nonumber && \hspace{-0.75cm} \mathbf{e}^{\parallel -} \cdot \, \mathbf{e}^{\phi} = \frac{\sin{\beta}}{L}  \hspace{0.15cm} 
\frac{_{| \sum_{b} (\p_{w_{b}}f) \, w_{b} |^{2}}}{^{\sqrt{\cos^{2}{\beta} \sum_{d} | \p_{y_{d}}f |^{2} + \sin^{2}{\beta} \sum_{d} | \p_{z_{d}}f |^{2}}}}  = -\cos{\mu} \sin{\nu}
\end{eqnarray}
from which it follows that $\nu = \beta$ in this case in which $k=1$. 
Therefore we can compute
\begin{equation} \hspace{-1.0cm} \label{direction-phi}
\boxed{\mathbf{e}^{\parallel} = -\cos{\mu} \hspace{0.15cm} \mathbf{e}^{\phi} \, \pm \, \sin{\mu} \hspace{0.15cm} \mathbf{e}^{\psi}}
\end{equation}
where we define 
\begin{equation} \nonumber
\mathbf{e}^{\psi} \equiv \csc{\mu}\left\{ \cos{\beta} \sqrt{1-\cos^{2}{\mu}\cos^{2}{\beta}} \hspace{0.15cm} \mathbf{e}^{\psi +} 
- \sin{\beta} \sqrt{1-\cos^{2}{\mu}\sin^{2}{\beta}} \hspace{0.15cm} \mathbf{e}^{\psi -}  \right\}
\end{equation}
the component of $\mathbf{e}^{\parallel}$ in $\mathrm{T}\Sigma$.  Also, (\ref{direction-phi-mu}) implies that
\begin{equation} \hspace{-1.0cm} \label{direction-Iphi}
\boxed{\mathrm{I} \, \mathbf{e}^{\phi} = \cos{\mu} \hspace{0.15cm} \mathbf{e}^{\perp} \, + \, \sin{\mu} \hspace{0.15cm} \mathbf{e}^{n}}
\end{equation}
This expression for $\mathrm{I} \, \mathbf{e}^{\phi}$  (which is only valid when $k=1$ and hence $\beta = \nu$) will be essential in the following analysis of the kappa symmetry conditions for worldvolume supersymmetry.

%---------------------------------------------------------------------------------------------
\subsection{Kappa symmetry conditions}  \label{subsection - holomorphic surfaces - kappa symmetry}
%---------------------------------------------------------------------------------------------

Let us now work in the coordinates
$x^{\mu} = (t,\rho,\varphi,\hat{R}) \cup (\sigma^{1},\sigma^{2},\sigma^{3},\sigma^{4}, x^{\phi}, x^{n},R,\tilde{R}) \cup (\sigma^{5})$ with $\sigma^{a}$ the spatial worldvolume coordinates\footnote{Notice that we make use of a coordinate system adapted to describe the D5-brane worldvolume.  Thus, the Dirac matrices $\gamma_{a}$ here are not identical to those used in appendix \ref{appendix - KSEs}.}.
The kappa symmetry condition can be written as 
\begin{eqnarray} \hspace{-1.0cm}
\nonumber &&  \hspace{-1.0cm} \Gamma  \, \varepsilon = - \,i (C\varepsilon)^{\ast} 
\hspace{0.5cm} \text{with} \hspace{0.35cm}
\Gamma = - \csc{\mu}  \, ( \gamma_{0} - \dot{\phi} \, \cos{\mu} \, \gamma_{8} ) \, \gamma_{4} \gamma_{5} \gamma_{6} \gamma_{7} \gamma_{12} \\
&& \hspace{-1.0cm} \hspace{4.38cm} = - \csc{\mu}  \, ( \gamma_{0} - \dot{\phi} \, \cos{\mu} \, \gamma_{8} ) \, 
(\gamma \gamma_{10} \gamma_{11} \gamma_{12}) \, \gamma_{8} \gamma_{9} \gamma_{10} \gamma_{11} \gamma_{12}
\end{eqnarray}
imposed on $\varepsilon = \varepsilon^{+} + \varepsilon^{-}$, the pullback of the Killing spinor (\ref{Killing-Spinor}) to the worldvolume $\mathbb{R}\times \Sigma \times S^{1}$, where $i(C\varepsilon^{\pm}) = \pm \, \varepsilon^{\pm}$. 
This kappa symmetry condition can be manipulated into the form
\begin{equation} \hspace{-1.0cm}
\gamma_{0} \gamma_{8} \left[ - \cos{\mu} \, \gamma_{10}  
\pm \sin{\mu} \, \gamma_{9} \, (\gamma \gamma_{10} \gamma_{11} \gamma_{12})   \right]  \gamma_{10} \,\, \varepsilon^{\pm}  = \varepsilon^{\pm}
\end{equation}
with $\hat{\gamma} = \gamma_{0} \gamma_{1} \gamma_{2}$ and $\gamma = \gamma_{4}\gamma_{5}\gamma_{6}\gamma_{7}\gamma_{8}\gamma_{9}$.
Now $\gamma_{3}\gamma_{10} \, \varepsilon^{\pm} = \varepsilon^{\pm}$ implies $\gamma_{10} \, \varepsilon^{\pm} = -\gamma_{3} \, \varepsilon^{\pm}$ from which it follows that
\begin{equation} \hspace{-1.0cm}
(\gamma_{0} \gamma_{3}) \, \gamma_{8} \, \left[ \cos{\mu} \, \gamma_{10} 
\mp \sin{\mu} \, \gamma_{9} \, (\gamma \gamma_{10} \gamma_{11} \gamma_{12}) \right] \, \varepsilon^{\pm} = \varepsilon^{\pm}.
\end{equation}

We can project
\begin{equation}
\varepsilon^{\pm} = \mathcal{P}_{\text{dilatino}}\, \mathcal{P}_{\text{Weyl}} \, \Psi^{\pm} \hspace{0.5cm}
\text{with} \hspace{0.25cm} \mathcal{P}_{\text{dilatino}} = \tfrac{1}{2}(1 + \gamma_{3}\gamma_{10}) \hspace{0.25cm} \text{and} \hspace{0.25cm}
\mathcal{P}_{\text{Weyl}} = \tfrac{1}{2}(1 - \hat{\gamma}\gamma\gamma_{12})
\end{equation} 
 out of the covariantly constant spinors $\Psi^{\pm}$
pulled back to $\mathbb{R}\times \mathcal{C} \times S^{1}$.  Here $i(C\Psi^{\pm}) = \pm\Psi^{\pm}$. 
Notice that the operator on the left-hand side of the above equation commutes with these projection operators. Thus, the kappa symmetry condition is satisfied if
\begin{equation} \hspace{-1.0cm}
(\gamma_{0} \gamma_{3}) \, \gamma_{8} \, \left[ \cos{\mu} \, \gamma_{10}
\mp \sin{\mu} \, \gamma_{9} \, (\gamma \gamma_{10} \gamma_{11} \gamma_{12})  \right] \, \Psi^{\pm} = \Psi^{\pm}.
\end{equation}
Hence, making use of the conditions (\ref{conditions-psi}), 
\begin{equation} \hspace{-1.0cm}
(\gamma_{0} \gamma_{3}) \, \gamma_{8} \,  \left(\cos{\mu} \,\, \gamma_{10} + \sin{\mu} \,\, \gamma_{9} \right)  \, \Psi^{\pm} = \Psi^{\pm}.
\end{equation}
We further decompose $\Psi^{\pm}$ into eigenstates of $\gamma_{0}\gamma_{3}$ and $\gamma_{1}\gamma_{2}$ with eigenvalues $i\,  r_{a}$, as well as into eigenstates of $\gamma^{y}_{4}\gamma^{y}_{6}$,  $\gamma^{y}_{5}\gamma^{y}_{7}$, $\gamma^{z}_{8}\gamma^{z}_{10}$ and $\gamma^{z}_{9}\gamma^{z}_{11}$ with eigenvalues $i \, s^{y}_{1}$, $i \, s^{y}_{2}$, $i \, s^{z}_{1}$  and $i \, s^{z}_{2}$.  Here $(r_{1},r_{2},s^{y}_{1},s^{y}_{2},s^{z}_{1},s^{z}_{2}) = (\pm,\pm,\pm,\pm,\pm,\pm)$ label the spinor $\Psi^{p}_{(r_{1},r_{2},s^{y}_{1},s^{y}_{2},s^{z}_{1},s^{z}_{2})}$ together with our original label $p = \pm$, where 
$i(C\Psi^{p})^{\ast} = p \, \Psi^{p}$. We note that the three conditions (\ref{conditions-psi}) imply
\begin{equation} \label{conditions-labels} \hspace{-0.5cm}
r_{1}r_{2} = -p \hspace{1.0cm} s^{y}_{1}s^{y}_{2} = p \hspace{1.0cm} \text{and} \hspace{1.0cm}
s^{z}_{1}s^{z}_{2} = p.
\end{equation}
The kappa symmetry condition now becomes
\begin{equation} \hspace{-1.0cm}
\boxed{\Gamma(\mathbf{e}^{\phi}) \hspace{0.1cm} \Gamma(\mathrm{I} \, \mathbf{e}^{\phi}) \hspace{0.15cm} \Psi^{p}_{(r_{1},r_{2},s^{y}_{1},s^{y}_{2},s^{z}_{1},s^{z}_{2})}
=  - i \,r_{1} \, \Psi^{p}_{(r_{1},r_{2},s^{y}_{1},s^{y}_{2},s^{z}_{1},s^{z}_{2})}}
\end{equation}
Here we note that $\Gamma(\mathbf{e}^{\phi}) = \gamma_{8}$, $\Gamma(\mathbf{e}^{n}) = \gamma_{9}$ and $\Gamma(\mathbf{e}^{\perp}) = \gamma_{10}$. We also make
use of (\ref{direction-Iphi}), which we obtained only for our particular choice of holomorphic surface $\mathcal{C}$, from which it follows that
$\Gamma(\mathbf{e}^{\phi}) \hspace{0.1cm} \Gamma(\mathrm{I} \, \mathbf{e}^{\phi}) = \gamma_{8} \, \left(\cos{\mu} \hspace{0.1cm} \gamma_{10} + \sin{\mu} \hspace{0.15cm} \gamma_{9} \right).$

It is possible from (\ref{directions-phi-Iphi}) to compute
\begin{eqnarray}
\nonumber && \hspace{-0.65cm} 
\Gamma(\mathbf{e}^{\phi}) \hspace{0.1cm} \Gamma(\mathrm{I} \, \mathbf{e}^{\phi}) 
= - \, i \, + \, \frac{ _{2 \hspace{0.015cm} i \, 
\sum_{a,b} [ \cos^{2}{\beta} \hspace{0.1cm} (\p_{y_{a}} f) \, \Gamma_{\bar{y}_{a}} + \sin^{2}{\beta} \hspace{0.1cm} (\p_{z_{a}} f) \, \Gamma_{\bar{z}_{a}} ] 
[ \cos^{2}{\beta} \hspace{0.1cm} (\p_{\bar{y}_{b}} \bar{f}) \, \Gamma_{y_{b}} + \sin^{2}{\beta} \hspace{0.1cm} (\p_{\bar{z}_{b}} \bar{f}) \, \Gamma_{z_{b}} ] }}
{ ^{[ \cos^{2}{\beta} \sum_{d} | \p_{y_{d}}f |^{2} + \sin^{2}{\beta} \sum_{d} | \p_{z_{d}} f |^{2} ]} } \\
\nonumber && \\
&& \hspace{-0.65cm} \hspace{2.285cm} = + \, i \, - \, \frac{_{2 \hspace{0.015cm} i \, 
\sum_{a,b} [\cos^{2}{\beta} \hspace{0.1cm} (\p_{\bar{y}_{a}} \bar{f}) \, \Gamma_{y_{a}} + \sin^{2}{\beta} \hspace{0.1cm} (\p_{\bar{z}_{a}} \bar{f}) \, \Gamma_{z_{a}} ]
[\cos^{2}{\beta} \hspace{0.1cm} (\p_{y_{b}} f) \, \Gamma_{\bar{y}_{b}} + \sin^{2}{\beta} \hspace{0.1cm} (\p_{z_{b}} f) \, \Gamma_{\bar{z}_{b}} ] } }
{ ^{[ \cos^{2}{\beta} \sum_{d} | \p_{y_{d}}f |^{2} + \sin^{2}{\beta} \sum_{d} | \p_{z_{d}}f |^{2} ]} }. \hspace{1.0cm}
\end{eqnarray}
We observe that these conditions are satisfied when the following conditions are imposed on the spinor labels:  If $\p_{y_{a}} f \neq 0$, then
\begin{eqnarray}
&& \hspace{-0.65cm} \Gamma_{y_{a}} \, \Psi^{p}_{(+,r_{2},s^{y}_{1},s^{y}_{2},s^{z}_{1},s^{z}_{2})} = 0 
\hspace{0.25cm} \Longrightarrow \hspace{0.2cm} s^{y}_{a} = -
\hspace{0.4cm} \text{and} \hspace{0.4cm} \Gamma_{\bar{y}_{a}} \, \Psi^{p}_{(-,r_{2},s^{y}_{1},s^{y}_{2},s^{z}_{1},s^{z}_{2})} = 0 
\hspace{0.25cm} \Longrightarrow \hspace{0.2cm} s^{y}_{a} = +  \hspace{1.0cm}
\end{eqnarray}
from which it follows that $r_{1}s^{y}_{a} = -$.  Similarly, if $\p_{z_{a}} f \neq 0$, then
\begin{eqnarray}
&& \hspace{-0.65cm} \Gamma_{z_{a}} \, \Psi^{p}_{(+,r_{2},s^{y}_{1},s^{y}_{2},s^{z}_{1},s^{z}_{2})} = 0 
\hspace{0.25cm} \Longrightarrow \hspace{0.2cm} s^{z}_{a} = -
\hspace{0.4cm} \text{and} \hspace{0.4cm} \Gamma_{\bar{z}_{a}} \, \Psi^{p}_{(-,r_{2},s^{y}_{1},s^{y}_{2},s^{z}_{1},s^{z}_{2})} = 0
\hspace{0.25cm} \Longrightarrow \hspace{0.2cm} s^{z}_{a} = + \hspace{1.0cm}
\end{eqnarray}
which gives $r_{1}s^{z}_{a} = -$.  Notice that, if $f$ is a function of both $y_{1}$ and $y_{2}$ (or both $z_{1}$ and $z_{2}$), then $p$ is no longer a free label, since $p = s^{y}_{1}s^{y}_{2} = s^{z}_{1} s^{z}_{2} = +$ is fixed by conditions (\ref{conditions-labels}).  These conditions also imply that $r_{2}$ is fixed by $p$ and $r_{1}$.

\emph{To summarize:}
\vspace{-0.2cm}

For a D5-brane giant graviton wrapping a 4-cycle $\Sigma$ in $S^{3}_{+} \times S^{3}_{-}$ and $S^{1}$, constructed with $\Sigma$ the intersection a general holomorphic surface $f(y_{1}z_{1},y_{1}z_{2},y_{2}z_{1},y_{2}z_{2}) = 0$ with the $S^{3}_{+}\times S^{3}_{-}$ submanifold, the kappa symmetry conditions on the spinor
\begin{equation} 
\varepsilon^{p}_{(r_{1},r_{2},s^{y}_{1},s^{y}_{2},s^{z}_{1},s^{z}_{2})}(\sigma^{a}) \, = \, \mathcal{P}_{\text{dilatino}}\, \mathcal{P}_{\text{Weyl}} \hspace{0.15cm} \Psi^{p}_{(r_{1},r_{2},s^{y}_{1},s^{y}_{2},s^{z}_{1},s^{z}_{2})}(\sigma^{a})
\end{equation} 
are satisfied if the labels obey the following relations:
\begin{equation} \hspace{-1.0cm}
-r_{1} = r_{2} = s^{y}_{1} = s^{y}_{2} = s^{z}_{1} = s^{z}_{2} \hspace{0.75cm} \text{and} \hspace{0.75cm} p = +
\end{equation}
indicating a $\frac{1}{8}$-BPS D5-brane giant graviton preserving 2 of 16 background supersymmetries.  

There is supersymmetry enhancement in the case of a holomorphic surface $f(y_{1}z_{1})=0$ with the function dependent only on a single composite complex coordinate.  The kappa symmetry conditions are satisfied, in this special case, if the spinor labels obey
 \begin{equation} \hspace{-1.0cm}
-r_{1} = p \,r_{2} = s^{y}_{1} = s^{z}_{1} \hspace{0.75cm} \text{and} \hspace{0.75cm} s^{y}_{2} = s^{z}_{2} = p.
\end{equation}
This indicates a $\tfrac{1}{4}$-BPS D5-brane giant graviton preserving 4 of the original 16 background supersymmetries.
The D5-brane giant graviton embedded into $\mathbb{R} \times S^{3}_{+}\times S^{3}_{-}\times S^{1}$, constructed explicitly in section \ref{section - examples}, is an example in this special $\tfrac{1}{4}$-BPS subclass of D5-brane giants.

%%%%%%%%%%%%%%%%%%%%%%%%%%%%%%%%%%%%%%%%%%%%%%%%%%%%%%%%%%%%%%%%%%%%%%%%%%%%%%%%%%%%%%%%%%%%%%
\section{Discussion}  \label{section - discussion}
%%%%%%%%%%%%%%%%%%%%%%%%%%%%%%%%%%%%%%%%%%%%%%%%%%%%%%%%%%%%%%%%%%%%%%%%%%%%%%%%%%%%%%%%%%%%%%

We have constructed various examples of $\tfrac{1}{4}$-BPS giant gravitons in the type IIB supergravity background $AdS_{3}\times S^{3}_{+}\times S^{3}_{-}\times S^{1}$ with pure R-R flux.  These probe D1 and D5-branes are supported by their coupling to the $C_{(2)}$ and $C_{(6)}$ R-R background potentials, and by  angular momenta  $\alpha \, P$ on one 3-sphere $S^{3}_{+}$ and $(1-\alpha) \, P$ on the other 3-sphere $S^{3}_{-}$, where $P$ is the total angular momentum and $\alpha \equiv \cos^{2}{\beta}$ controls the relative size of the 3-spheres.  We found two examples of D1-brane giant gravitons wrapping a circle in $AdS_{3}$ and a 1-cycle on a torus in $S_{+}^{3} \times S_{-}^{3}$, respectively. These D1-brane giant gravitons reduce to $\tfrac{1}{2}$-BPS AdS and sphere giants on $AdS_{3}\times S^{3} \times T^{4}$ in the $\alpha \to 0$ or $\alpha \to 1$ limits. As for giants in $AdS_{3}\times S^{3}$ \cite{McGreevy:2000,Hashimoto:2000}, the energy and angular momentum of these D1-brane giant gravitons
\[ H = \pm \, P = \frac{Q_{5}^{+} Q_{5}^{-}}{(Q_{5}^{+} + Q_{5}^{-})} \]
are independent of the size of the 1-cycle in $AdS_{3}$ or $S_{+}^{3} \times S_{-}^{3}$.  The potential is flat and there is no stringy exclusion principle.  This indicates that these $\tfrac{1}{4}$-BPS D1-branes have separated, at no cost in energy, from the D1-D5-D5$^{\prime}$-branes setting up the geometry.
We also found a $\tfrac{1}{4}$-BPS D5-brane giant graviton wrapping a 4-cycle in $S_{+}^{3}\times S_{-}^{3}$ and the $S^{1}$. We do not expect this D5-brane solution to survive the $\alpha \to 0$ or $\alpha \to 1$ limits (except for the maximal giant) and subsequent compactification to $AdS_{3}\times S^{3} \times T^{4}$.
Moreover, this D5-brane giant graviton does not exhibit the perculiar properties of AdS and sphere giants in $AdS_{3}\times S^{3}$.  The energy and angular momentum
\[ H = \pm \, P = Q_{1} \sin^{2}{\theta} \]
 does depend on a parameter $\theta$ which controls the size and shape of the 4-cycle in $S_{+}^{3} \times S_{-}^{3}$.  The maximal giant graviton has energy $H = \pm P = Q_{1}$, indicating a stringy exclusion principle, and consists of two $\tfrac{1}{2}$-BPS D5-branes, each wrapping an $S_{\pm}^{3}\times S^{1} \times S^{1}$.

We have further constructed a general class of $\tfrac{1}{8}$-BPS D5-brane giant gravitons wrapping a 4-cycle $\Sigma$ in $S^{3}_{+} \times S^{3}_{-}$ and 
the $S^{1}$ at fixed time $t=0$. Here $\Sigma$ is the intersection of a holomorphic surface $\mathcal{C}$ in $\mathbb{C}^{2}_{+} \times \mathbb{C}^{2}_{-}$, defined by
\begin{equation} \hspace{-1.0cm} \nonumber
\mathcal{C}: \hspace{0.2cm} f(y_{1}z_{1},y_{1}z_{2},y_{2}z_{1},y_{2}z_{2}) = 0,
\end{equation}
with the submanifold $S^{3}_{+}\times S^{3}_{-}$, where $y_{a}$ and $z_{a}$ are the $\mathbb{C}^{2}_{+}$ and $\mathbb{C}^{2}_{-}$ complex coordinates.  This surface is boosted into motion along the preferred direction $\mathbf{e}^{\parallel} = \cos{\beta}\hspace{0.15cm} \mathbf{e}^{\parallel +} + \sin{\beta} \hspace{0.15cm} \mathbf{e}^{\parallel -}$ in $\mathrm{T}(S^{3}_{+}\times S^{3}_{-})$, with $\mathbf{e}^{\parallel\pm} = \mathrm{I} \, \mathbf{e}^{\perp \pm}$  the natural preferred directions induced by the action of the complex structure $\mathrm{I}$ on the directions $\mathbf{e}^{\perp \pm}$ orthogonal to $\mathrm{T}S^{3}_{\pm}$ in $\mathrm{T}\mathbb{C}^{2}_{\pm}$.  This choice of $\alpha$-dependent preferred direction is crucial to this holomorphic surface construction.

This work leads naturally to a consideration of the maximal D1-brane giant graviton in $\mathbb{R} \times S^{3}_{+} \times S^{3}_{-}$ and either half of the maximal D5-brane giant graviton in $\mathbb{R} \times S^{3}_{+} \times S^{3}_{-} \times S^{1}$ as integrable boundary conditions for open IIB superstrings on $AdS_{3}\times S^{3}_{+}\times S^{3}_{-}\times S^{1}$.  These are described by a coset model \cite{Babichenko:2010}. We leave the study of open string integrability in this $AdS_{3}\times S^{3}_{+} \times S^{3}_{-}\times S^{1}$ background for future research.

This new class of $\tfrac{1}{8}$-BPS D5-brane giant gravitons will be dual to $\tfrac{1}{8}$-BPS operators in the dual gauge theory, which is conjectured to arise from the $\mathcal{N}=(0,4)$ supersymmetric gauge theory of \cite{Tong:2014} when it flows in the infrared to a CFT$_{2}$ with enhanced $\mathcal{N}=(4,4)$ supersymmetry.  We expect that it should be possible to build these protected operators in the gauge theory of \cite{Tong:2014} from equal numbers $n \sim O(Q_{1})$ of the $Y_{a}$ and $Z_{a}$ complex scalar fields, which carry charges under different $SU(2)_{\pm}$'s and transform in the adjoint of the $U(Q_{1})$ gauge group. It seems natural to associate these $Y_{a}$ and $Z_{a}$ fields with the even and odd sites of the alternating $\mathfrak{d}(2,1;\alpha)^{2}$ spin chain of \cite{Sax-Stefanski:2011,Borsato-et-al:2012,Sax-et-al:2013} in the limit in which the number of fields $n \ll Q_{1}$ is small in comparison to the rank of the $U(Q_{1})$ gauge group. It was shown in $\mathcal{N}=4$ SYM theory that $\tfrac{1}{4}$-BPS operators can be built using representations of Brauer algebras \cite{Kimura:2010}. We conjecture that similar protected operators may be built out of the $Y_{1}$ and $Z_{1}$ scalars in the field theory \cite{Tong:2014} dual to our D5-brane giant graviton example in section \ref{subsection - examples - D5}. 

It would be interesting to compute the spectrum of small fluctuation about our $\tfrac{1}{4}$-BPS giant graviton solutions.  We expect an $\alpha$-dependent spectrum.  The fluctuation spectrum of the D5-brane giant graviton example should also exhibit a dependence on the parameter $\theta$, due to the changing shape of the worldvolume with $\theta$.  This fluctuation spectrum will be dual to the anomalous dimension spectrum of excitations of a $\tfrac{1}{4}$-BPS operator in the CFT$_{2}$.  We leave this as a topic for future investigation.

%%%%%%%%%%%%%%%%%%%%%%%%%%%%%%%%%%%%%%%%%%%%%%%%%%%%%%%%%%%%%%%%%%%%%%%%%%%%%%%%%%%%%%%%%%%%%%
\section*{Acknowledgements}
%%%%%%%%%%%%%%%%%%%%%%%%%%%%%%%%%%%%%%%%%%%%%%%%%%%%%%%%%%%%%%%%%%%%%%%%%%%%%%%%%%%%%%%%%%%%%%

I would like to thank Jan Gutowski, Antonio Pittelli, Vidas Regelskis, Bogdan Stefa\'{n}ski, Jr. and Alessandro Torrielli for useful discussions.

\appendix

%%%%%%%%%%%%%%%%%%%%%%%%%%%%%%%%%%%%%%%%%%%%%%%%%%%%%%%%%%%%%%%%%%%%%%%%%%%%%%%%%%%%%%%%%%%%%%
\section{Killing spinor equations} \label{appendix - KSEs}
%%%%%%%%%%%%%%%%%%%%%%%%%%%%%%%%%%%%%%%%%%%%%%%%%%%%%%%%%%%%%%%%%%%%%%%%%%%%%%%%%%%%%%%%%%%%%%

%---------------------------------------------------------------------------------------------
\subsection{Covariantly constant spinor $\Psi$ on $\mathbb{R}^{2+2} \times \mathbb{C}_{+}^{2} \times \mathbb{C}_{-}^{2} \times S^{1}$}  \label{subappendix - KSEs - 13D}
%---------------------------------------------------------------------------------------- -----

We begin by introducing a covariantly constant spinor $\Psi = \Psi_{1} + i \Psi_{2}$ in the 2+11 dimensional $\mathbb{R}^{2+2} \times \mathbb{C}^{2} \times \mathbb{C}^{2} \times S^{1}$ spacetime.  The 32-component spinors $\Psi_{k}$ satisfy the Majorana condition $(C \Psi_{k})^{*} = \Psi_{k}$ and, since $\gamma_{0} \cdots \gamma_{12} = \mathbf{1}$, there is no Weyl condition in odd dimensions. Here
\begin{equation} \hspace{-0.5cm} \nonumber
\mathcal{D}_{\mu}\Psi = 0 \hspace{1.2cm} \text{with} \hspace{0.6cm} \mathcal{D}_{\mu} = \p_{\mu} + \tfrac{1}{4} \, (\omega^{\alpha\beta})_{\mu} \, \gamma_{\alpha\beta}
\end{equation} 
are the supercovariant derivatives, and $\Gamma_{\mu} = (\hat{e}^{\alpha})_{\mu} \gamma_{\alpha}$ the curved gamma matrices written in terms of the vielbiens $\hat{e}^{\alpha}$ and the flat gamma matrices\footnote{We make use of conventions in which $-(\gamma_{0})^{2} = -(\gamma_{3})^{2} = (\gamma_{1})^{2} = (\gamma_{2})^{2} = (\gamma_{4})^{2} = \cdots = (\gamma_{12})^{2} = \mathbf{1}$. We choose a representation of the 2+11 dimensional Lorentz group such that $\gamma_{8}$, $\gamma_{9}$, $\gamma_{10}$ and $\gamma_{12}$ are purely imaginary, while $\gamma_{0}$, $\gamma_{1}$, $\gamma_{2}$, $\gamma_{3}$, $\gamma_{4}$, $\gamma_{5}$, $\gamma_{6}$, $\gamma_{7}$ and $\gamma_{11}$ are real.  The charge conjugation matrix is $C = \gamma_{8}\gamma_{9}\gamma_{10}\gamma_{12}$.} $\gamma_{\alpha}$. 
We shall solve explicitly for $\Psi(x^{\mu}) = \mathcal{M}(x^{\mu}) \, \Psi_{0}$, with $\Psi_{0}$ a constant spinor, making use of various coordinates.

%%%%%%%%%%
\subsubsection*{$AdS_{3} \times S_{+}^{3} \times S_{-}^{3} \times S^{1}$ coordinates with extra radial coordinates}
%%%%%%%%%%

We consider $AdS_{3} \times S_{+}^{3} \times S_{-}^{3} \times S^{1}$ coordinates, together with the extra radial coordinates $\hat{R}$, $R_{+}$ and $R_{-}$ shown in the metric (\ref{metric-flat}).

\vspace{-0.1cm}
$\boxed{\mathbb{R}^{2+2}}$
\vspace{-0.2cm}

In the global $AdS_{3}$ coordinates $(t,\rho,\varphi)$ and the radial coordinate $\hat{R}$, the vielbiens are
\begin{eqnarray}
\nonumber && \hspace{-1.0cm} \hat{e}^{0} = \hat{R} \, \cosh{\rho} \hspace{0.15cm} dt \hspace{1.2cm} 
\hat{e}^{1} = \hat{R} \hspace{0.15cm} d\rho \hspace{1.2cm} 
\hat{e}^{2} = \hat{R} \, \sinh{\rho} \hspace{0.15cm} d\varphi \hspace{1.2cm} 
\hat{e}^{3} = d\hat{R}.
\end{eqnarray}
The spin connections $\hat{\omega}^{\alpha}_{\beta}=(\omega^{\alpha\beta} )_{\mu} \, dx^{\mu}$, satisfying $d \hat{e}^{\alpha} + \omega^{\alpha}_{\hspace{0.15cm}\beta} \wedge \hat{e}^{\beta} = 0$, can be separated into the $AdS_{3}$ spin connections
\begin{eqnarray}
\nonumber && \hspace{-1.0cm} \hat{\omega}^{01} = - \hspace{0.05cm} \hat{\omega}^{10} = \sinh{\rho} \hspace{0.15cm} dt \hspace{1.51cm} 
 \hat{\omega}^{12} = - \hspace{0.05cm}  \hat{\omega}^{21}=-\cosh{\rho}\hspace{0.15cm} d\varphi  
\end{eqnarray}
and the additional $\mathbb{R}^{2+2}$ spin connections $\hat{\omega}^{\alpha 3} = - \hspace{0.05cm}  \hat{\omega}^{3\alpha} = - \hspace{0.05cm}  \hat{R}^{-1} \hspace{0.1cm} \hat{e}^{\alpha}$ for  $\alpha = 0,1,2$. 
The  conditions for a covariantly constant spinor $\Psi$ in $\mathbb{R}^{2+2}$ therefore take the form
\begin{eqnarray} \label{covariantly-constant-R22}
\nonumber && \hspace{-1.0cm} \mathcal{D}_{t} \Psi = \nabla_{t} \Psi - \tfrac{1}{2} \cosh{\rho} \hspace{0.1cm} (\gamma_{0} \gamma_{3}) \, \Psi = 0 
\hspace{0.86cm} \text{with} \hspace{0.5cm} \nabla_{t} = \p_{t} + \tfrac{1}{2} \, \sinh{\rho} \hspace{0.1cm} (\gamma_{0}\gamma_{1}) \\
\nonumber && \hspace{-1.0cm} \mathcal{D}_{\rho} \Psi = \nabla_{\rho} \Psi - \tfrac{1}{2} \, (\gamma_{1}\gamma_{3})\, \Psi = 0 
\hspace{1.875cm} \text{with} \hspace{0.5cm} \nabla_{\rho} = \p_{\rho} \\
\nonumber && \hspace{-1.0cm} \mathcal{D}_{\varphi} \Psi =  \nabla_{\varphi} \Psi - \tfrac{1}{2} \sinh{\rho} \hspace{0.1cm} (\gamma_{2} \gamma_{3}) \, \Psi = 0 
\hspace{0.75cm} \text{with} \hspace{0.5cm} \nabla_{\varphi} = \p_{\varphi} - \tfrac{1}{2} \, \cosh{\rho} \hspace{0.1cm} (\gamma_{1}\gamma_{2}) \\
&& \hspace{-1.0cm} \mathcal{D}_{\hat{R}} \Psi = 0. 
\end{eqnarray}
Here $\nabla_{t}$, $\nabla_{\rho}$ and $\nabla_{\varphi}$ are the supercovariant derivatives in $AdS_{3}$.

$\boxed{\mathbb{C}_{+}^{2}}$
\vspace{-0.2cm} 

In the $S_{+}^{3}$ coordinates $(\theta_{+},\chi_{+},\phi_{+})$ and radial coordinate $R_{+}$, the vielbiens are
\begin{eqnarray} 
\nonumber && \hspace{-0.65cm} \hat{e}^{4} = R_{+} \, \sec{\beta} \hspace{0.15cm} d\theta_{+}  \hspace{0.75cm} 
\hat{e}^{5} = R_{+} \, \sec{\beta} \, \cos{\theta_{+}}\hspace{0.1cm} d\chi_{+} \hspace{0.75cm} 
\hat{e}^{6} = R_{+} \, \sec{\beta} \, \sin{\theta_{+}} \hspace{0.1cm} d\phi_{+}  \hspace{0.75cm} 
\hat{e}^{7} = \sec{\beta} \hspace{0.1cm} dR_{+}. \hspace{1.0cm}
\end{eqnarray}
The spin connections can be separated into the $S_{+}^{3}$ spin connections
\begin{eqnarray}
\nonumber &&  \hspace{-1.0cm} \hat{\omega}^{45} = -\hat{\omega}^{54} = \sin{\theta_{+}} \hspace{0.1cm} d\chi_{+}  \hspace{1.2cm} \hat{\omega}^{46} = -\hat{\omega}^{64} = - \cos{\theta_{+}} \hspace{0.1cm} d\phi_{+} 
\end{eqnarray}
with the additional $\mathbb{C}_{+}^{2}$ spin connections 
$\hat{\omega}^{\alpha 7} = -\hat{\omega}^{7 \alpha} = -R_{+}^{-1} \, \cos{\beta} \hspace{0.2cm} \hat{e}^{\alpha}$ for $\alpha = 4,5,6$. 
The  conditions for a covariantly constant spinor $\Psi$ in $\mathbb{C}_{+}^{2}$ are
\begin{eqnarray} \label{covariantly-constant-C2+}
\nonumber && \hspace{-1.0cm} \mathcal{D}_{\theta_{+}} \Psi = \nabla_{\theta_{+}} \Psi + \tfrac{1}{2} \, (\gamma_{4} \gamma_{7}) \, \Psi = 0 
\hspace{01.875cm} \text{with} \hspace{0.5cm} \nabla_{\theta_{+}} = \p_{\theta_{+}}  \\
\nonumber && \hspace{-1.0cm} \mathcal{D}_{\chi_{+}} \Psi = \nabla_{\chi_{+}} \Psi + \tfrac{1}{2} \cos{\theta_{+}} \, (\gamma_{5}\gamma_{7})\, \Psi = 0 
\hspace{0.725cm} \text{with} \hspace{0.5cm} \nabla_{\chi_{+}} = \p_{\chi_{+}} + \tfrac{1}{2} \, \sin{\theta_{+}} \hspace{0.1cm} (\gamma_{4}\gamma_{5}) \\
\nonumber && \hspace{-1.0cm} \mathcal{D}_{\phi_{+}} \Psi =  \nabla_{\phi_{+}} \Psi + \tfrac{1}{2} \sin{\theta_{+}} \hspace{0.1cm} (\gamma_{6} \gamma_{7}) \, \Psi 
= 0 
\hspace{0.75cm} \text{with} \hspace{0.5cm} \nabla_{\phi_{+}} = \p_{\phi_{+}} - \tfrac{1}{2} \, \cos{\theta_{+}} \hspace{0.1cm} (\gamma_{4}\gamma_{6}) \hspace{0.5cm}\\
&& \hspace{-1.0cm} \mathcal{D}_{R_{+}} \Psi = 0. 
\end{eqnarray}
Here $\nabla_{\theta_{+}}$, $\nabla_{\chi_{+}}$ and $\nabla_{\phi_{+}}$ are the supercovariant derivatives of $S_{+}^{3}$.

\vspace{-0.1cm}
$\boxed{\mathbb{C}_{-}^{2}}$
\vspace{-0.2cm}

The manifold $\mathbb{C}_{-}^{2}$ in the $S_{-}^{3}$ coordinates $(\theta_{-},\chi_{-},\phi_{-})$ and radial coordinate $R_{-}$ has vielbiens
\begin{eqnarray} 
&& \nonumber \hspace{-0.65cm} \hat{e}^{8} = R_{-} \, \csc{\beta} \hspace{0.15cm} d\theta_{-}  \hspace{0.75cm} 
\hat{e}^{9} = R_{-} \, \csc{\beta} \, \cos{\theta_{-}}\hspace{0.1cm} d\chi_{-} \hspace{0.75cm} 
\hat{e}^{10} = R_{-} \, \csc{\beta} \, \sin{\theta_{-}} \hspace{0.1cm} d\phi_{-}  \hspace{0.75cm} 
\hat{e}^{11} = \csc{\beta} \hspace{0.1cm} dR_{-} \hspace{1.0cm}
\end{eqnarray}
while the spin connections can be separated into the $S_{+}^{3}$ spin connections
\begin{eqnarray}
\nonumber &&  \hspace{-1.0cm} \hat{\omega}^{45} = -\hat{\omega}^{54} = \sin{\theta_{-}} \hspace{0.1cm} d\chi_{-}  \hspace{1.2cm} \hat{\omega}^{46} = -\hat{\omega}^{64} = - \cos{\theta_{-}} \hspace{0.1cm} d\phi_{-} 
\end{eqnarray}
with the additional $\mathbb{C}_{-}^{2}$ spin connections 
$\hat{\omega}^{\alpha 7} = -\hat{\omega}^{3k} = -R_{-}^{-1} \, \sin{\beta} \hspace{0.2cm} \hat{e}^{\alpha}$ for $\alpha = 8,9,10$.
The  conditions for a covariantly constant spinor $\Psi$ in $\mathbb{C}_{-}^{2}$ are
\begin{eqnarray} \label{covariantly-constant-C2-}
\nonumber && \hspace{-1.0cm} \mathcal{D}_{\theta_{-}} \Psi = \nabla_{\theta_{-}} \Psi + \tfrac{1}{2} \, (\gamma_{8} \gamma_{11}) \, \Psi = 0 
\hspace{2.025cm} \text{with} \hspace{0.5cm} \nabla_{\theta_{-}} = \p_{\theta_{-}}  \\
\nonumber && \hspace{-1.0cm} \mathcal{D}_{\chi_{-}} \Psi = \nabla_{\chi_{-}} \Psi + \tfrac{1}{2} \cos{\theta_{-}} \, (\gamma_{9}\gamma_{11})\, \Psi = 0 
\hspace{0.85cm} \text{with} \hspace{0.5cm} \nabla_{\chi_{-}} = \p_{\chi_{-}} + \tfrac{1}{2} \, \sin{\theta_{-}} \hspace{0.1cm} (\gamma_{8}\gamma_{9}) \\
\nonumber && \hspace{-1.0cm} \mathcal{D}_{\phi_{-}} \Psi =  \nabla_{\phi_{-}} \Psi + \tfrac{1}{2} \sin{\theta_{-}} \hspace{0.1cm} (\gamma_{10} \gamma_{11}) \, \Psi 
= 0 
\hspace{0.75cm} \text{with} \hspace{0.5cm} \nabla_{\varphi} = \p_{\phi_{-}} - \tfrac{1}{2} \, \cos{\theta_{-}} \hspace{0.1cm} (\gamma_{8}\gamma_{10}) \hspace{0.5cm}\\
&& \hspace{-1.0cm} \mathcal{D}_{R_{-}} \Psi = 0. 
\end{eqnarray}
Here $\nabla_{\theta_{-}}$, $\nabla_{\chi_{-}}$ and $\nabla_{\phi_{-}}$ are the supercovariant derivatives of $S_{-}^{3}$.

\vspace{-0.1cm}
$\boxed{S^{1}}$
\vspace{-0.2cm}

The vielbien is $ \hat{e}^{12} = \ell \hspace{0.15cm} d\xi$ and the curved spacetime gamma matrix is $\Gamma_{\xi} = \ell \hspace{0.1cm} \gamma_{12}$.  The condition for a convariantly constant spinor $\Psi$ in $S^{1}$ is simply
\begin{equation} \hspace{-1.0cm} \label{covariantly-constant-S1}
\mathcal{D}_{\xi} \Psi = \nabla_{\xi} \Psi = 0 \hspace{1.0cm} \text{with} \hspace{0.5cm} \nabla_{\xi} = \p_{\xi}.
\end{equation}

\begin{tabular}{|p{\textwidth}|}
\hline
\vspace{-0.1cm}
The covariantly constant spinor in $\mathbb{R}^{2+2} \times \mathbb{C}_{+}^{2} \times \mathbb{C}_{-}^{2} \times S^{1}$ can be written as $\Psi(x^{\mu}) = \mathcal{M}(x^{\mu}) \, \Psi_{0}$ in terms of
\begin{eqnarray}
\nonumber && \hspace{-1.0cm} \mathcal{M}(x^{\mu}) = 
e^{\frac{1}{2} (\gamma_{1} \gamma_{3}) \, \rho} \hspace{0.15cm}  e^{ \frac{1}{2} (\gamma_{0} \gamma_{3}) \, t} \hspace{0.15cm} 
e^{\frac{1}{2} (\gamma_{1} \gamma_{2}) \, \varphi}  \hspace{0.15cm} \\
&& \hspace{-1.0cm} \hspace{1.65cm} 
\times \, e^{ - \frac{1}{2} (\gamma_{4}\gamma_{7}) \, \theta_{+}} \hspace{0.15cm} e^{ - \frac{1}{2} (\gamma_{5}\gamma_{7}) \, \chi_{+} } \hspace{0.15cm}
e^{\frac{1}{2} (\gamma_{4}\gamma_{6})  \, \phi_{+} }  \hspace{0.15cm}
e^{- \frac{1}{2} (\gamma_{8}\gamma_{11}) \, \theta_{-}} \hspace{0.15cm} e^{-\frac{1}{2} (\gamma_{9}\gamma_{11}) \, \chi_{-} } \hspace{0.15cm}
e^{\frac{1}{2} (\gamma_{8}\gamma_{10}) \, \phi_{-} }  \hspace{0.5cm}
\end{eqnarray}
with $\Psi_{0}$ a constant spinor. 
\vspace{0.2cm}
\\
\hline
\end{tabular}

%%%%%%%%%%
\subsubsection*{The complex manifolds $\mathbb{C}_{\pm}^{2}$ in terms of the radii and phases}
%%%%%%%%%%

Let us consider the $\mathbb{R}^{2+2}$ and $S^{1}$ coordinates of the previous subsection ($AdS_{3}$ coordinates with the extra radial coordinate $\hat{R}$, together with $\xi$), but let us now make use of the radii and phases of the complex $\mathbb{C}_{\pm}^{2}$ coordinates $y_{a}$ and $z_{a}$ shown in the metric (\ref{metric-flat-radii-and-phases}).

\vspace{-0.1cm}
$\boxed{\mathbb{C}_{+}^{2}}$
\vspace{-0.2cm}

In the radii and phases of the $\mathbb{C}^{2}_{+}$ complex coordinates $y_{a} = r_{y,a} \, e^{i \hspace{0.05cm} \psi_{y,a}}$, the vielbiens are
\begin{eqnarray} \nonumber
&& \hspace{-0.65cm} (\hat{e}^{y})^{4} = \sec{\beta} \hspace{0.15cm} dr_{y,1}  \hspace{0.75cm} 
(\hat{e}^{y})^{5} = \sec{\beta} \hspace{0.15cm} dr_{y,2} \hspace{0.75cm} 
(\hat{e}^{y})^{6} = \sec{\beta} \hspace{0.15cm} r_{y,1} \hspace{0.1cm} d\psi_{y,1}  \hspace{0.75cm} 
(\hat{e}^{y})^{7} = \sec{\beta} \hspace{0.15cm} r_{y,2} \hspace{0.1cm} d\psi_{y,2} \hspace{1.0cm}
\end{eqnarray}
and the spin connections take the form
\begin{equation}  \hspace{-0.5cm}
\nonumber (\hat{\omega}^{y})^{46} = - \, (\hat{\omega}^{y})^{64} = - \, d\psi_{y,1} \hspace{1.0cm} 
(\hat{\omega}^{y})^{57} = - \, (\hat{\omega}^{y})^{75} = - \, d\psi_{y,2}.
\end{equation}
The covariantly constant spinor conditions are
\begin{eqnarray}
\p_{r_{y,1}} \Psi = \p_{r_{y,2}} \Psi = 0 \hspace{1.0cm}
\left[ \p_{\psi_{y,1}} - \tfrac{1}{2} \, (\gamma^{y}_{4} \gamma^{y}_{6}) \right] \Psi = 0\hspace{1.0cm}
\left[ \p_{\psi_{y,2}} - \tfrac{1}{2} \, (\gamma^{y}_{5} \gamma^{y}_{7}) \right] \Psi = 0.
\end{eqnarray}
Here we have exchanged the flat gamma matrices $\gamma_{4}$, $\gamma_{5}$, $\gamma_{6}$ and $\gamma_{7}$ of the previous subsection for new gamma matrices 
$\gamma^{y}_{4}$, $\gamma^{y}_{5}$, $\gamma^{y}_{6}$ and $\gamma^{y}_{7}$, but with $\gamma_{4}\gamma_{5}\gamma_{6}\gamma_{7} = \gamma^{y}_{4} \gamma^{y}_{5}\gamma^{y}_{6}\gamma^{y}_{7}$ invariant.

$\boxed{\mathbb{C}_{-}^{2}}$
\vspace{-0.1cm}

In the radii and phases of the $\mathbb{C}^{2}_{-}$ complex coordinates $z_{a} = r_{z,a} \, e^{i\psi_{z,a}}$, the vielbiens are
\begin{eqnarray} \nonumber
&& \hspace{-0.65cm} (\hat{e}^{z})^{8} = \csc{\beta} \hspace{0.15cm} dr_{z,1}  \hspace{0.75cm} 
(\hat{e}^{z})^{9} = \csc{\beta} \hspace{0.15cm} dr_{z,2} \hspace{0.75cm} 
(\hat{e}^{z})^{10} = \csc{\beta} \hspace{0.15cm} r_{z,1} \hspace{0.1cm} d\psi_{z,1}  \hspace{0.75cm} 
(\hat{e}^{z})^{11} = \csc{\beta} \hspace{0.15cm} r_{z,2} \hspace{0.1cm} d\psi_{z,2} \hspace{1.0cm}
\end{eqnarray}
and the spin connections are
\begin{equation}  \hspace{-0.5cm}
\nonumber (\hat{\omega}^{z})^{8\hspace{0.05cm}10} = - \, (\hat{\omega}^{z})^{10 \hspace{0.05cm} 8} = - \, d\psi_{z,1} \hspace{1.0cm} 
(\hat{\omega}^{z})^{9\hspace{0.05cm}11} = - \, (\hat{\omega}^{z})^{11 \hspace{0.05cm} 8} = -\, d\psi_{z,2}.
\end{equation}
The covariantly constant spinor conditions are
\begin{eqnarray}
\p_{r_{z,1}} \Psi = \p_{r_{z,2}} \Psi = 0 \hspace{1.0cm}
\left[ \p_{\psi_{z,1}} - \tfrac{1}{2} \gamma^{z}_{8} \gamma^{z}_{10} \right] \Psi = 0 \hspace{1.0cm}
\left[ \p_{\psi_{z,2}} - \tfrac{1}{2} \gamma^{z}_{9} \gamma^{z}_{11} \right] \Psi = 0.
\end{eqnarray}
Here we have exchanged the flat gamma matrices $\gamma_{8}$, $\gamma_{9}$, $\gamma_{10}$ and $\gamma_{11}$ of the previous subsection for new gamma matrices 
$\gamma^{y}_{8}$, $\gamma^{y}_{9}$, $\gamma^{y}_{10}$ and $\gamma^{y}_{11}$, with $\gamma_{8}\gamma_{9}\gamma_{10}\gamma_{11} = \gamma^{y}_{8} \gamma^{y}_{9}\gamma^{y}_{10}\gamma^{y}_{11}$ invariant.

\begin{tabular}{|p{\textwidth}|}
\hline
\vspace{-0.1cm}
The covariantly constant spinor $\Psi(x^{\mu}) = \mathcal{M}(x^{\mu}) \, \Psi_{0}$ can be written in terms of
\begin{eqnarray}
&& \hspace{-0.65cm} \mathcal{M}(x^{\mu}) = 
e^{\frac{1}{2} (\gamma_{1} \gamma_{3}) \, \rho} \hspace{0.15cm}  e^{ \frac{1}{2} (\gamma_{0} \gamma_{3}) \, t} \hspace{0.15cm} 
e^{\frac{1}{2} (\gamma_{1} \gamma_{2}) \, \varphi}  \hspace{0.15cm}
e^{\frac{1}{2} (\gamma^{y}_{4}\gamma^{y}_{6}) \, \psi_{y,1}} \hspace{0.15cm} e^{\frac{1}{2} (\gamma^{y}_{5}\gamma^{y}_{7}) \, \psi_{y,2} } \hspace{0.15cm}
e^{\frac{1}{2} (\gamma^{z}_{8}\gamma^{z}_{10}) \, \psi_{z,1}} \hspace{0.15cm} e^{\frac{1}{2} (\gamma^{z}_{9}\gamma^{z}_{11}) \, \psi_{z,2} } \hspace{0.8cm}
\end{eqnarray} 
with $\Psi_{0}$ a constant spinor.
\vspace{0.2cm}
\\
\hline
\end{tabular}

%---------------------------------------------------------------------------------------------
\subsection{Type IIB Killing spinor $\varepsilon$ on $AdS_{3}\times S_{+}^{3}\times S_{-}^{3}\times S^{1}$}  \label{subappendix - KSEs - 10D}
%---------------------------------------------------------------------------------------- -----

An infinitesimal supersymmetry transformation in type IIB supergravity is parametrized by two 16-component Weyl-Majorana spinors $\varepsilon_{1}$ and $\varepsilon_{2}$, which can be arranged into single Weyl spinor $\varepsilon = \varepsilon_{1} + i\varepsilon_{2}$.  The dilatino and gravitino Killing Spinor equations (KSEs) are \cite{Schwarz:1983,Becker-Becker-Schwarz}
\begin{eqnarray} 
\nonumber && \hspace{-0.25cm} \delta \lambda = \frac{1}{2} \left[ (\p_{\mu}\Phi) \, \Gamma^{\mu} -i \, e^{\Phi} \, \mathbf{F}_{(1)} \right]\varepsilon +
\frac{1}{4} \left[ \, i\, e^{\Phi} \, \tilde{\mathbf{F}}_{(3)} - \mathbf{H}_{(3)} \right] (C\varepsilon)^{*} = 0  \\
\nonumber && \hspace{-0.25cm} \delta \Psi_{\mu} = \left\{ \nabla_{\mu} +\frac{1}{8} \, i \, e^{\Phi} \, \mathbf{F}_{(1)} \, \Gamma_{\mu} 
+ \frac{1}{16} \, i \, e^{\Phi} \, \tilde{\bf{F}}_{(5)} \, \Gamma_{\mu}  \right\} \varepsilon 
- \frac{1}{8} \left[ \mathbf{H}_{(3) \, \mu}  + i\, e^{\Phi} \, \tilde{\mathbf{F}}_{(3)} \, \Gamma_{\mu} \hspace{0.1cm} \right]  (C\varepsilon)^{*} = 0,
\hspace{0.65cm}
\end{eqnarray}
with  $\nabla_{\mu} \equiv \p_{\mu} + \tfrac{1}{4} \, (\omega^{\alpha\beta})_{\mu} \, \gamma_{\alpha\beta}$ the supercovariant derivatives\footnote{
We make use of the conventions:
\, $\mathbf{H}_{(3)} \equiv \frac{1}{3!} \, H_{(3) \, \mu\nu\rho} \, \Gamma^{\mu\nu\rho}$ \, and \,
$\mathbf{H}_{(3) \, \mu} \equiv \frac{1}{2!} \, H_{(3) \, \mu\nu\rho} \, \Gamma^{\nu\rho}$.
}.  Here
\begin{eqnarray}
\nonumber && \hspace{-0.5cm} \tilde{F}_{(3)} \equiv F_{(3)} - C_{(0)}H_{(3)} \hspace{1.0cm} \\
\nonumber && \hspace{-0.5cm} \tilde{F}_{(5)} \equiv F_{(5)} - \tfrac{1}{2} C_{(2)} \wedge H_{(3)} + \tfrac{1}{2} B_{(2)} \wedge F_{(3)} 
= d \left[C_{(4)} - \tfrac{1}{2} C_{(2)} \wedge B_{(2)} \right] = \ast \tilde{F}_{(5)}
\end{eqnarray}
with fluxes $F_{(1)} = dC_{(0)}$, $F_{(2)} = dC_{(3)}$ and $F_{(5)} = dC_{(4)}$. When only the metric $g_{\mu\nu}$ and the 3-form R-R flux $F_{(3)}$ are non-vanishing, the dilatino and gravitino KSEs simplify:
\begin{eqnarray} 
&& \hspace{-0.5cm} \delta \lambda 
= \frac{i}{4} \, {\bf F}_{(3)}  \, (C\varepsilon)^{\ast} = 0 \hspace{1.2cm}
\delta \Psi_{\mu} = \nabla_{\mu} \, \varepsilon -  \frac{i}{8} \, {\bf F}_{(3)}  \Gamma_{\mu} \, (C\varepsilon)^{\ast} = 0. \hspace{0.75cm}
\end{eqnarray}

%%%%%%%%%%
\subsubsection*{The type IIB $AdS_{3}\times S_{+}^{3}\times S_{-}^{3} \times S^{1}$ KSEs on $\varepsilon$}
%%%%%%%%%% 

Let us write down the KSEs on the 16-component Weyl spinor $\varepsilon$ for the $AdS_{3} \times S^{3}_{+} \times S^{3}_{-} \times S^{1}$ background.
We can compute
\begin{equation}
\hspace{-0.5cm} {\bf F}_{(3)} \equiv \frac{1}{3!} \, F^{(3)}_{\mu_{1}\mu_{2}\mu_{3}} \, \Gamma^{\mu_{1}\mu_{2}\mu_{3}}
= \frac{2}{L} \, \left( -\hat{\gamma} + \cos{\beta} \, \gamma_{+} + \sin{\beta} \, \gamma_{-} \right)
\end{equation}
where we define $\hat{\gamma} \equiv \gamma_{0}\gamma_{1}\gamma_{2}$, $\gamma_{+} \equiv \gamma_{4}\gamma_{5}\gamma_{6}$ and $\gamma_{-} \equiv \gamma_{8} \gamma_{9} \gamma_{10}$.  The dilatino KSE is hence
\begin{equation} \label{dilatino-KSE}
\hspace{-1.0cm} 
\mathcal{O} \, \varepsilon \equiv \left[ \cos{\beta} \,   (\hat{\gamma} \gamma_{+})   + \sin{\beta} \, (\hat{\gamma} \gamma_{-})   \right] \varepsilon = \varepsilon
\end{equation}
and the gravitino KSEs become
\begin{eqnarray} \label{gravitino-KSEs-AdS3}
\nonumber & \hspace{-7.0cm} \boxed{AdS_{3}} \hspace{1.0cm} & \nabla_{t} \, \varepsilon + \tfrac{i}{2} \hspace{0.05cm} \cosh{\rho} \hspace{0.1cm} (\gamma_{0} \hat{\gamma}) \, (C\varepsilon)^{\ast} = 0 \hspace{1.1cm} \\
\nonumber &&  \nabla_{\rho} \, \varepsilon + \tfrac{i}{2} \, (\gamma_{1}\hat{\gamma})\, (C\varepsilon)^{\ast} = 0  \\
&&  \nabla_{\varphi} \, \varepsilon + \tfrac{i}{2} \hspace{0.05cm} \sinh{\rho} \hspace{0.1cm} (\gamma_{2} \hat{\gamma}) \, (C\varepsilon)^{\ast} = 0  
\end{eqnarray}
\vspace{-0.75cm}
\begin{eqnarray} \label{gravitino-KSEs-S3+}
\nonumber & \hspace{-7.5cm} \boxed{S_{+}^{3}} \hspace{1.0cm} &  \nabla_{\theta_{+}} \varepsilon - \tfrac{i}{2} \, (\gamma_{4} \gamma_{+}) \, (C\varepsilon)^{\ast} = 0 \hspace{1.75cm} \\
\nonumber &&  \nabla_{\chi_{+}} \varepsilon - \tfrac{i}{2} \cos{\theta_{+}} \, (\gamma_{5}\gamma_{+})\, (C\varepsilon)^{\ast} = 0 \\
&& \nabla_{\phi_{+}} \varepsilon- \tfrac{i}{2} \sin{\theta_{+}} \hspace{0.1cm} (\gamma_{6} \gamma_{+}) \, (C\varepsilon)^{\ast} = 0 
\end{eqnarray}
\vspace{-0.75cm}
\begin{eqnarray} \label{gravitino-KSEs-S3-}
\nonumber & \hspace{-7.5cm} \boxed{S_{-}^{3}} \hspace{1.0cm} & \nabla_{\theta_{-}} \varepsilon - \tfrac{i}{2} \, (\gamma_{8} \gamma_{-}) \, (C\varepsilon)^{\ast} = 0 \hspace{1.75cm} \\
\nonumber && \nabla_{\chi_{-}} \varepsilon - \tfrac{i}{2} \cos{\theta_{-}} \, (\gamma_{9}\gamma_{-})\, (C\varepsilon)^{\ast} = 0 \\
&& \nabla_{\phi_{-}} \varepsilon - \tfrac{i}{2} \sin{\theta_{-}} \hspace{0.1cm} (\gamma_{10} \gamma_{-}) \, (C\varepsilon)^{\ast} = 0 \hspace{0.5cm}
\end{eqnarray}
\vspace{-0.75cm}
\begin{eqnarray}  \label{gravitino-KSEs-S1}
& \hspace{-8.25cm} \boxed{S^{1}} \hspace{4.0cm} & \nabla_{\xi} \, \varepsilon = 0. \hspace{1.0cm}
\end{eqnarray}
with supercovariant derivatives $\nabla_{\mu}$ as before. The charge conjugation matrix is $C = \gamma_{-} \gamma_{12}$. 

\begin{tabular}{|p{\textwidth}|}
\hline
\vspace{-0.1cm}
The solution to the gravitino KSEs takes the form
\begin{equation} \hspace{-0.5cm} \label{Killing-Spinor}
\varepsilon(x^{\mu}) \, = \, \varepsilon^{+}(x^{\mu}) + \varepsilon^{-}(x^{\mu})
\, = \,  \mathcal{M}^{+}(x^{\mu}) \hspace{0.15cm} \varepsilon^{+}_{0} + \mathcal{M}^{-}(x^{\mu}) \hspace{0.15cm} \varepsilon^{-}_{0} 
\end{equation}
where
\begin{eqnarray}
\nonumber && \hspace{-0.65cm} \mathcal{M}^{\pm}(x^{\mu}) = 
e^{\pm \frac{1}{2} (\gamma_{0} \gamma_{2}) \, \rho} \hspace{0.15cm}  e^{\pm \frac{1}{2} (\gamma_{1} \gamma_{2}) \, t} \hspace{0.15cm} 
e^{\frac{1}{2} (\gamma_{1} \gamma_{2}) \, \varphi} \hspace{0.15cm} \\
&& \hspace{-0.65cm} \hspace{1.57cm} \times \hspace{0.1cm} e^{\pm \frac{1}{2} (\gamma_{5}\gamma_{6}) \, \theta_{+}} \hspace{0.15cm} e^{\mp \frac{1}{2} (\gamma_{4}\gamma_{6}) \, \chi_{+} } \hspace{0.15cm}
e^{\frac{1}{2} (\gamma_{4}\gamma_{6})  \, \phi_{+} }  \hspace{0.15cm}
e^{\pm \frac{1}{2} (\gamma_{9}\gamma_{10}) \, \theta_{-}} \hspace{0.15cm} e^{\mp \frac{1}{2} (\gamma_{8}\gamma_{10}) \, \chi_{-} } \hspace{0.15cm}
e^{\frac{1}{2} (\gamma_{8}\gamma_{10}) \, \phi_{-} } \hspace{1.0cm}
\end{eqnarray}
with $\varepsilon^{\pm}_{0}$ constant Weyl spinors satisfying $ i \, (C\varepsilon^{\pm}_{0})^{\ast} = \pm \, \varepsilon^{\pm}_{0}$.  
\vspace{0.2cm}
\\
\hline
\end{tabular}

Here $\varepsilon$ has been decomposed into $\varepsilon^{\pm}$ which separately satisfy the KSEs and $ i \, (C\varepsilon^{\pm})^{\ast} = \pm \, \varepsilon^{\pm}$.   The dilatino KSE implies 
$\mathcal{O} \, \varepsilon^{\pm}_{0} = \varepsilon^{\pm}_{0}$, since $\mathcal{O}$ and $\mathcal{M}^{\pm}$ compute.
The operator $\mathcal{O}$ is traceless and squares to unity, so half the degrees of freedom contained in 
$ \varepsilon^{\pm}_{0}$ satisfy this constraint. 

We have thus shown that the type IIB supergravity background $AdS_{3} \times S_{+}^{3} \times S_{-}^{3} \times S^{1}$ with pure R-R flux preserves 16 of 32 supersymmetries and is therefore $\tfrac{1}{2}$-BPS.  $AdS_{3} \times S_{+}^{3} \times S_{-}^{3} \times S^{1}$ backgrounds with pure NS-NS flux or mixed flux can be found by S-duality transformations.

%%%%%%%%%%
\subsubsection*{Projecting $\varepsilon$ out of a covariantly constant spinor $\Psi$}
%%%%%%%%%%

Let us impose the following conditions on the covariantly constant spinor $\Psi$:
\begin{eqnarray} \label{conditions-psi} 
(\hat{\gamma} \gamma_{3}) \, \Psi = -i \, (C\Psi)^{\ast}  \hspace{0.75cm} (\gamma_{+} \gamma_{7}) \, \Psi = i \, (C\Psi)^{\ast} 
\hspace{0.75cm} (\gamma_{-} \gamma_{11}) \, \Psi = i \, (C\Psi)^{\ast} \hspace{0.75cm} \gamma_{12} \, \Psi = -i \, (C\Psi)^{\ast} \hspace{0.3cm}
\end{eqnarray}
Three of these conditions are independent, while the last condition then follows automatically. 

It is now possible to project the 16-component Weyl spinor $\varepsilon = \varepsilon_{1} + i \varepsilon_{2}$ out of this 32-component spinor $\Psi = \Psi_{1} + i \Psi_{2}$ in such a way that $\varepsilon$ satisfies both the $AdS_{3}\times S_{+}^{3} \times S_{-}^{3} \times S^{1}$ gravitino KSEs and the dilatino KSE.  Here we define

\begin{tabular}{|p{\textwidth}|}
\hline
\vspace{-0.5cm}
\begin{equation}
 \hspace{-0.25cm}  \varepsilon(x^{\mu}) 
 = \tfrac{1}{2} \left[ 1 + \gamma_{3} \left( \cos{\beta} \, \gamma_{7} + \sin{\beta} \, \gamma_{11}\right) \right] \hspace{0.1cm}
 \tfrac{1}{2} \left[ 1 + \hat{\gamma} \gamma_{+} \gamma_{-} \gamma_{12} \right] \hspace{0.1cm} \Psi(x^{\mu})
 \, \equiv \, \mathcal{P}_{\text{dilatino}} \hspace{0.05cm} \mathcal{P}_{\text{Weyl}} \hspace{0.1cm} \Psi(x^{\mu})
\end{equation}
\vspace{-0.5cm}
\\
\hline
\end{tabular}

We see that $(\hat{\gamma} \gamma_{+} \gamma_{-} \gamma_{12}) \, \varepsilon = \varepsilon$ is Weyl and also satisfies $\gamma_{3} \left( \cos{\beta} \, \gamma_{7} + \sin{\beta} \, \gamma_{11}\right) \, \varepsilon = \varepsilon$ which will lead to the dilatino condition once additional conditions (\ref{conditions-psi}) have been imposed on $\Psi$.
Note that these 3 independent conditions which we impose on $\Psi$ are consistent with our projecting out 16 real degrees of freedom contained in the 32-component Weyl spinor $\varepsilon$ from 128 real degrees of freedom contained in the spinor $\Psi$.  

We observe that $\varepsilon$ satisfies the dilatino KSE (\ref{dilatino-KSE}) by applying $\mathcal{P}_{\text{dilatino}} \, \mathcal{P}_{\text{Weyl}}$ to the following equation, which is a direct result of the conditions (\ref{conditions-psi}):
\begin{equation} \nonumber
\left[\hat{\gamma} \left( \cos{\beta} \, \gamma_{+} + \sin{\beta} \, \gamma_{-}\right) \right] \Psi 
= \left[\gamma_{3} \left( \cos{\beta} \, \gamma_{7} + \sin{\beta} \, \gamma_{11}\right) \right] \Psi
\end{equation}
and using $\gamma_{3} \left( \cos{\beta} \, \gamma_{7} + \sin{\beta} \, \gamma_{11}\right) \, \varepsilon = \varepsilon$, which arises from our choice of  projector $\mathcal{P}_{\text{dilatino}}$.
We see that $\varepsilon$ satisfies the gravitino KSEs (\ref{gravitino-KSEs-AdS3}), (\ref{gravitino-KSEs-S3+}), (\ref{gravitino-KSEs-S3-}) and (\ref{gravitino-KSEs-S1}) by applying $\mathcal{P}_{\text{dilatino}} \, \mathcal{P}_{\text{Weyl}}$ to the left-hand side of the covariantly constant conditions (\ref{covariantly-constant-R22}), (\ref{covariantly-constant-C2+}), (\ref{covariantly-constant-C2-}) and (\ref{covariantly-constant-S1}), taking into account also the conditions (\ref{conditions-psi}).  Again, we find $\varepsilon$ contains 16 real degrees of freedom implying $AdS_{3}\times S^{3}_{+}\times S^{3}_{-} \times S^{1}$ is a $\tfrac{1}{2}$-BPS type IIB supergravity background.

%%%%%%%%%%%%%%%%%%%%%%%%%%%%%%%%%%%%%%%%%%%%%%%%%%%%%%%%%%%%%%%%%%%%%%%%%%%%%%%%%%%%%%%%%%%%%%
\section{Kappa symmetry conditions} \label{appendix - kappa symmetry}
%%%%%%%%%%%%%%%%%%%%%%%%%%%%%%%%%%%%%%%%%%%%%%%%%%%%%%%%%%%%%%%%%%%%%%%%%%%%%%%%%%%%%%%%%%%%%%

The D$p$-brane action is
\begin{eqnarray}
&& \hspace{-1.0cm} S_{\text{D}p}  = - \frac{1}{(2\pi)^{p}} \int_{\mathcal{W}} d^{p+1}\sigma \hspace{0.2cm} e^{-\Phi} \sqrt{-\det  g_{ab} } \hspace{0.2cm} \pm \hspace{0.1cm} \frac{1}{(2\pi)^{p}} \int_{\mathcal{W}} \hspace{0.15cm} \sum_{\ell} C_{(\ell)}  \hspace{1.0cm}
\end{eqnarray}
where we have turned off all gauge fields on the worldvolume $\mathcal{W}$.  Supersymmetry on the worldvolume of a D$p$-brane requires the existence of kappa symmetry on the worldvolume, which halves the number of fermionic degrees of freedom to match the number of bosonic degrees of freedom. The kappa symmetry condition is \cite{Bergshoeff-Townsend:1997,Bergshoeff-et-al:1997,Simon:2011}
\begin{equation}
\Gamma \, \varepsilon = \mp \, i (C \varepsilon)^{\ast}
\hspace{0.75cm} \text{with} \hspace{0.5cm}
\Gamma = \frac{1}{(p+1)!} \, \frac{\epsilon^{a_{0} \hspace{0.025cm}\ldots \hspace{0.05cm} a_{p}}}{\sqrt{-\det{g}}} \hspace{0.1cm}  (\p_{a_{0}}X^{\mu_{0}}) \cdots (\p_{a_{p}}X^{\mu_{p}}) \, \, \Gamma_{\mu_{0} \hspace{0.025cm}\ldots \hspace{0.05cm}\mu_{p}},
\end{equation}
imposed on the pullback $\varepsilon(\sigma^{a})$ of the type IIB supergravity Killing spinor $\varepsilon(x^{\mu})$, given by (\ref{Killing-Spinor}), to the D$p$-brane worldvolume $\mathcal{W}$.  The $\pm$ distinguishes between branes and anti-branes.

%---------------------------------------------------------------------------------------------
\subsection{Kappa symmetry of the D1-brane giant graviton examples}  \label{subappendix - kappa symmetry - D1}
%---------------------------------------------------------------------------------------- -----

%%%%%%%%%%
\subsubsection*{D1-brane giant graviton in $AdS_{3}$}
%%%%%%%%%%

Let us consider the D1-brane giant graviton wrapping the $\varphi$ circle in $AdS_{3}$, which was constructed in section (\ref{subsection - examples - D1}).
The kappa symmetry condition $\Gamma \, \varepsilon = i (C \varepsilon)^{\ast}$ on the pullback $\varepsilon(t,\varphi) = \varepsilon^{+}(t,\varphi) + \varepsilon^{-}(t,\varphi)$ of the Killing spinor to the D1-(anti-)brane worldvolume is  
\begin{eqnarray} 
&& \hspace{-1.0cm}
\Gamma \, \varepsilon^{\pm}(t,\varphi) = \pm \,\varepsilon^{\pm}(t,\varphi) \hspace{0.75cm} \text{with} \hspace{0.5cm} \Gamma 
=  \frac{\left( \cosh{\rho} \,\, \gamma_{0} + \dot{\chi} \, \cos{\beta} \,\, \gamma_{5} + \dot{\chi} \, \sin{\beta} \, \, \gamma_{9}\right)  \gamma_{2} }{\sinh{\rho}},   \hspace{0.5cm}
\end{eqnarray}
where it follows from (\ref{Killing-Spinor}) that $\varepsilon^{\pm}(t,\varphi) = \mathcal{M}^{\pm}(t,\varphi) \hspace{0.1cm} \varepsilon^{\pm}_{0}$ can be written in terms of
\begin{eqnarray} && \hspace{-0.65cm} 
 \mathcal{M}^{\pm}(t,\varphi)  =  
e^{\pm\frac{1}{2} (\gamma_{0} \gamma_{2})  \, \rho} \hspace{0.15cm}  
e^{\pm \frac{1}{2} (\gamma_{4}\gamma_{6}) \, \chi_{+}} \hspace{0.15cm} 
e^{\pm \frac{1}{2} (\gamma_{8}\gamma_{10}) \, \chi_{-}} \hspace{0.15cm} 
e^{ \frac{1}{2} (\gamma_{1} \gamma_{2}) \, (\varphi \, \pm \, t)} \hspace{0.15cm} 
\end{eqnarray}
with $\rho$ constant, and $\chi_{+} = \alpha \, \dot{\chi} \, t$ and $\chi_{-} = (1-\alpha) \, \dot{\chi}\, t$. Here $\dot{\chi} = \pm \hspace{0.05cm} 1$ indicates the direction of motion of the anti-brane.  The dilatino KSE (\ref{dilatino-KSE}) also implies  $\mathcal{O} \, \varepsilon^{\pm}(t,\varphi)  = \varepsilon^{\pm}(t,\varphi)$.  Clearly, an additional consistency condition, which arises from insisting that both the kappa symmetry and dilatino conditions be satisfied simultaneously, is
\begin{equation}  
[\Gamma, \mathcal{O}] \hspace{0.15cm} \varepsilon^{\pm}(t,\varphi) = \pm \hspace{0.1cm}  \dot{\chi} \hspace{0.15cm} \frac{2 \cos{\beta} \sin{\beta}}{\sinh{\rho}} \hspace{0.15cm}  \gamma_{0} \gamma_{1} \gamma_{5} \gamma_{9}\, (\gamma_{4}\gamma_{6} - \gamma_{8}\gamma_{10}) \hspace{0.15cm} \varepsilon^{\pm}(t,\varphi) = 0.
\end{equation}
This is satisfied if we set $\gamma_{4}\gamma_{6} \hspace{0.1cm} \varepsilon^{\pm}_{0} =  \gamma_{8}\gamma_{10} \hspace{0.1cm} \varepsilon^{\pm}_{0}$.  The dilatino condition then becomes
\begin{equation} \nonumber \hspace{-0.5cm}
\left(\cos{\beta} \, \gamma_{5} + \sin{\beta} \, \gamma_{9}\right) \, \varepsilon^{\pm}(t,\varphi) =  \hat{\gamma} \, \gamma_{4} \gamma_{6} \hspace{0.15cm} \varepsilon^{\pm}(t,\varphi),
\end{equation}
which simplifies the kappa symmetry condition to
\begin{equation} \nonumber \hspace{-0.5cm}
\frac{(\cosh{\rho} \,\, \gamma_{0}\gamma_{2} - \dot{\chi} \hspace{0.1cm} \gamma_{0} \gamma_{1} \gamma_{4}\gamma_{6})}{\sinh{\rho}} \hspace{0.15cm} \varepsilon^{\pm}(t,\varphi) 
= \varepsilon^{\pm}(t,\varphi).
\end{equation}
We observe that, while the operator $\mathcal{O}$ in the dilatino KSE commutes through $\mathcal{M}^{\pm}(t,\varphi)$, this is not true of the operator above.  The kappa symmetry conditions on $\varepsilon^{\pm}_{0}$ becomes
\begin{equation} 
\nonumber \hspace{-0.5cm}
\mathcal{M}^{\pm}(t,\varphi)^{-1} \hspace{0.15cm} \frac{(\cosh{\rho} \,\, \gamma_{0}\gamma_{2} - \dot{\chi} \hspace{0.1cm} \gamma_{0} \gamma_{1} \gamma_{4}\gamma_{6}) }{\sinh{\rho}} \hspace{0.15cm} \mathcal{M}^{\pm}(t,\varphi) \hspace{0.2cm} \varepsilon^{\pm}_{0} 
=  \varepsilon^{\pm}_{0},
\end{equation} 
which can be manipulated into the form
\begin{eqnarray}  \nonumber
&&  \hspace{-0.65cm} \pm \hspace{0.1cm} \frac{\cosh{\rho}}{\sinh{\rho}} \hspace{0.15cm} 
e^{-\frac{1}{2} (\gamma_{1} \gamma_{2}) \, (\varphi \, \pm \, t) } \hspace{0.1cm}
 \gamma_{0} \gamma_{2} \left( \mathbf{1} + \dot{\chi} \hspace{0.1cm} \gamma_{1}\gamma_{2} \gamma_{4}\gamma_{6} \right) \hspace{0.1cm} \varepsilon^{\pm}_{0}
\, - \, \dot{\chi} \hspace{0.1cm} \gamma_{1}\gamma_{2} \gamma_{4}\gamma_{6} \hspace{0.15cm} \varepsilon^{\pm}_{0} = \varepsilon^{\pm}_{0}. \hspace{1.0cm}
\end{eqnarray}
This condition is satisfied if
$\dot{\chi} \hspace{0.1cm} \gamma_{1}\gamma_{2} \hspace{0.1cm}\varepsilon^{\pm}_{0} = \gamma_{4}\gamma_{6}\hspace{0.1cm}\varepsilon^{\pm}_{0}$.  This allows us to further simplify the dilatino condition to
\begin{equation} \nonumber
\dot{\chi} \hspace{0.1cm} \gamma_{0} \left(\cos{\beta} \, \gamma_{5} + \sin{\beta} \, \gamma_{9}\right) \hspace{0.1cm} \varepsilon^{\pm}(t,\varphi)
= \varepsilon^{\pm}(t,\varphi).
\end{equation}
We thus conclude that the spinors
$\varepsilon^{\pm}_{0}$ must satisfy three independent conditions for kappa symmetry:
\begin{eqnarray}
\nonumber && \hspace{-0.2cm} 
\, \dot{\chi} \hspace{0.1cm} \gamma_{1}\gamma_{2} \hspace{0.1cm} \varepsilon^{\pm}_{0} = \gamma_{4}\gamma_{6} \hspace{0.1cm} \varepsilon_{0}^{\pm} 
= \gamma_{8}\gamma_{10} \hspace{0.1cm} \varepsilon^{\pm}_{0}   
\hspace{1.2cm}  \gamma_{0} \left(\cos{\beta} \, \gamma_{5} + \sin{\beta \, \gamma_{9}}\right) \, \varepsilon^{\pm}_{0} 
= \dot{\chi} \hspace{0.1cm} \varepsilon^{\pm}_{0},   \hspace{1.0cm}
\end{eqnarray}
each of which halves the number of degrees of freedom contained in $\varepsilon_{0}$. Hence this D1-brane giant graviton in $AdS_{3}$ preserves 4 of the original 16 background supersymmetries on the worldvolume.  It is therefore a $\tfrac{1}{4}$-BPS configuration.

%%%%%%%%%%
\subsubsection*{D1-brane giant graviton in $\mathbb{R} \times S^{3}_{+} \times S^{3}_{-}$}
%%%%%%%%%%

We now consider the D1-brane giant graviton wrapping the curve $\gamma$ (parameterized by $\phi$) in $S^{3}_{+} \times S^{3}_{-}$ constructed in section (\ref{subsection - examples - D1}).
The kappa symmetry condition $\Gamma \, \varepsilon = - \hspace{0.05cm} \dot{\chi} \hspace{0.1cm} i \hspace{0.05cm} (C \varepsilon)^{\ast}$ on the pullback $\varepsilon(t,\varphi) = \varepsilon^{+}(t,\varphi) + \varepsilon^{-}(t,\varphi)$ of the Killing spinor to the worldvolume is
\begin{eqnarray}
\nonumber && \hspace{-0.7cm} 
\Gamma \, \varepsilon^{\pm}(t,\phi) = \mp \, \dot{\chi} \hspace{0.1cm} \varepsilon^{\pm}(t,\phi) \hspace{0.4cm} \text{with} \hspace{0.3cm} 
 \Gamma =  \frac{\left( \gamma_{0} + \dot{\chi} \, \cos{\theta} \cos{\beta} \, \gamma_{5} + \dot{\chi} \, \cos{\theta}\sin{\beta} \, \, \gamma_{9}\right)  \left( \cos{\beta} \,\, \gamma_{6} + \sin{\beta} \,\, \gamma_{10} \right)}{\sin{\theta}}, \\
\end{eqnarray}
where it follows from (\ref{Killing-Spinor}) that $\varepsilon^{\pm}(t,\phi) = \mathcal{M}^{\pm}(t,\phi) \hspace{0.1cm} \varepsilon^{\pm}_{0}$ 
in terms of
\begin{eqnarray} &&   
\mathcal{M}^{\pm}(t,\phi)  =  
e^{\pm \frac{1}{2}(\gamma_{5}\gamma_{6}) \, \theta} \hspace{0.15cm} 
e^{\pm \frac{1}{2}(\gamma_{9}\gamma_{10}) \, \theta} \hspace{0.15cm} 
e^{\pm \frac{1}{2} (\gamma_{1} \gamma_{2}) \, t } \hspace{0.15cm}
e^{\frac{1}{2} ( \gamma_{4}\gamma_{6})  \, (\phi_{+} \, \mp \, \chi_{+}) }  \hspace{0.15cm}
e^{\frac{1}{2} ( \gamma_{8}\gamma_{10})  \, (\phi_{-} \, \mp \, \chi_{-})} \hspace{0.5cm}
\hspace{1.0cm}
\end{eqnarray}
with $\theta$ constant, $\chi_{+} \hspace{-0.1cm} = \alpha \, \dot{\chi} \hspace{0.05cm} t$ and $\chi_{-} \hspace{-0.1cm} = (1-\alpha) \, \dot{\chi} \hspace{0.05cm} t$,  $\phi_{+} \hspace{-0.1cm} = \alpha \, \phi$ and $\phi_{-} \hspace{-0.1cm} = (1-\alpha) \, \phi$.
Here $\dot{\chi} = \pm \hspace{0.05cm} 1$ gives the direction of motion of the brane/anti-brane. Once more, $\varepsilon^{\pm}(t,\phi)$ satisfies $\mathcal{O} \, \varepsilon^{\pm}(t,\phi) = \varepsilon^{\pm}(t,\phi)$ from the dilatino KSE (\ref{dilatino-KSE}).  The consistency condition is now
\begin{eqnarray}
&& \hspace{-0.75cm} [\Gamma,\mathcal{O}] \hspace{0.1cm} \varepsilon^{\pm}(t,\phi) = \pm \, 2 \cos{\beta}\sin{\beta} \, \csc{\theta} \,\, \gamma_{1}\gamma_{2} \\
\nonumber && \hspace{-0.75cm} \hspace{2.4cm} \times 
 \left[ \left( \gamma_{8}\gamma_{9}  - \gamma_{4} \gamma_{5} \right) \gamma_{6} \gamma_{10}
+ \dot{\chi} \, \cos{\theta} \left( \cos{\beta} \,\, \gamma_{0}\gamma_{4} - \sin{\beta} \,\, \gamma_{0}\gamma_{8} \right) 
\left( \gamma_{5}\gamma_{9} + \gamma_{6}\gamma_{10}  \right)  \right] \,\varepsilon^{\pm}(t,\phi) = 0,
\end{eqnarray}
which, rewritten in terms of the constant spinors $\varepsilon^{\pm}_{0}$, is
\begin{eqnarray}
\nonumber && \hspace{-0.75cm}  
\mathcal{M}^{\pm}(t,\phi)^{-1} \hspace{0.1cm} [\Gamma,\mathcal{O}] \hspace{0.15cm} \mathcal{M}^{\pm}(t,\phi) \hspace{0.2cm} \varepsilon^{\pm}_{0} \\
\nonumber && \hspace{-0.75cm} = 2\cos{\beta}\sin{\beta}  \, \left\{\, \cos{\theta} \,\,
 \gamma_{1}\gamma_{2} \gamma_{5}\gamma_{9} \left(\gamma_{8}\gamma_{10} - \gamma_{4}\gamma_{6}\right)  \right.  \\  
\nonumber && \hspace{-0.75cm} \hspace{1.5cm} \pm \hspace{0.05cm}  \cot{\theta} \,\, e^{-(\gamma_{4}\gamma_{6}) \, (\phi_{+} \mp \chi_{+})} \,
\gamma_{6}\gamma_{8}\gamma_{9}\gamma_{10} \,
\left[ \mathbf{1} - \dot{\chi} \hspace{0.05cm} \cos{\beta} \,\, \gamma_{0}\gamma_{5} \, (-\gamma_{4}\gamma_{6}\gamma_{8}\gamma_{10}) 
- \dot{\chi} \hspace{0.05cm} \sin{\beta} \,\, \gamma_{0} \gamma_{9} \right]   \\
\nonumber && \hspace{-0.75cm}  \hspace{1.5cm}  \mp \hspace{0.05cm} \cot{\theta} \,\, e^{-(\gamma_{8}\gamma_{10}) \, (\phi_{-} \mp \chi_{-})} 
\left. \gamma_{4}\gamma_{5}\gamma_{6}\gamma_{10} \left[ \mathbf{1} - \dot{\chi} \hspace{0.05cm} \cos{\beta} \,\, \gamma_{0}\gamma_{5}
- \dot{\chi} \hspace{0.05cm} \sin{\beta} \,\, \gamma_{0} \gamma_{9} \, (-\gamma_{4}\gamma_{6}\gamma_{8}\gamma_{10}) \right] \, \right\} \hspace{0.1cm} \varepsilon^{\pm}_{0} = 0,
\end{eqnarray}
which is satisfied if $\gamma_{4}\gamma_{6} \hspace{0.1cm} \varepsilon^{\pm}_{0} = \gamma_{8} \gamma_{10} \hspace{0.1cm} \varepsilon^{\pm}_{0}$ and 
$\dot{\chi} \hspace{0.05cm} \left(\cos{\beta} \, \gamma_{0}\gamma_{5} + \sin{\beta \, \gamma_{0} \gamma_{9}}\right) \varepsilon^{\pm}_{0} = \varepsilon^{\pm}_{0}$.
The dilatino condition becomes equivalent to $\dot{\chi} \hspace{0.1cm} \gamma_{1}\gamma_{2} \hspace{0.1cm} \varepsilon^{\pm}_{0} = \gamma_{4} \gamma_{6} \hspace{0.1cm} \varepsilon^{\pm}_{0}$ and the kappa symmetry condition is
\begin{eqnarray}
\nonumber && \hspace{-0.75cm} \mathcal{M}^{\pm}(t,\phi)^{-1} \hspace{0.1cm} \Gamma \hspace{0.15cm} \mathcal{M}^{\pm}(t,\phi)   \hspace{0.15cm} \varepsilon^{\pm}_{0} 
=-  \left(\cos{\beta} \, \gamma_{0}\gamma_{5} + \sin{\beta \, \gamma_{0} \gamma_{9}}\right) \hspace{0.15cm} \varepsilon^{\pm}_{0}  \\
\nonumber && \hspace{-0.75cm} \hspace{4.69cm} \pm \hspace{0.025cm} \cot{\theta} \, [ \cos{\beta} \,\, e^{-(\gamma_{4}\gamma_{6}) \, (\phi_{+} \mp \chi_{+})} \hspace{0.1cm}\gamma_{0}\gamma_{6}
+ \sin{\beta} \,\, e^{-(\gamma_{8}\gamma_{10}) \, (\phi_{-} \mp \chi_{-})} \hspace{0.1cm} \gamma_{0}\gamma_{10} ] \\
\nonumber && \hspace{-0.75cm} \hspace{5.05cm} \times \, [ \mathbf{1} -  \dot{\chi} \hspace{0.05cm} 
\left(\cos{\beta} \, \gamma_{0}\gamma_{5} + \sin{\beta \, \gamma_{0} \gamma_{9}}\right)] \hspace{0.15cm} \varepsilon^{\pm}_{0} \\
\nonumber && \hspace{-0.75cm} \hspace{4.69cm} = - \, \dot{\chi} \hspace{0.1cm} \varepsilon^{\pm}_{0}.
\end{eqnarray}
The kappa symmetry and dilatino conditions, together with the consistency condition, are therefore satisfied if 
\begin{eqnarray} 
\nonumber && \hspace{-0.2cm} \dot{\chi} \hspace{0.1cm} \gamma_{1}\gamma_{2} \hspace{0.1cm} \varepsilon^{\pm}_{0}  = \gamma_{4}\gamma_{6} \hspace{0.1cm} \varepsilon^{\pm}_{0} 
= \gamma_{8}\gamma_{10} \hspace{0.1cm} \varepsilon^{\pm}_{0} \hspace{1.0cm} 
\gamma_{0} \left(\cos{\beta} \, \gamma_{5} + \sin{\beta \,  \gamma_{9}}\right) \hspace{0.1cm} \varepsilon^{\pm}_{0} 
= \dot{\chi} \hspace{0.1cm}  \varepsilon^{\pm}_{0}.   \hspace{1.0cm}
\end{eqnarray}
This D1-brane giant graviton in $\mathbb{R} \times S^{3}_{+}\times S^{3}_{-}$ therefore preserves 4 of the original 16 background supersymmetries on the worldvolume and is thus $\tfrac{1}{4}$-BPS.

%---------------------------------------------------------------------------------------------
\subsection{Kappa symmetry of the D5-brane giant graviton example}  \label{subappendix - kappa symmetry - D5}
%---------------------------------------------------------------------------------------- -----

%%%%%%%%%%
\subsubsection*{D5-brane giant graviton in $\mathbb{R} \times S^{3}_{+} \times S^{3}_{-} \times S^{1}$}
%%%%%%%%%%

Let us consider the D5-brane giant graviton, parameterized by $\sigma^{a} = (t,\theta_{+},\tilde{\chi},\phi,\tilde{\phi},\xi)$, which wraps a 4-cycle in $S^{3}_{+} \times S^{3}_{-}$ and the $S^{1}$.  This was constructed in section (\ref{subsection - examples - D5}).
The kappa symmetry condition  $\Gamma \, \varepsilon = - \hspace{0.05cm} \dot{\chi} \hspace{0.1cm} i \hspace{0.05cm} (C \varepsilon)^{\ast}$ on the pullback $\varepsilon(\sigma^{a}) = \varepsilon^{+}(\sigma^{a}) + \varepsilon^{-}(\sigma^{a})$ of the Killing spinor to the worldvolume of the D5-brane is
\begin{eqnarray} 
\nonumber && \hspace{-0.7cm}
\Gamma \, \varepsilon^{\pm}(\sigma^{a}) = \mp \, \dot{\chi} \hspace{0.1cm} \varepsilon^{\pm}(\sigma^{a}) \hspace{0.4cm}
\text{with} \hspace{0.3cm}
\Gamma = \left[\left( \sin{\beta} \cos{\theta_{+}} \, \gamma_{0}\gamma_{5} - \cos{\beta} \cos{\theta_{-}} \, \gamma_{0}\gamma_{9} - \dot{\chi} \, \cos{\theta_{+}}\cos{\theta_{-}} \, \gamma_{5}\gamma_{9}\right) \right. \\
&& \hspace{-0.7cm} \hspace{6.0cm} \frac{ \times \left( \sin{\beta}\cos{\theta_{+}} \sin{\theta_{-}} \, \gamma_{4} + \cos{\beta} \cos{\theta_{-}} \sin{\theta_{+}}\, \gamma_{8} \right) \, \gamma_{6} \gamma_{10} \gamma_{12} ]}
{\left[ (\sin^{2}{\beta}\cos^{2}{\theta}_{+} + \cos^{2}{\beta}\cos^{2}{\theta_{-}}) - \cos^{2}{\theta_{+}} \, \cos^{2}{\theta_{-}} \right]}  \hspace{1.0cm}
\end{eqnarray}
where it follows from (\ref{Killing-Spinor}) that $\varepsilon^{\pm}(\sigma^{a}) = \mathcal{M}^{\pm}(\sigma^{a}) \hspace{0.1cm} \varepsilon^{\pm}_{0}$ is written in terms of
\begin{eqnarray} &&  \hspace{-0.2cm}
\mathcal{M}^{\pm}(t,\phi)  =  
e^{\pm \frac{1}{2}(\gamma_{5}\gamma_{6}) \, \theta_{+}} \hspace{0.15cm} 
e^{\pm \frac{1}{2}(\gamma_{9}\gamma_{10}) \, \theta_{-}} \hspace{0.15cm} 
e^{\pm \frac{1}{2} (\gamma_{1} \gamma_{2}) \, t } \hspace{0.15cm}
e^{\frac{1}{2} ( \gamma_{4}\gamma_{6})  \, (\phi_{+} \, \mp \, \chi_{+}) }  \hspace{0.15cm}
e^{\frac{1}{2} ( \gamma_{8}\gamma_{10})  \, (\phi_{-} \, \mp \, \chi_{-})} \hspace{0.15cm}
\hspace{1.0cm}
\end{eqnarray}
with $\theta_{-}(\theta_{+})$ defined through (\ref{D5-ansatz}), 
$\chi_{+} = \alpha \, \dot{\chi} \hspace{0.05cm} t - \tilde{\chi}$ and $\chi_{-} = (1-\alpha) \, \dot{\chi} \hspace{0.05cm} t + \tilde{\chi}$, and 
$\phi_{+} = \alpha \, \phi - \tilde{\phi}$ and $\phi_{-} = (1-\alpha) \, \phi + \tilde{\phi}$. Here $\dot{\chi} = \pm \hspace{0.05cm} 1$ indicates the direction of motion of the brane/anti-brane.
The dilatino KSE (\ref{dilatino-KSE}) implies $\mathcal{O} \, \varepsilon^{\pm}(\sigma^{a}) = \varepsilon^{\pm}(\sigma^{a})$. The additional consistency condition is 
\begin{eqnarray}
&& \hspace{-0.7cm} [\Gamma,\mathcal{O}] \hspace{0.1cm} \varepsilon^{\pm}(\sigma^{a}) = \mp \hspace{0.1cm} \frac{2 \, \cos{\beta} \, \sin{\beta} \, \cos{\theta_{+}} \cos{\theta_{-}}}
{\left[ (\sin^{2}{\beta}\cos^{2}{\theta}_{+} + \cos^{2}{\beta}\cos^{2}{\theta_{-}}) - \cos^{2}{\theta_{+}} \, \cos^{2}{\theta_{-}} \right]} 
\hspace{0.15cm} (\gamma_{1}\gamma_{2}\gamma_{12}) \\
\nonumber && \hspace{-0.7cm} \hspace{2.28cm} \times 
\left[ \cos{\beta} \sin{\theta_{+}} \, \gamma_{4}\gamma_{8}\gamma_{10} 
+ \sin{\beta} \sin{\theta_{+}} \, \gamma_{5}\gamma_{9}\gamma_{6} + \cos{\beta}\sin{\theta_{-}} \, \gamma_{5}\gamma_{9}\gamma_{10}  \right.  \\
\nonumber && \hspace{-0.7cm} \hspace{2.37cm} \left. 
+ \hspace{0.01cm} \sin{\beta}\sin{\theta_{-}} \, \gamma_{4}\gamma_{8}\gamma_{6} 
+ \dot{\chi} \, \cos{\theta_{+}} \sin{\theta_{-}} \, \gamma_{0} \gamma_{9} \gamma_{10} 
+ \dot{\chi} \, \cos{\theta_{-}} \sin{\theta_{+}} \, \gamma_{0} \gamma_{5} \gamma_{6} \right]\, \varepsilon^{\pm}(\sigma^{a}) = 0,
\end{eqnarray}
which, rewritten in terms of the constant spinors $\varepsilon^{\pm}_{0}$, is given by
\begin{eqnarray}
\nonumber && \hspace{-0.75cm}
\mathcal{M}^{\pm}(\sigma^{a})^{-1} \, [\Gamma,\mathcal{O}] \hspace{0.15cm} \mathcal{M}^{\pm}(\sigma^{a}) \hspace{0.2cm} \varepsilon^{\pm}_{0}\\
\nonumber && \hspace{-0.75cm} = \mp \hspace{0.1cm} \frac{2 \, \cos{\beta} \, \sin{\beta} \, \cos{\theta_{+}} \cos{\theta_{-}}}
{\left[ (\sin^{2}{\beta}\cos^{2}{\theta}_{+} + \cos^{2}{\beta}\cos^{2}{\theta_{-}}) - \cos^{2}{\theta_{+}} \, \cos^{2}{\theta_{-}} \right]} 
\hspace{0.15cm} (\gamma_{1}\gamma_{2}\gamma_{12}) \\ 
\nonumber && \hspace{-0.75cm} \hspace{0.375cm} \times \, \Big\{ \, \cos{\theta_{-}} \sin{\theta_{+}} \,\, 
e^{-(\gamma_{4}\gamma_{6}) \, (\phi_{+} \mp \, \chi_{+})} \hspace{0.1cm}  
 \gamma_{0}\gamma_{5} \gamma_{6} \, \left[\cos{\beta} \,\, \gamma_{0}\gamma_{5} \, (-\gamma_{4}\gamma_{6} \gamma_{8} \gamma_{10}) + \sin{\beta} \, \gamma_{0}\gamma_{9} - \dot{\chi} \, \mathbf{1} \right] \\
\nonumber && \hspace{-0.75cm} \hspace{0.75cm}  - \hspace{0.05cm} \cos{\theta_{+}} \sin{\theta_{-}} \,\, 
e^{-(\gamma_{8}\gamma_{10}) (\phi_{-} \mp \, \chi_{-} )} \hspace{0.1cm}
 \gamma_{0}\gamma_{9}\gamma_{10} \left[ \cos{\beta} \,\, \gamma_{0}\gamma_{5} + \sin{\beta} \, \gamma_{0}\gamma_{9} \, (-\gamma_{4}\gamma_{6}\gamma_{8}\gamma_{10}) - \dot{\chi} \, \mathbf{1} \right]  \\
\nonumber && \hspace{-0.75cm} \hspace{0.75cm} 
+ \hspace{0.05cm}\sin{\theta_{+}}\sin{\theta_{-}} \,\, e^{-(\gamma_{4}\gamma_{6}) \, (\phi_{+} \mp \, \chi_{+})}  \,\,
e^{-(\gamma_{8}\gamma_{10}) (\phi_{-} \mp \, \chi_{-} )} \hspace{0.1cm} \gamma_{5}\gamma_{6}\gamma_{8} \left( \cos{\beta} \,\, \gamma_{5}\gamma_{9} + \sin{\beta} \,\, \mathbf{1} \right)\left( \gamma_{4}\gamma_{6} -\gamma_{8}\gamma_{10} \right)  \Big\} \hspace{0.15cm} \varepsilon^{\pm}_{0} \, = \, 0.
\end{eqnarray}
The kappa symmetry condition can be computed to be
\begin{eqnarray}
\nonumber && \hspace{-0.75cm}  \mathcal{M}^{\pm}(\sigma^{a})^{-1} \,\, \Gamma \,\, \mathcal{M}^{\pm}(\sigma^{a}) \hspace{0.2cm} \varepsilon^{\pm}_{0}
= \frac{1}
{\left[ (\sin^{2}{\beta}\cos^{2}{\theta}_{+} + \cos^{2}{\beta}\cos^{2}{\theta_{-}}) - \cos^{2}{\theta_{+}} \, \cos^{2}{\theta_{-}} \right]}  \\  
\nonumber && \hspace{-0.75cm} \times \Big\{
  (\hat{\gamma}\gamma_{+}\gamma_{-}\gamma_{12}) 
\left( \sin^{2}{\beta} \cos^{2}{\theta_{+}} \sin^{2}{\theta_{-}}  \,\, \gamma_{1}\gamma_{2}\gamma_{8}\gamma_{10} 
+ \cos^{2}{\beta} \cos^{2}{\theta_{-}} \sin^{2}{\theta_{+}} \,\, \gamma_{1}\gamma_{2}\gamma_{4}\gamma_{6}\right) \\
\nonumber && \hspace{-0.75cm} \hspace{0.4cm} \pm \,
\left[
\cos{\beta} \cos{\theta_{-}} \sin{\theta_{+}} \,\, e^{-(\gamma_{4}\gamma_{6}) \, (\phi_{+} \mp \, \chi_{+})}\hspace{0.1cm}\gamma_{4} + \hspace{0.025cm}  
\sin{\beta} \cos{\theta_{+}}\sin{\theta_{-}} \,\, e^{-(\gamma_{8}\gamma_{10}) ( \phi_{-} \mp \, \chi_{-} )} \hspace{0.1cm} \gamma_{8}  \right] \\
\nonumber && \hspace{-0.75cm} \hspace{0.9cm} \times \, \cos{\theta_{+}} \cos{\theta_{-}} 
\hspace{0.1cm} \gamma_{4} \gamma_{5}\gamma_{9}\gamma_{6}\gamma_{10}\gamma_{8}\gamma_{12} \, \left( \cos{\beta} \,\, \gamma_{0}\gamma_{5} + \sin{\beta} \,\, \gamma_{0} \gamma_{9} - \dot{\chi} \, \mathbf{1} \right)  \\
\nonumber && \hspace{-0.75cm} \hspace{0.4cm} \pm \cos{\beta} \sin{\beta} \cos{\theta_{+}} \cos{\theta_{-}} \,\, 
e^{-(\gamma_{4}\gamma_{6}) \, ( \phi_{+} \mp \, \chi_{+})} \,\, e^{-(\gamma_{8}\gamma_{10}) ( \phi_{-} \mp \, \chi_{-} )} \,\,
\gamma_{0} \gamma_{4} \gamma_{5}\gamma_{9}\gamma_{10}\gamma_{12} \, \left(\gamma_{4}\gamma_{6}\gamma_{8}\gamma_{10} + \mathbf{1} \right) \Big\} \hspace{0.1cm} \varepsilon^{\pm}_{0} \\
\nonumber && \hspace{-0.75cm} \, = \, - \dot{\chi} \hspace{0.1cm} \varepsilon^{\pm}_{0}.
\end{eqnarray}
The kappa symmetry, dilatino and consistency conditions are satisfied if
\begin{eqnarray} 
\nonumber && \hspace{-0.2cm} \dot{\chi} \hspace{0.1cm} \gamma_{1}\gamma_{2} \hspace{0.1cm}\varepsilon^{\pm}_{0}  = \gamma_{4}\gamma_{6} \hspace{0.1cm} \varepsilon^{\pm}_{0} 
= \gamma_{8}\gamma_{10} \hspace{0.1cm} \varepsilon^{\pm}_{0} \hspace{1.2cm}   
\gamma_{0}  \left(\cos{\beta} \, \gamma_{5} + \sin{\beta \,  \gamma_{9}}\right)  \, \varepsilon^{\pm}_{0} 
= \dot{\chi} \hspace{0.1cm}  \varepsilon^{\pm}_{0}.   \hspace{1.0cm}
\end{eqnarray}
This D5-brane giant graviton in $\mathbb{R} \times S^{3}_{+}\times S^{3}_{-} \times S^{1}$ thus preserves 4 of the original 16 background supersymmetries and is therefore $\tfrac{1}{4}$-BPS.

%%%%%%%%%%%%%%%%%%%%%%%%%%%%%%%%%%%%%%%%%%%%%%%%%%%%%%%%%%%%%%%%%%%%%%%%%%%%%%%%%%%%%%%%%%%%%%

\end{document}